\documentclass[%
reprint,
superscriptaddress,
 amsmath,amssymb,
 aps,
]{revtex4-2}

\usepackage{graphicx}
\usepackage{dcolumn}
\usepackage{bm}

\usepackage{amsfonts}
\newcommand{\dr}[0]{\mathrm{d}^3\mathbf{r}~}
\newcommand{\rJ}[0]{\mathbf{(r \cdot J)}}
\newcommand{\rxJ}[0]{\mathbf{(r \times J)}}
\newcommand{\alp}[0]{\boldsymbol{\alpha}}
\newcommand{\apn}[2]{\alpha^{\mathrm{#1}}_{\mathrm{#2}}}
\newcommand{\apnm}[4]{(\apn{#1}{#3})^{#2}_{#4}}
\newcommand{\splitatcommas}[1]{%
    \begingroup
        \ifnum\mathcode`,="8000
        \else
            \begingroup\lccode`~=`, \lowercase{\endgroup
            \edef~{\mathchar\the\mathcode`, \penalty0 \noexpand\hspace{0pt plus 1em}}%
        }\mathcode`,="8000
        \fi
        #1%
    \endgroup
}
\newcommand{\dF}[2]{\frac{\mathop{\mathrm{d}#1}}{\mathop{\mathrm{d}#2}}}
\newcommand{\ddF}[2]{\frac{\mathop{\mathrm{d^2}#1}}{\mathop{\mathrm{d}#2}^2}}
\newcommand{\deF}[3]{\frac{\mathop{\mathrm{d^2}#1}}{\mathop{\mathrm{d}#2}\mathop{\mathrm{d}#3}}}
\newcommand{\inalign}[2]{\scriptsize\begin{aligned}[]#1&\\[-0.4em]#2&\end{aligned}}

\begin{document}

\preprint{APS/123-QED}

\title{Describing meta-atoms using the exact higher-order polarizability tensors}

\author{Jungho Mun}
\affiliation{%
    Department of Chemical Engineering, Pohang University of Science and Technology (POSTECH), 77 Chungam-ro, Republic of korea
}%
\author{Sunae So}
\affiliation{%
  Department of Mechanical Engineering, Pohang University of Science and Technology (POSTECH), 77 Chungam-ro, Republic of korea
}%
\author{Jaehyuck Jang}
\affiliation{%
    Department of Chemical Engineering, Pohang University of Science and Technology (POSTECH), 77 Chungam-ro, Republic of korea
}%
\author{Junsuk Rho}%
\email{jsrho@postech.ac.kr}
\affiliation{%
    Department of Chemical Engineering, Pohang University of Science and Technology (POSTECH), 77 Chungam-ro, Republic of korea
}%
\affiliation{%
  Department of Mechanical Engineering, Pohang University of Science and Technology (POSTECH), 77 Chungam-ro, Republic of korea
}%

\date{\today}

\begin{abstract}
    In nanophotonics, multipole framework has become an indispensable theoretical tool for analyzing subwavelength meta-atoms and their radiation properties. 
    This work presents higher-order exact dynamic polarizability (\(\alp\)) tensors, which can fully represent anisotropic meta-atoms with higher-order multipole transitions. 
    By using the irreducible exact Cartesian multipoles and field components as the basis, the exact \(\alp\)-tensor rigorously reflects symmetry information of particles including reciprocity. In addition, the exact \(\alp\)-tensor can be obtained from \(\mathbf{T}\)-matrix simply using basis transformation. 
    Finally, we show that description of meta-atoms using \(\alp\)-tensors incorporated with multiple-scattering theory vastly extends the applicability of the multipole framework in nanophotonics, allowing accurate and efficient depiction of complicated, random, multi-scale systems.
\begin{description}
\item[Usage]
Preprint.
\end{description}
\end{abstract}

\maketitle


\section{Introduction}
    Under the paradigm of metamaterials, their constituent meta-atoms and their configurations determine the material properties. The meta-atoms have been efficiently analyzed using the multipole decomposition technique \cite{Muhlig2011, Grahn2012}, because a few low-order multipole moments efficiently reconstruct the electromagnetic radiation and the relevant physics from a subwavelength localized current-charge source. Due to this feature, the multipole framework has become an useful and indispensable tool for nanophotonics \cite{Liu2017}. 
    Manipulation of light in the nanoscale has been facilitated by interference of multipole radiations, which provides the underlying principles behind many optical phenomena and relevant applications. Notably, the multipole framework has given insights on directional scattering \cite{Liu2018}, lattice Kerker effects \cite{Babicheva2017}, non-radiating anapoles \cite{Gurvitz2019, Baryshnikova2019}, lattice invisiblity effects \cite{Terekhov2019}, Fano-like resonances \cite{Gallinet2011a, Suryadharma2019}, optical anti-ferromagnetism \cite{Liu2017}, optical nonlinearity \cite{Smirnova2016}, radiative heat transfer, weak localization \cite{Mishchenko2008}, photonic topological insulators \cite{Pocock2018}, and bound states in the continuum \cite{Sadrieva2019}. 
    
    Formulation of multipole radiation is a textbook problem \cite{Jackson1999}, but given the importance of the multipole framework, expressions for multipoles are still under research \cite{Alaee2018, Fruhnert2017, Grahn2012} with possibility on toroidal multipoles as an extra multipole family \cite{Savinov2019, Gurvitz2019}.
    In general, the multipoles under discussion are excited at a specific illumination, but they do not provide complete information of highly anisotropic meta-atoms. On the other hand, dynamic polarizability \(\alp\) tensor (or transition \(\mathbf{T}\) matrix) maps the induced multipole modes at arbitrary incident fields, and has been used to treat scattering objects in many different fields including optics, acoustics, and astrophysics \cite{DeVries1998, Mishchenko2008, Mishchenko2010}. In nanophotonics, analysis of meta-atoms based on their \(\mathbf{T}\)-matrix started to become remarked rather recently \cite{Fruhnert2017, Suryadharma2017}.
    It has been pointed out that complicated coupled configurations involving multiple meta-atoms can be efficiently studied by describing the meta-atoms in terms of \(\alp\)-tensor (or \(\mathbf{T}\)-matrix) and using the multiple-scattering theory (MST) \cite{Fruhnert2017}. Electromagnetically coupled discrete scattering objects can be self-consistently treated to describe for collective responses of multiple particles \cite{DeAbajo1999, Stout2008, Stout2011} and periodic particle arrays \cite{DeAbajo2007, Baur2018, Evlyukhin2010, Babicheva2018, Babicheva2019, MahdiSalary2017, Watson2017}, and this framework may significantly reduce the calculation loads for complicated, random \cite{Rahimzadegan2019, Jenkins2018}, or multi-scale systems \cite{Pattelli2018, Govorov2010, Wu2015}.
    
    In the following work, we first discuss induced multipoles in different expressions: approximate Cartesian, exact Cartesian, and spherical multipoles, where exact Cartesian and spherical multipoles are essentially identical with different choice of basis \cite{Alaee2018, Alaee2019}. In the next section, we present expression of local fields and field gradients in terms of spherical multipoles. This naturally leads us to obtain transformation between \(\alp\)-tensor and \(\mathbf{T}\)-matrix. This basis transformation is used to analyze meta-atoms based on their \(\alp\)-tensors, whose properties can be intuitively interpreted due to the Cartesian basis. Finally, we show that the MST allows efficient and accurate description of electromagnetically coupled meta-atoms, and that analytic scattering objects can be implemented under the multipole framework. 

\section{Exact Cartesian multipoles and field components}
    
    In standard electrodynamic textbooks, the spherical multipoles appear from the multipole decomposition of electromagnetic fields using the vector spherical wave functions (VSWFs) as the basis \cite{Jackson1999}. Because the VSWFs span the vector fields satisfying the transverse Helmholtz type equations, the electromagnetic fields in a homogeneous media can be exactly reconstructed, and the renowned Mie theory is also based on this expansion. Because of the difficulty in interpreting the spherical basis, the spherical multipoles are not directly analyzed per se, but their associated scattering power or radiation fields are.
    
    Therefore, multipole framework in nanophotonics most frequently utilizes the approximate expressions for localized charge-current density multipoles in the Cartesian basis, which are sufficiently straightforward and resemble the expressions in electrostatics and magnetostatics. Although the approximate multipoles sometimes give better convergence \cite{Evlyukhin2019}, they cannot exactly reconstruct the electrodynamic radiation fields and the related scattering phenomena \cite{Alaee2018}. Scattering from subwavelength nanoparticles with moderate refractive-index generally shows good agreement, but the error grows for larger particles and high-refractive-index particles. 
    This error has been corrected by toroidal multipoles, which appear from multipole decomposition of the localized current sources \cite{Gurvitz2019, Talebi2018, Baryshnikova2019, Evlyukhin2016}. However, it has been pointed out that the radiation fields from toroidal multipoles do not have independent (orthogonal) basis to those from electric and magnetic multipoles \cite{Alaee2018}. Therefore, it is controversial whether to treat the toroidal multipoles as the third multipole family \cite{Savinov2019}, or as a correction to the basic Cartesian multipoles \cite{Fernandez-Corbaton2017}. 
    
    Recently, exact expressions for the localized charge-current density multipoles in the Cartesian basis up to MQ were developed without relying on the toroidal multipoles \cite{Alaee2018, Fernandez-Corbaton2015}; we present expressions of the exact Cartesian multipoles up to MO in the Methods section for completeness. 
    It is important to note that the exact Cartesian multipoles and the spherical multipoles have identical physical meaning \cite{Alaee2018}. They are just expressed in different basis and are transformed to each other as
    \begin{equation}
        \mathbf{v}^p_n = \frac{c_n E_0}{k^3} \bar{\bar{\mathrm{V}}}_n\mathbf{b}^p_n,
    \label{eqn:multipole}
    \end{equation}
    where \(\mathbf{v}^p_n\) is irreducible Cartesian multipole of order \(n\) with superscript \(p\) = \(\mathrm{e}\) or \(\mathrm{m}\) denoting electric or magnetic multipoles, respectively. \(\mathbf{b}^{p}_{n} = [b^{p}_{n,-n},b^{p}_{n,-n+1},\cdots,b^{p}_{n,n-1},b^{p}_{n,n}]^\top\) is a vector containing the spherical multipoles. \(c_1 = \sqrt{6 \pi}\), \(c_2 = \sqrt{20 \pi}\), \(c_3 = \sqrt{105 \pi /2}\) are normalization constants, and \(\bar{\bar{\mathrm{V}}}_n\) is basis transformation matrix given in the Methods section.
    The transformations between the spherical multipoles and reducible Cartesian multipoles for ED, MD, and EQ can be found in previous work \cite{Muhlig2011, Grahn2012}, although the expression for exact Cartesian multipoles has only been published recently \cite{Alaee2018, Fernandez-Corbaton2015, Alaee2019, Evlyukhin2019}. 
    It should be noted that \(\mathbf{b}^p_n\) is irreducible and has \(2n+1\) components, whereas basic Cartesian multipoles are symmetric and traceless \cite{Gurvitz2019}. 
    The redundancy due to symmetricalness can be removed noting that 
    \(Q^p_{\alpha\beta} = Q^p_{\beta\alpha}\) and \(O^p_{\alpha\beta\gamma}=O^p_{\alpha\gamma\beta}=O^p_{\beta\alpha\gamma}=O^p_{\beta\gamma\alpha}=O^p_{\gamma\alpha\beta}=O^p_{\gamma\beta\alpha}\), where \(Q^p\) and \(O^p\) are quadrupole and octupole, respectively, and \(\alpha, \beta, \gamma = x, y, z\).
    Excluding the symmetric components, quadrupole and octupole have 6 and 10 components, respectively. The extra redundancy can be removed using \(Q^p_{xx}+Q^p_{yy}+Q^p_{zz}=0\) and \(O^p_{\alpha xx}+O^p_{\alpha yy}+O^p_{\alpha zz}=0\).
    In this work, we use irreducible Cartesian multipoles \(\mathbf{v}^\mathrm{p}_n\) given in the Methods section.
    
    Analogously, the Cartesian local field components and the spherical multipoles are transformed to each other as
    \begin{equation}
        \mathbf{u}^p_n = \frac{i E_0}{c_n} \bar{\bar{\mathrm{U}}}_n\mathbf{a}^p_n,
    \label{eqn:field}
    \end{equation}
    where \(\mathbf{u}^p_n\) is the irreducible local Cartesian field components given in the Methods section, and \(\bar{\bar{\mathrm{U}}}_n = (\bar{\bar{\mathrm{V}}}_n^{-1})^\dag\) is the basis transformation matrix for the field components, where the superscript \(\dag\) denotes Hermitian conjugate. 

\section{Extended point polarizability and T-matrix}
    Many optical phenomena have been successfully resolved from induced multipoles, but they are given at a specific illumination and are generally not invariant under the interaction with other particles or under different external excitation fields. Strongly anisotropic meta-atoms exhibit different induced multipoles depending on its environment and excitation fields, so the induced moments obtained from a specific situation do not consistently represent the inherent properties of the meta-atoms.
    
    Hence, a different quantity is required to consistently describe an identical particle in an isolated state, nearby other particles, or in a lattice, and \(\alp\)-tensor serves for this purpose. \(\alp\)-tensor is defined as a response tensor linearly relating the local fields to the induced multipoles, so it is irrelevant of excitation conditions and allows us to calculate the induced moments at arbitrary incident fields. This feature allows us to calculate the collective responses of coupled particles using MST, which we discuss in later sections. However, \(\alp\)-tensor is usually truncated at dipole order \cite{Arango2013, Asadchy2014, Liu2016}, and higher-order \(\alp\)-tensor, which includes higher-order multipole moments and field gradients \cite{Arango2014}, is rarely utilized due to the complicated retrieval process if not for spheres with isotropic responses \cite{Babicheva2019}.
    
    These higher-order multipole transitions have been systematically treated using \(\mathbf{T}\)-matrix, which linearly relates the spherical multipoles of incident field to those of scattered field as: \(\mathbf{b}^{p'}_{n'} = \mathbf{T}^{np}_{n'p'} \cdot \mathbf{a}^{p}_{n}\), where $\mathbf{T}^{np}_{n'p'}$ is a $(2n'+1) \times (2n+1)$ matrix corresponding to the transition from multipole order $n$ of mode $p$ to $n'$ of $p'$, and $\mathbf{a}^p$ and $\mathbf{b}^p$ are vectors containing spherical multipoles of incident and scattered fields, respectively \cite{Mishchenko1996}. The combined \(\mathbf{T}\)-matrix is expressed as
    \begin{subequations}
    \begin{equation}
        \begin{bmatrix}
        \mathbf{b}^\mathrm{e} \\ \mathbf{b}^\mathrm{m}
        \end{bmatrix}
        =\begin{bmatrix}
        \mathbf{T}^\mathrm{e}_\mathrm{e} & \mathbf{T}^\mathrm{m}_\mathrm{e} \\
        \mathbf{T}^\mathrm{e}_\mathrm{m} & \mathbf{T}^\mathrm{m}_\mathrm{m}
        \end{bmatrix}
        \begin{bmatrix}
        \mathbf{a}^\mathrm{e} \\ \mathbf{a}^\mathrm{m}
        \end{bmatrix}
    \end{equation}
    \begin{equation}
        \mathbf{T}^{p}_{p'}
        =\begin{bmatrix}
        \mathbf{T}^{p1}_{p'1} & \mathbf{T}^{p1}_{p'2} & \cdots \\
        \mathbf{T}^{p2}_{p'1} & \mathbf{T}^{p2}_{p'2} &  \\
        \vdots & & \ddots
        \end{bmatrix}
    \end{equation}
    \end{subequations}
    
    In this work, we define \(\alp\)-tensor as: \(\mathbf{v}^{p'}_{n'} = \alpha^{np}_{n'p'} \cdot \mathbf{u}^{p}_{n}\), where $\alpha^{np}_{n'p'}$ as a $(2n'+1) \times (2n+1)$ matrix corresponding to the transition from \(\mathbf{u}^{p}_{n}\) to \(\mathbf{v}^{p'}_{n'}\). The combined \(\alp\)-tensor is expressed as
    \begin{subequations}
    \begin{equation}
        \begin{bmatrix}
        \mathbf{v}^\mathrm{e} \\ \mathbf{v}^\mathrm{m}
        \end{bmatrix}
        =\begin{bmatrix}
        \alp^\mathrm{e}_\mathrm{e} & \alp^\mathrm{m}_\mathrm{e} \\
        \alp^\mathrm{e}_\mathrm{m} & \alp^\mathrm{m}_\mathrm{m}
        \end{bmatrix}
        \begin{bmatrix}
        \mathbf{u}^\mathrm{e} \\ \mathbf{u}^\mathrm{m}
        \end{bmatrix}
    \end{equation}
    \begin{equation}
        \alp^{p}_{p'}
        =\begin{bmatrix}
        \alp^{p1}_{p'1} & \alp^{p1}_{p'2} & \cdots \\
        \alp^{p2}_{p'1} & \alp^{p2}_{p'2} &  \\
        \vdots & & \ddots
        \end{bmatrix}
    \end{equation}
    \end{subequations}
    
    Analogous to the induced multipoles, \(\alp\)-tensor and \(\mathbf{T}\)-matrix are identical but with different choice of basis. 
    Using the basis transformations (Eqns.~\ref{eqn:field} and \ref{eqn:multipole}), \(\alp\)-tensor is obtained from \(\mathbf{T}\)-matrix as 
    \begin{equation}
        \alpha^{pn}_{p'n'} = \frac{c_nc_{n'}}{ik^3} \bar{\bar{\mathrm{V}}}_{n'} \mathbf{T}^{pn}_{p'n'} \bar{\bar{\mathrm{U}}}_{n}^{-1}
        \label{eqn:T2a}
    \end{equation}
    Note that the expressions for isotropic (scalar) dipolar \cite{Arango2013, Liu2016} and quadrupolar objects \cite{Babicheva2019} are found similar.
    \(\mathbf{T}\)-matrix is inversely obtained from \(\alp\)-tensor as
    \begin{equation}
        \mathbf{T}^{pn}_{p'n'} = \frac{ik^3}{c_nc_{n'}} \bar{\bar{\mathrm{V}}}_{n'}^{-1} \alpha^{pn}_{p'n'} \bar{\bar{\mathrm{U}}}_{n}
        \label{eqn:a2T}
    \end{equation}
    
    \(\alp\)-tensor and \(\mathbf{T}\)-matrix contains information on particle symmetries and conservation laws, as well as complete information on scattering by a particle. A \(\alp\)-tensor based on \emph{exact} multipoles rigorously reflects several symmetries; that is, \(\alp\)-tensor retrieved using approximate multipoles does not rigorously satisfy the symmetries. 
    First, \(\alp\)-tensor of Onsager reciprocal particles satisfy 
    \begin{equation}
        \alpha^{pn}_{p'n'} = (\alpha^{p'n'}_{pn})^\top
    \end{equation}
    This expression for dipolar particles has been known \cite{Sersic2011}, but extension to higher-order requires suitable choice of normalization and irreducible basis, which are introduced in the Methods section of this work. \(\mathbf{T}\)-matrix elements of Onsager reciprocal particles satisfy the following relationship:
    $(T^{pn}_{p'n'})^{m}_{m'}=(-1)^{m+m'}(T^{p'n'}_{pn})^{-m'}_{-m}$. 
    This expression has been used to check the accuracy of numerically calculated \(\mathbf{T}\)-matrix in literature \cite{Mishchenko1996}. 
    A lossless particle has no intrinsic absorption, so its extinction equals to scattering, where the extinction and scattering power from a particle are given as  
    \begin{subequations}
    \begin{equation}
        P_\mathrm{ext} = \frac{k}{2\eta}\sum_p\sum_n\mathrm{Im}[(\mathbf{u}^p_n)^\dag\mathbf{v}^p_n]
    \end{equation}
    \begin{equation}
        P_\mathrm{sca} = \frac{k^4}{2\eta}\sum_p\sum_n\frac{1}{c_n^2}[(\mathbf{v}^p_n)^\dag\bar{\bar{\mathrm{U}}}_n\bar{\bar{\mathrm{U}}}_n^\dag\mathbf{v}^p_n]
    \end{equation}
    \label{eqn:power}
    \end{subequations}
    \(\alp\)-tensor of a lossless dipolar particle satisfies: 
    $\frac{k^3}{6\pi} \alpha^\dag \alpha = \frac{1}{2i}(\alpha^\dag - \alpha)$,
    which reduces to the optical theorem 
    $\frac{k^3}{6\pi}|\alpha|^2 = \mathrm{Im}(\alpha)$ 
    for dipolar scalar $\alpha$ \cite{Sersic2011}.
    \(\mathbf{T}\)-matrix of a lossless particle satisfies: 
    $\mathbf{T}^{\dag}\mathbf{T}=-\frac{1}{2}(\mathbf{T}^{\dag}+\mathbf{T})$ \cite{Waterman1971}. This expression has also been used to check the accuracy of \(\mathbf{T}\)-matrix for lossless particles in literature \cite{Mishchenko1996}. 
    
    Some geometric symmetries visually appear in \(\alp\)-tensors.
    The parity operation is given as $\mathbf{r}\rightarrow-\mathbf{r}$. Upon the parity operation, $\mathbf{E}\rightarrow-\mathbf{E}$, $\mathbf{H}\rightarrow\mathbf{H}$, $\nabla\rightarrow-\nabla$. Also, electric multipoles and field components ($\mathbf{u}^\mathrm{e}$ and $\mathbf{v}^\mathrm{e}$) have parity of $(-1)^n$, and the magnetic counterparts ($\mathbf{u}^\mathrm{m}$ and $\mathbf{v}^\mathrm{m}$) have $(-1)^{n-1}$. From which, we see that $\alpha^{\mathrm{e}n}_{\mathrm{e}n'}$ and $\alpha^{\mathrm{m}n}_{\mathrm{m}n'}$ have parity of $(-1)^{n+n'}$, and $\alpha^{\mathrm{e}n}_{\mathrm{m}n'}$ and $\alpha^{\mathrm{m}n}_{\mathrm{e}n'}$ has $(-1)^{n+n'+1}$. The corresponding \(\mathbf{T}\)-matrix components have the same parity.
    Also, parity is closely related to the concept of true chirality, or reciprocal parity-odd \cite{Barron1986}. Note that chirality of chiral molecules has been embedded in the magneto-electric coupling term \cite{Govorov2010}, which attributes to the reciprocal parity-odd property. By breaking the reciprocity, it is possible to undergo false chirality, which is nonreciprocal, parity-odd \cite{Barron1986, Asadchy2014, Asadchy2019}.
    Other kinds of symmetries including rotational symmetry, \(N\)-fold rotational symmetry, mirror symmetry, and time-reversal symmetry can be confirmed by checking if \(\mathbf{T}\)-matrix or \(\alp\)-tensor is invariant under the transformations.
    
\section{Meta-atoms and metaphotonics}
    In metamaterials and metaphotonics, manipulation of light at the nanoscale utilizes optically resonant subwavelength meta-atoms, whose properties have generally been analyzed by dipolar \(\alp\)-tensor \cite{Arango2013, Asadchy2014, Liu2016}. 
    However, recently emerged high-refractive-index particles \cite{Kuznetsov2012, Evlyukhin2010, Babicheva2017, Babicheva2018, Terekhov2019} and coupled plasmonic systems \cite{Fruhnert2017, Arango2014} often involve higher-order multipole transitions.
    It should be noted that \(\mathbf{T}\)-matrix of meta-atoms can be systematically retrieved for arbitrary multipole order \cite{Fruhnert2017}, but detailed analysis on their properties from their \(\mathbf{T}\)-matrix is difficult due to the spherical basis, while retrieval of higher-order \(\alp\)-tensor can be cumbersome \cite{Arango2014}.
    In this section, we analyze several meta-atoms from their \(\alp\)-tensors, which are transformed from \(\mathbf{T}\)-matrix using Eqn.~\ref{eqn:T2a}. We will show that higher-order \(\alp\)-tensor are necessary to describe anisotropic meta-atoms, whose properties can be more intuitively analyzed due to the Cartesian basis. In addition, \(\alp\)-tensor allows analysis on several particle properties including anisotropy, symmetries, spectral modal resonances, and origin of chirality and optical magnetism.

\begin{figure}[t]
        \centering
        \includegraphics{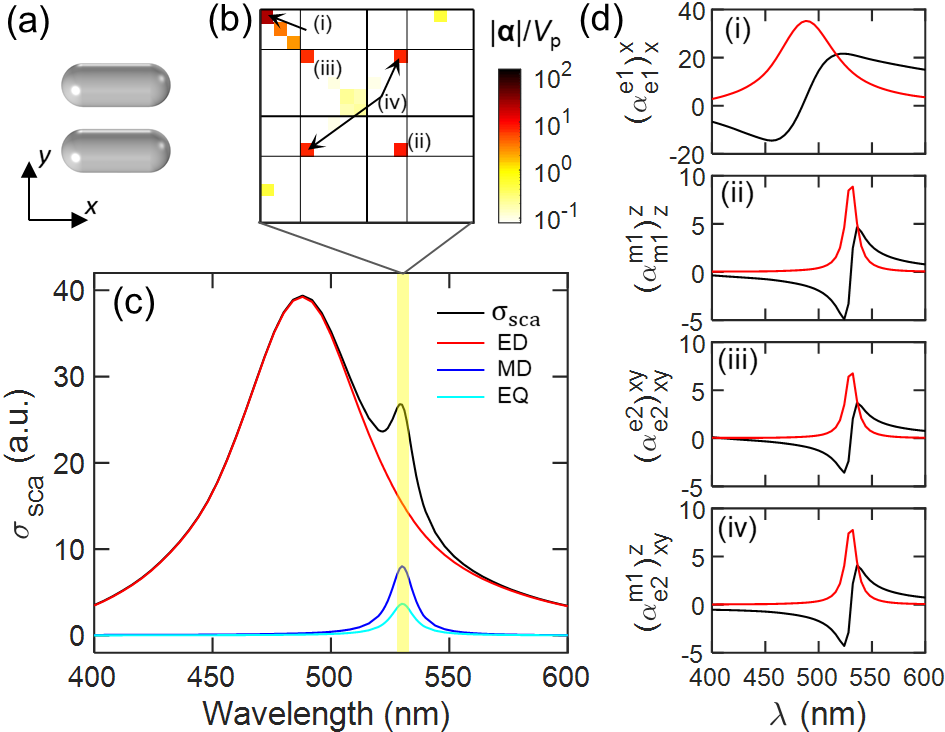}
        \caption{Plasmonic double-bars. (a) Schematics. (b) Retrieved \(\alp\)-tensor at \(\lambda=530~\text{nm}\). (c) Multipole-decomposed scattering cross-section at x-polarized planewave incidence propagating in y-direction. (d) Spectra of \(\apnm{e1}{x}{e1}{x}\), \(\apnm{m1}{z}{m1}{z}\), \(\apnm{e2}{xy}{e2}{xy}\), and \(\apnm{m1}{z}{e2}{xy}\). Geometrical parameters are: radius 20~nm, length 100~nm, and gap distance 20~nm.}
        \label{fig:PDB}
    \end{figure}

    Hybridized plasmonic structures can exhibit higher-order multipole modes even with subwavelength feature sizes. Among them, plasmonic double bars (PDB) has exhibited strong MD mode even at the visible regime \cite{Dolling2005}. In principle, higher-order \(\alp\)-tensors can be directly retrieved \cite{Arango2014}, but we obtained them by basis transformation (Eq.~\ref{eqn:T2a}) using the retrieved \(\mathbf{T}\)-matrix \cite{Fruhnert2017}. 
    Multipole-decomposed scattering cross-section exhibits broad ED resonance at 490~nm and sharp EQ and MD resonances at 530~nm (Fig.~\ref{fig:PDB}c), and the origins of the multipolar modes can be analyzed from retrieved \(\alp\)-tensor (Fig.~\ref{fig:PDB}b) and spectra of its components (Fig.~\ref{fig:PDB}d).
    
    The broad ED resonance is easily attributed to \(\apnm{e1}{x}{e1}{x}\), and the sharp EQ and MD resonances are spectrally attributed to \(\apnm{m1}{z}{m1}{z}\), \(\apnm{e2}{xy}{e2}{xy}\), and \(\apnm{m1}{z}{e2}{xy}\).
    Analysis from \(\alp\)-tensor allows us to see that PDB has different origin of optical magnetism from split-ring resonators (\(\apn{e1}{m1}\)) \cite{Arango2013, Liu2016} or high-refractive-index spheres (\(\apn{m1}{m1}\)) (see Appendix~\ref{sec:tmatrix} for more details).
    Importantly, only a few components are dominant in the retrieved \(\alp\)-tensor, making the analysis easier (Fig.~\ref{fig:PDB}b). This simplification partly comes from the particle symmetry. Notably, the parity symmetry removes the half of the components: \(\apn{e2}{e1}\), \(\apn{e1}{e2}\), \(\apn{m1}{e1}\), \(\apn{e1}{m1}\), \(\apn{m2}{e2}\), \(\apn{e2}{m2}\), \(\apn{m2}{m1}\), and \(\apn{m1}{m2}\). 
    The reciprocity rigorously enforces \(\apnm{e2}{xy}{m1}{z}\) = \(\apnm{m1}{z}{e2}{xy}\). 
    Because the meta-atom is strong anisotropic, its \(\alp\)-tensor is simplified in the Cartesian basis.
    Interestingly, \(\apnm{m1}{z}{m1}{z}\), \(\apnm{e2}{xy}{e2}{xy}\), and \(\apnm{e2}{xy}{m1}{z}\) have very similar spectral feature resembling Lorentzian resonances, indicating that some components may additionally be coupled together possibly using singular value decomposition technique \cite{Suryadharma2019} or modular analysis \cite{Asadchy2019}.
    
    Plasmonic chiral particles have exhibited chiral responses far-exceeding those from natural materials. Among them, a twisted double-bars (TDB) has been widely used to generate exceptionally strong chiral responses \cite{Auguie2011}, which we assess from circular dichroism (CD).
    
    \begin{figure}[t]
        \centering
        \includegraphics{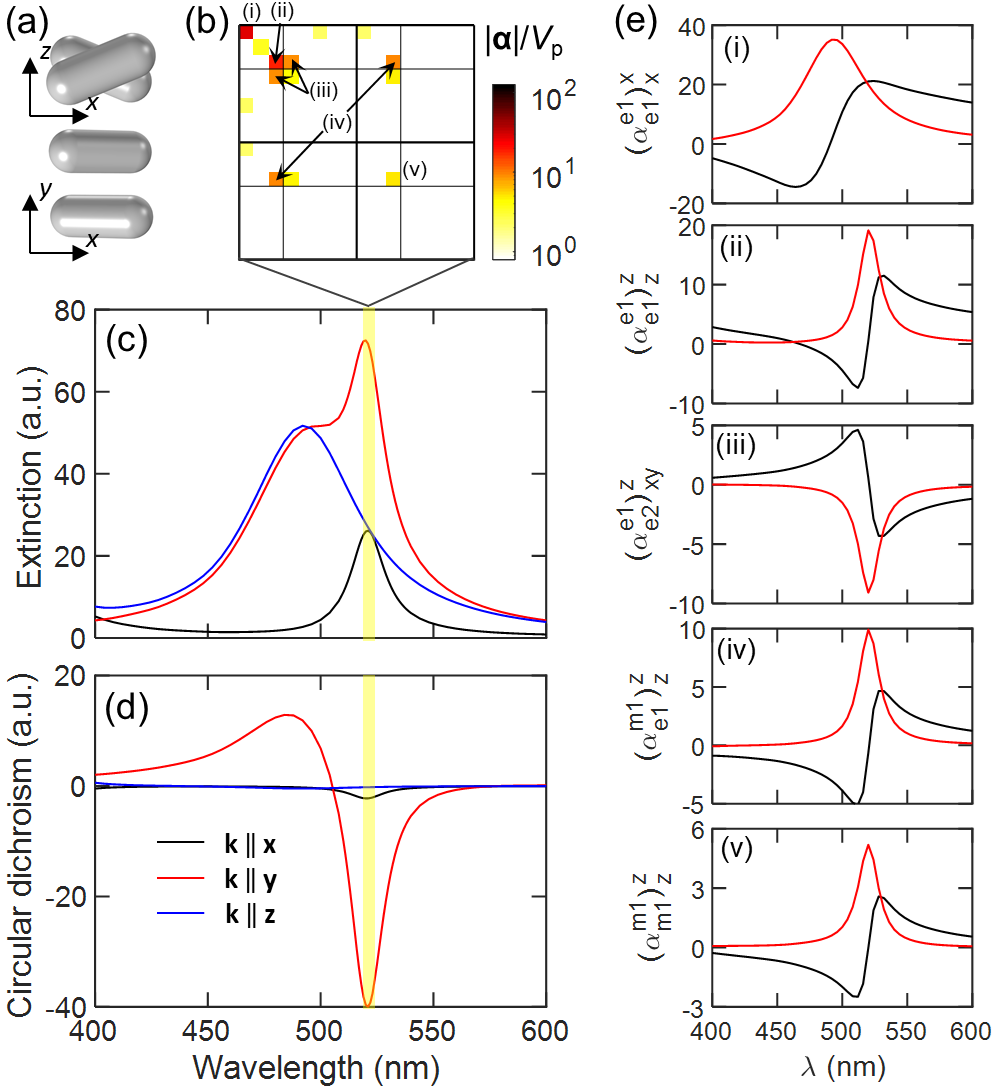}
        \caption{(a) The schematics of twisted double-bars. (b) Retrieved \(\alp\)-tensor at $\lambda=520~\text{nm}$. (c) Extinction and (d) circular dichroism at planewave incidence propagating in \(x\)-, \(y\)-, and \(z\)-directions. (e) Spectra of \(\apnm{e1}{x}{e1}{x}\), \(\apnm{e1}{z}{e1}{z}\), \(\apnm{e1}{z}{e2}{xy}\), \(\apnm{e1}{z}{m1}{z}\), and \(\apnm{m1}{z}{m1}{z}\). Geometrical parameters are: radius 20~nm, length 100~nm, gap distance 20~nm, and twist angle 45\(^\circ\).}
        \label{fig:TDB}
    \end{figure}
    
    Different from PDB, TDB is geometrically chiral, and therefore parity-odd, and \(\apn{e1}{m1}\), \(\apn{m1}{e1}\), \(\apn{e2}{e1}\), and \(\apn{e1}{e2}\) transition components are now allowed. 
    Again, \(\apnm{e1}{z}{e2}{xy} = \apnm{e2}{xy}{e1}{z}\) and \(\apnm{e1}{z}{m1}{z} = \apnm{m1}{z}{e1}{z}\) are rigorously enforced due to reciprocity, so we will only mention one of each.
    
    An important property of TDB is strongly anisotropic CD, which is visible for the light propagation parallel to the twist-axis (\(\hat{y}\)), but not for \(\mathbf{k \parallel x}\) nor \(\mathbf{k \parallel z}\). Interestingly, this anisotropic CD cannot be explained by dipolar \(\alp\)-tensor alone, because \(\apnm{e1}{z}{m1}{z}\) contributes to chiral response for both \(\mathbf{k \parallel x}\) and \(\mathbf{k \parallel y}\). 
    For this anisotropic CD, \(\apnm{e1}{z}{e2}{xy}\) is essential; they constructively contribute to CD for \(\mathbf{k \parallel y}\) (red line, Fig.~\ref{fig:TDB}d), whereas destructively for \(\mathbf{k \parallel x}\) (black line, Fig.~\ref{fig:TDB}d). Noting Eq.~\ref{eqn:power}a, it is convienient to see that \(\apnm{e1}{z}{e2}{xy} \approx -\apnm{e1}{z}{m1}{z}\) eliminating the CD response. Similarly, \(\apnm{e1}{x}{m1}{x}\) and \(\apnm{e1}{x}{e2}{yz}\) are responsible for smaller CD response around 480~nm, resulting asymmetric CD response (red line, Fig.~\ref{fig:TDB}d).
    TDB clearly shows that higher-order multipole transition is necessary for describing plasmonic meta-atoms and reconstructing ansotropic chiral responses even in far-fields, as well as in near-fields \cite{Mun2019}.

\section{Multiple-scattering theory and electromagnetically coupled systems}
    In the previous section, we have shown that higher-order \(\alp\)-tensor is necessary for analyzing isolated meta-atoms and interpreting their interaction with light. Additionally, \(\alp\)-tensor can be used to model interacting meta-atoms for further research. In literature, self-consistent coupled multipole equations have been formulated using the Green's tensor to illustrate periodic 2D arrays of plasmonic \cite{DeAbajo2007, Auguie2008, Swiecicki2017, Babicheva2018} and dielectric \cite{Evlyukhin2010, Babicheva2019} spheres, and finite \cite{Martikainen2017, Draine1994} or random \cite{Watson2017} systems. 
    In this framework, electromagnetic interactions between scattering objects are taken into account without any approximation, while single scattering objects are described by \(\alp\)-tensors approximated to low-order multipole orders. This framework is simplified version of superposition \(\mathbf{T}\)-matrix method (STMM), which has been extensively studied for coupled spheres \cite{DeAbajo1999, Stout2008, Stout2011}. 
    Earlier studies have usually incorporated small spheres, whose \(\alp\) can be easily obtained using the quasistatic approximation (see Appendix~\ref{sec:quasistatic} for more details), to study their interaction with light and with nearby scattering objects or molecules \cite{Govorov2010, Wu2015}. 
    However, meta-atoms with complicated multipolar transitions can also be modelled into \(\alp\)-tensors, which are then inserted into the MST \cite{Watson2017}, potentially allowing studies on more complicated physics, e.g., Fano resonances \cite{Gallinet2011a, Suryadharma2019} and hybridization of particle and lattice resonances in 2D \cite{MahdiSalary2017, Baur2018, Terekhov2019, Kwadrin2014} and 3D \cite{Liu2008, Kim2017} arrays.
    In this section, we reconstruct several physical phenomena arising in electromagnetically coupled meta-atoms simply by implementing \(\alp\)-tensors into the MST, and discuss the advantages of this method. See Appendix~\ref{sec:Green} for expressions of Green's tensors up to MO and their symmetries due to our choice of irreducible basis.

    \begin{figure}[t]
        \centering
        \includegraphics{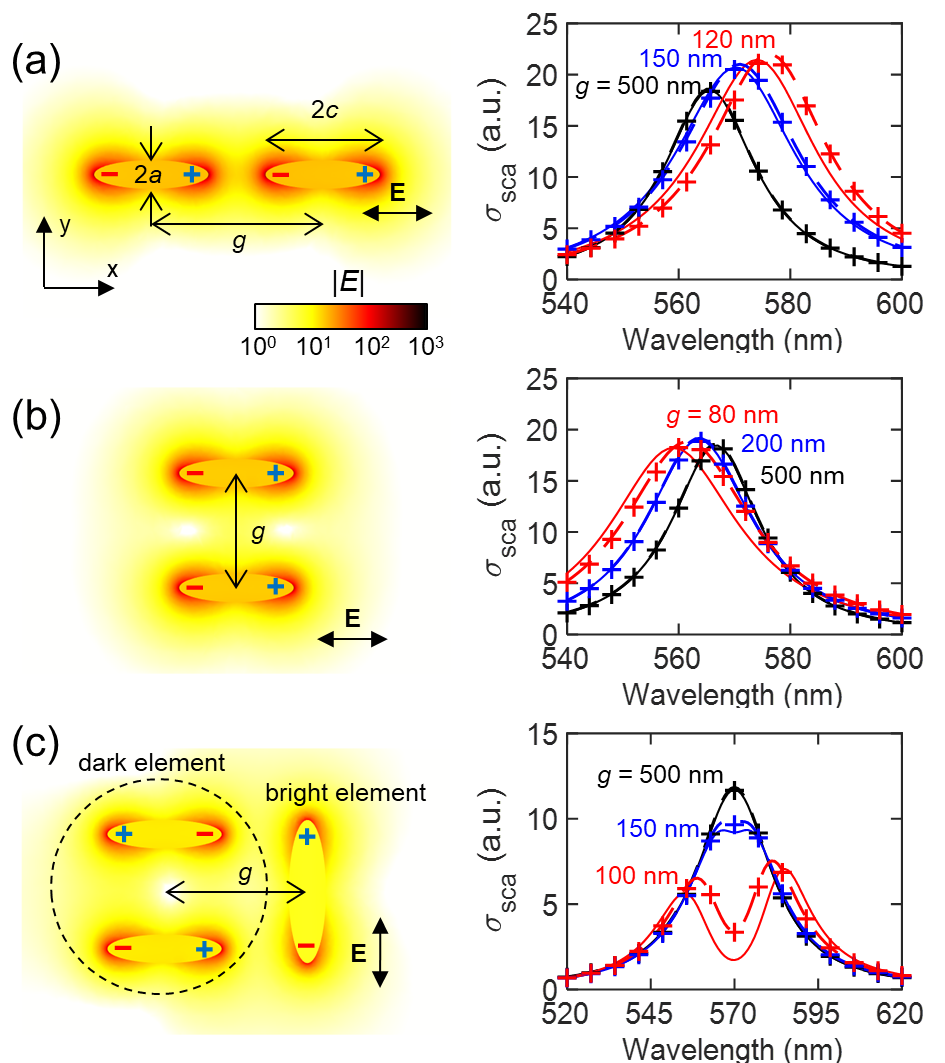}
        \caption{Reconstructed optically coupled systems using \(\alp\)-tensors and MST. (a,b) Plasmon-coupling between two coupled nanorods at different coupling configurations, and (c) Fano resonance between dark and bright elements. The nanorods in (a,b) are approximated as \(\alp\)-tensor with only \(\apnm{e1}{x}{e1}{x}\); the bright element in (c), as \(\apnm{e1}{y}{e1}{y}\); the dark element in (c), as \(\apnm{e1}{x}{e1}{x}\), \(\apnm{m1}{z}{m1}{z}\), \(\apnm{m1}{z}{e2}{xy}\), \(\apnm{e2}{xy}{m1}{z}\), and \(\apnm{e2}{xy}{e2}{xy}\) (see Fig.~\ref{fig:PDB}b). The configurations are illustrated in the schematics on the left sides. The scattering cross-sections calculated at different center-to-center distances \(g\) are on the right sides. Dashed-lines and cross-marks are the reference solutions calculated using STMM and FEM, respectively. The nanorods in (a,b) and the dark element in (c) have \(a\) = 10~nm and \(c\) = 40~nm, and the bright element in (c) has \(a\) = 13~nm and \(c\) = 50~nm.}
        \label{fig:toymodel}
    \end{figure}
    
    First, electromagnetically coupled two plasmonic nanorods are illustrated using MST (Fig.~\ref{fig:toymodel}a and b), where only \(\apnm{e1}{x}{e1}{x}\) component is considered in their \(\alp\)-tensors. Such closely situated plasmonic particles are strongly coupled, resulting strong spectral resonance shift. 
    Spectral red-shift (Fig.~\ref{fig:toymodel}a) and blue-shift (Fig.~\ref{fig:toymodel}b) can be reconstructed depending on the configuration of the coupled nanorods. This phenomenon can be intuitively interpreted by plasmon hybridization theory (PHT) \cite{Nordlander2004}; the induced charge density configuration in Fig.~\ref{fig:toymodel}a becomes stable by the hybridization redshifting the resonance, whereas the configuration in Fig.~\ref{fig:toymodel}b becomes unstable blue-shifting the resonance. However, PHT is based on the quasistatic approximation, so quantitative analysis is difficult for large, complicated systems. Another widely used theoretical framework for interpreting coupled optical systems is the coupled mode theory (CMT), which approximates the scattering objects as harmonic oscillators that are coupled to each other. However, CMT relies on fitting procedure to retrieve the relevant parameters and requires experimental or simulated results to begin with, so the CMT cannot be used to provide new information. In addition, it is of question whether the fitted parameters from the simple coupled harmonic oscillators can reliably represent the vectorial nature of electromagnetic coupling.
    Another widely studied phenomenon arising in electromagnetically coupled systems is Fano-like resonance, where a dark element is coupled to a bright element. The dark element cannot be directly excited by the incident field, but the coupling between the dark mode and the bright mode allows the dark mode to be indirectly excited. To reconstruct this phenomenon, we mimicked the dolmen configuration \cite{Gallinet2011a} using a dark element with two horizontal nanorods and a bright element with a vertical nanorod (Fig.~\ref{fig:toymodel}c). 
    Only \(\apnm{e1}{y}{e1}{y}\) component is considered for the bright element, and \(\apnm{e1}{x}{e1}{x}\), \(\apnm{m1}{z}{m1}{z}\), \(\apnm{e2}{xy}{m1}{z}\), \(\apnm{m1}{z}{e2}{xy}\), and \(\apnm{e2}{xy}{e2}{xy}\) components are considered for the dark element as Fig.~\ref{fig:PDB}. 
    The calculated scattering cross-section shows a dip near 570~nm, where the dark element has resonance, and this dip grows larger as \(g\) becomes smaller due to the stronger coupling between the dark and bright elements. 
    
    The spectra calculated using multipole methods using the truncated \(\alp\)-tensors (solid lines) show excellent quantitative agreement with the reference (dashed-lines and cross-marks), but the error grows larger as \(g\) decreases (Fig.~\ref{fig:toymodel}). This is because multipoles are efficient in describing long-range interactions but not in describing strong coupling between plasmonic particles in near-field \cite{Park2014}, which requires an increasingly large number of multipole order for accurate description \cite{Stout2008}. 
    Still, the collective responses between plasmonic particles situated in a reasonably far distance and dielectric particles \cite{DeAbajo1999} can be efficiently and accurately described under the multipole framework.
    
    Importantly, the multipole framework has superior computational efficiency compared to the traditional numerical methods, such as finite-difference time-domain and finite-element methods. Noticeably, this framework has shown significant potentials for rigorously studying electromagnetic problems involving disordered, aperiodic \cite{Rahimzadegan2019, Jenkins2018, Pinheiro2017}, and multi-scale systems with a large number of particles (\(N\)\textgreater10,000) over a large volume \cite{Pattelli2018}. 
    Especially, aperiodic metasurfaces \cite{MahdiSalary2017, Rahimzadegan2019, Jenkins2018} and random media \cite{Pinheiro2017} could be accurately studied using this method, and optimization \cite{Forestiere2012} and dataset construction for deep-learning neural networks \cite{So2019} would significantly benefit from this framework. 
    In addition, the multipole framework can implement localized shaped beams from simple Gaussian beams \cite{Novotny2009} to highly focused \cite{AlvaroRanhaNeves2006} and helical beams \cite{Wu2015, Wang2017}, providing a versatile framework to rigorously study spin-orbit interactions. Therefore, the multipole framework could be used to rigorously study electromagnetic phenomena arising in complex, disordered media consisting of discrete scattering objects.

\section{Toy models}

    \begin{figure}[t]
        \centering
        \includegraphics{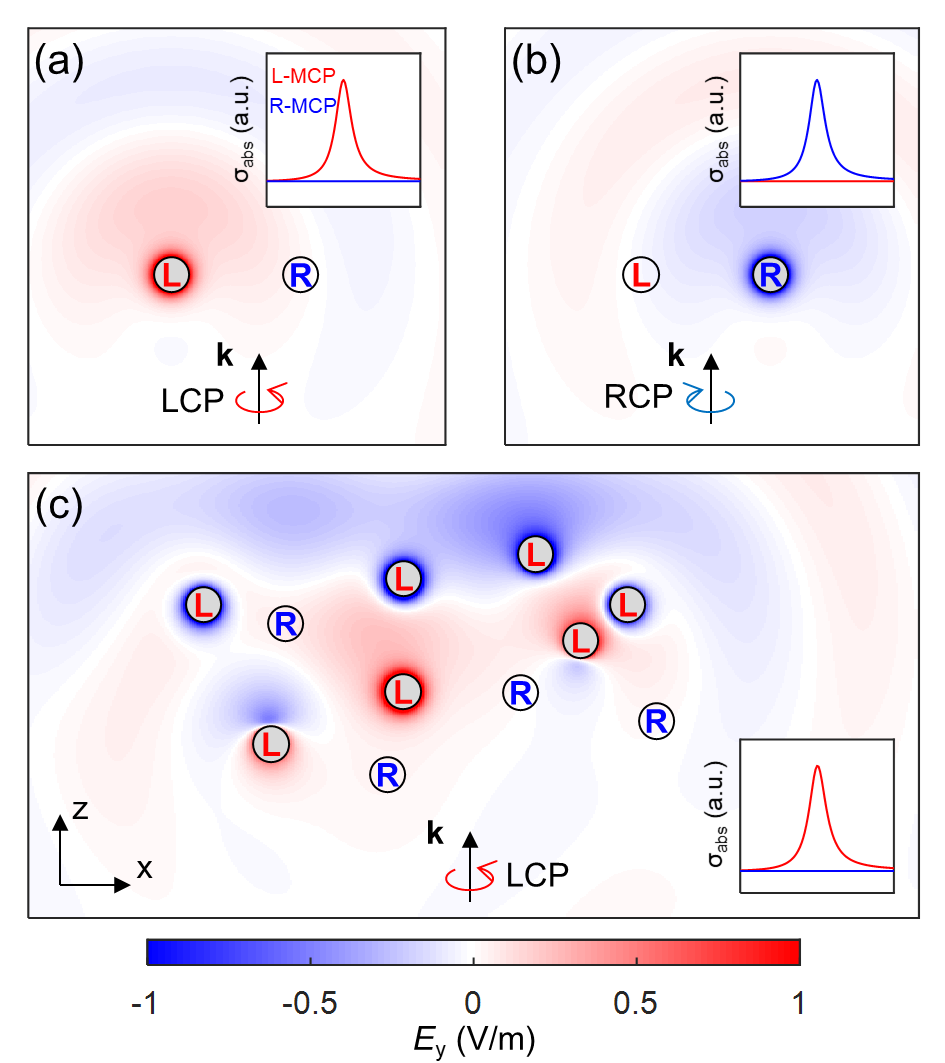}
        \caption{(a,b) Completely decoupled dipolar MCPs at (a) LCP and (b) RCP incidences. Inset spectra are absorption by L-MCP (red) and R-MCP (blue). (c) Randomly dispersed dual particles at LCP incidence. Inset spectra are the sums of absorption by L-MCP (red) and R-MCP (blue). The positions of L-MCPs and R-MCPs are denoted by circled letters L and R, respectively.}
        \label{fig:dual}
    \end{figure}
    In previous examples, \(\alp\)-tensors of realistic particles were considered, but it is also possible to consider \(\alp\)-tensors of arbitrary particles without information on their physical geometric parameters. Recently, Fernandez-Corbaton, et al. proposed the concept of maximally-chiral particles (MCPs), which are excited by light with one helicity, re-radiates light with the same helicity, and are completely transparent to light with the opposite helicity. They also proposed that two opposite MCPs are completely uncoupled to each other, and a media consisting of MCPs with a single handedness will be completely transparent to one helicity and opaque to the other helicity \cite{Fernandez-Corbaton2016}. Demonstration of this concept is difficult, because purely MCPs have not been discovered, although approximately MCPs have been studied \cite{Fernandez-Corbaton2016, Fruhnert2017}. 
    Still, we can theoretically investigate the concepts above using \(\alp\)-tensors and MST, because arbitrary \(\alp\)-tensors without physical parameters can be devised, which cannot be done using the traditional numerical methods. Dual particles at dipole approximation have \(\apn{e1}{e1} = \apn{m1}{m1} = \pm\apn{e1}{m1} = \pm\apn{m1}{e1}\), where \(\pm\) corresponds to left- (L-MCP) and right-MCPs (R-MCP), which interact with left- (LCP) and right-circularly-polarized lights (RCP), respectively. 

    In this section, we implemented MCPs using isotropic (scalar), dipolar \(\alp\)-tensors with Lorentzian resonance. We confirmed that two opposite MCPs are completely uncoupled with each other, and only L-MCP is excited at LCP incidence (Fig.~\ref{fig:dual}a), and R-MCP is excited at RCP incidence (Fig.~\ref{fig:dual}b). This is because L-MCP is excited by LCP and re-radiates LCP, which cannot excite R-MCP. In a mixture of L-MCPs and R-MCPs, R-MCPs are completely transparent upon LCP incidence, and L-MCPs are electromagnetically coupled.

    This section shows \(\alp\)-tensors as a powerful method to treat symmetries and conservation laws at the microscopic level (single scattering object) without dealing with the geometrical parameters.
    In fact, implementation of \(\alp\)-tensors without their relevant geometrical structures has been performed in literature to treat realistic molecules at weak excitations \cite{Novotny2009}, where the molecules were approximated as \(\alp\)-tensors and implemented in the framework of MST to study plasmon-enhanced circular dichroism \cite{Govorov2010} and helical dichroism \cite{Wu2015}.

\section{Conclusions}
    In summary, we have introduced the systematic transformation between \(\alp\)-tensor in the irreducible Cartesian basis and \(\mathbf{T}\)-matrix in the spherical basis using the basis transformation between the exact Cartesian and the spherical multipoles (Eqn.~\ref{eqn:multipole}) and between the local field components and the spherical multipoles (Eqn.~\ref{eqn:field}). 
    In general, characterization of meta-atoms using \(\alp\)-tensor has been limited to dipolar regime, but recent advances in nanophotonics and metamaterials utilize higher-order multipole transitions coming from coupled plasmonic and high-index dielectric nanoantennas. 
    The exact higher-order \(\alp\)-tensors introduced in this work can rigorously describe these scattering systems with higher-order multipole transitions retaining their symmetry information, and we present how to obtain exact \(\alp\)-tensors from \(\mathbf{T}\)-matrix using the facile basis transformation .
    Incorporated with the MST, the multipole framework can be a versatile theoretical framework in nanophotonics to rigorously investigate optical phenomena arising in coherently coupled multi-body systems \cite{Mishchenko2008}.
    In this framework, the well-defined symmetries and conservation laws can be treated at the microscopic level (single scattering object) using \(\alp\)-tensors, and the electromagnetic interaction between them are treated using the Green's tensors presented in Appendix~\ref{sec:Green}. 
    It is worthwhile to note that the potential applicability of \(\alp\)-tensor (or \(\mathbf{T}\)-matrix) into the MST has been mentioned in many previous papers, but only spherical particles have been generally considered. By simply taking nonspherical structured meta-atoms into account, the multipole framework can be extended into many different applications. 

    The multipole framework is especially advantageous for complicated, random, multi-scale problems due to computational efficiency, and analytic scattering objects, such as realistic molecules and dual particles, can also be implemented. 
    We hope this study may serve as a fundamental reference for the multipole framework in nanophotonics. Moreover, the uniquitity of the multipole framework allows this work applicable to other fields including acoustics, astronomy, and remote sensing. 

\section*{Acknolwledgement}
   This work was financially supported by the National Research Foundation (NRF) grants (NRF-2019R1A2C3003129, CAMM-2019M3A6B3030637, NRF-2019R1A5A8080290, and NRF-2018M3D1A1058998) funded by the Ministry of Science and ICT (MSIT), Korea. S.S. acknowledges global Ph.D. fellowship (NRF-2017H1A2A1043322) from the NRF-MSIT, Korea. J.J. acknowledges a fellowship from Hyundai Motor Chung Mong-Koo Foundation and a NRF grant (NRF-2019R1A6A3A13091132) by Ministry of Education, Korea. The authors thank Dr. Ivan Fernandez-Corbaton for helpful advice when preparing the manuscript. 

%

\pagebreak
\appendix

\section{Irreducible Cartesian basis}\label{sec:irreducible}
    The irreducible Cartesian multipoles are given as:
    \begin{subequations}
    \begin{equation} 
        \mathbf{v}^\mathrm{e}_1 = \frac{1}{\epsilon}[\splitatcommas{D^\mathrm{e}_x, D^\mathrm{e}_y, D^\mathrm{e}_z}]^\top,
    \end{equation}
    \begin{equation} 
        \mathbf{v}^\mathrm{e}_2 = \frac{k}{\epsilon}[\splitatcommas{Q^\mathrm{e}_{xy}, Q^\mathrm{e}_{xz}, Q^\mathrm{e}_{yz}, Q^\mathrm{e}_{xx}, Q^\mathrm{e}_{yy}}]^\top,
    \end{equation}
    \begin{equation} 
        \mathbf{v}^\mathrm{e}_3 = \frac{k^2}{\epsilon}[\splitatcommas{O^\mathrm{e}_{xyy}, O^\mathrm{e}_{xzz}, O^\mathrm{e}_{yzz}, O^\mathrm{e}_{yxx}, O^\mathrm{e}_{zxx}, O^\mathrm{e}_{zyy}, O^\mathrm{e}_{xyz}}]^\top.
    \end{equation}
    \label{eqn:irreducible_multipoles}
    \end{subequations}
    \(\mathbf{v}^\mathrm{m}_n\) has the same form, except that \(\epsilon^{-1}\) is exchanged with \(i\eta\) (i.e. \(\mathbf{v}^\mathrm{m}_1 = i\eta[D^\mathrm{m}_x,D^\mathrm{m}_y,D^\mathrm{m}_z]^\top\)). 
    Irreducible basis of Cartesian field components are given as:
    \begin{subequations}
    \begin{equation} 
        \mathbf{u}^\mathrm{e}_1 = [E_x,E_y,E_z]^\top,
    \end{equation}
    \begin{equation} 
        \mathbf{u}^\mathrm{e}_2 = \frac{1}{k}
        \begin{bmatrix}
        \dF{E_x}{y}+\dF{E_y}{x}\\
        \dF{E_x}{z}+\dF{E_z}{x}\\
        \dF{E_y}{z}+\dF{E_z}{y}\\
        \dF{E_x}{x}-\dF{E_z}{z}\\
        \dF{E_y}{y}-\dF{E_z}{z}
        \end{bmatrix},
    \end{equation}
    \begin{equation} 
        \mathbf{u}^\mathrm{e}_3
        =\frac{1}{k^2}
        \begin{bmatrix}
        \ddF{E_x}{y}-\ddF{E_x}{x}+2\deF{E_y}{x}{y} \\
        \ddF{E_x}{z}-\ddF{E_x}{x}+2\deF{E_z}{x}{z} \\
        \ddF{E_y}{z}-\ddF{E_y}{y}+2\deF{E_z}{y}{z} \\
        \ddF{E_y}{x}-\ddF{E_y}{y}+2\deF{E_x}{x}{y} \\
        \ddF{E_z}{x}-\ddF{E_z}{z}+2\deF{E_x}{x}{z} \\
        \ddF{E_z}{y}-\ddF{E_z}{z}+2\deF{E_y}{y}{z} \\
        2(\deF{E_x}{y}{z}+\deF{E_y}{x}{z}+\deF{E_z}{x}{y})
        \end{bmatrix},
    \end{equation}
    \label{eqn:irreducible_field}
    \end{subequations}
    and their magnetic counterparts \(\mathbf{u}^\mathrm{m}_n\) have the same form, except that \(i\eta\) factor is multiplied (i.e. \(\mathbf{u}^\mathrm{m}_1 = {i\eta}[H_x,H_y,H_z]^\top\)). 
    For simplicity, basis components are represented using indices as \(\footnotesize{x/y/z}\), \(\footnotesize{xy/xz/yz/xx/yy}\), and \(\footnotesize{xyy/xzz/yzz/yxx/zxx/zyy/xyz}\) for dipole, quadrupole, and octupole. specific \(\alp\)-tensor components are interpreted based on these indices; e.g., \(\apnm{e1}{z}{e2}{xy}\) relates \((u^\mathbf{e}_1)_z\) to \((v^\mathbf{e}_2)_{xy}\).
    
    In SI unit, Cartesian multipoles and field components in different multipole orders and types have different units, so different components of \(\alp\)-tensors have different units. 
    By using the normalized irreducible Cartesian multipoles (Eq.~\ref{eqn:irreducible_multipoles}) and field components (Eq.~\ref{eqn:irreducible_field}), multipoles have units of [V\(\cdot\)m\(^2\)], the field components have units of [V/m], and the \(\alp\)-tensors have units of [m\(^3\)]. 
    
    The basis transformation matrices that transform between spherical basis and irreducible Cartesian basis are introduced as
    \begin{subequations}
    \begin{equation} 
        \bar{\bar{\mathrm{V}}}_1
        =\frac{1}{\sqrt{2}}
        \begin{bmatrix}
        1 & 0 & -1 \\
        -i & 0 & -i \\
        0 & \sqrt{2} & 0
        \end{bmatrix}
    \end{equation}
    \begin{equation} 
        \bar{\bar{\mathrm{V}}}_2=
        \frac{1}{\sqrt{4}}
        \begin{bmatrix}
        -i & 0 & 0 & 0 & i \\
        0 & 1 & 0 & 1 & 0 \\
        0 & -i & 0 & -i & 0 \\
        1 & 0 & -\sqrt{\frac{2}{3}} & 0 & 1 \\
        -1 & 0 & -\sqrt{\frac{2}{3}} & 0 & -1
        \end{bmatrix}
    \end{equation}
    \begin{equation} 
        \bar{\bar{\mathrm{V}}}_3=
        \frac{1}{\sqrt{8}}
        \begin{bmatrix}
        -1 & 0 & -\frac{1}{\sqrt{15}} & 0 & \frac{1}{\sqrt{15}} & 0 & 1 \\
        0 & 0 & \frac{4}{\sqrt{15}} & 0 & -\frac{4}{\sqrt{15}} & 0 & 0 \\
        0 & 0 & -\frac{4i}{\sqrt{15}} & 0 & -\frac{4i}{\sqrt{15}} & 0 & 0 \\
        -i & 0 & \frac{i}{\sqrt{15}} & 0 & \frac{i}{\sqrt{15}} & 0 & -i \\
        0 & \sqrt{\frac{2}{3}} & 0 & -\frac{2}{\sqrt{5}} & 0 & \sqrt{\frac{2}{3}} & 0 \\
        0 & -\sqrt{\frac{2}{3}} & 0 & -\frac{2}{\sqrt{5}} & 0 & -\sqrt{\frac{2}{3}} & 0 \\
        0 & -\sqrt{\frac{2}{3}}i & 0 & 0 & 0 & \sqrt{\frac{2}{3}}i & 0
        \end{bmatrix}
    \end{equation}
    \end{subequations}

\vspace{30mm}
\begin{widetext}
\vspace{30mm}
\section{Exact Cartesian multipoles}\label{sec:exact}
    For completeness, we present expressions of the exact Cartesian multipoles up to MO as:
    \begingroup
    \allowdisplaybreaks
    \begin{subequations}
    \begin{align}
        \begin{split} 
            D^\mathrm{e}_\alpha &= 
            -\frac{1}{i\omega}
            \int{\dr
            \Bigg\{
            J_\alpha j_0(kr)
            +\frac{k^2}{2}\Big[3 \rJ r_\alpha-r^2J_\alpha\Big]
            \frac{j_2(kr)}{(kr)^2}
            \Bigg\}}
        \end{split}\\
        \begin{split} 
            D^\mathrm{m}_\alpha &= 
            \frac{3}{2}
            \int{\dr\rxJ_\alpha \frac{j_1(kr)}{kr}}
        \end{split}\\
        \begin{split} 
            Q^\mathrm{e}_{\alpha\beta} &= 
            -\frac{3}{2i\omega}
            \int\dr
            {\Bigg\{}
            \Big[r_\alpha J_\beta + r_\beta J_\alpha
            -\frac{2}{3}\delta_{\alpha\beta}\rJ\Big]
            \frac{j_1(kr)}{kr}\\
            &\hspace{5mm}
            +2k^2\Big[5\rJ r_\alpha r_\beta 
            -r^2(r_\alpha J_\beta+r_\beta J_\alpha)
            -r^2\delta_{\alpha\beta}\rJ\Big]
            \frac{j_3(kr)}{(kr)^3}
            {\Bigg\}}
        \end{split}\\
        \begin{split} 
            Q^\mathrm{m}_{\alpha\beta} &= \frac{5}{2}
            \int{\dr
            {\Big[}
            r_\alpha \rxJ_\beta + r_\beta \rxJ_\alpha
            {\Big]}
            \frac{j_2(kr)}{(kr)^2}
            }
        \end{split}\\
        \begin{split} 
            O^\mathrm{e}_{\alpha\beta\gamma} &= 
            -\frac{5}{2i\omega}
            \int\dr
            {\Bigg\{}
            \Big[r_\alpha r_\beta J_\gamma +r_\beta r_\gamma J_\alpha +r_\gamma r_\alpha J_\beta \\
            &\hspace{10mm}
            -\frac{1}{5}\delta_{\alpha\beta}(r^2J_\gamma+2\rJ r_\gamma)
            -\frac{1}{5}\delta_{\beta\gamma}(r^2J_\alpha+2\rJ r_\alpha)
            -\frac{1}{5}\delta_{\gamma\alpha}(r^2J_\beta+2\rJ r_\beta)\Big]
            \frac{j_2(kr)}{(kr)^2}\\
            &\hspace{5mm}
            +\frac{3k^2}{4}
            \Big[7\rJ r_\alpha r_\beta r_\gamma
            -r^2(r_\alpha r_\beta J_\gamma +r_\beta r_\gamma J_\alpha +r_\gamma r_\alpha J_\beta)\\
            &\hspace{10mm}
            +\frac{1}{5}\delta_{\alpha\beta}r^2(r^2J_\gamma+5\rJ r_\gamma)
            +\frac{1}{5}\delta_{\beta\gamma}r^2(r^2J_\alpha+5\rJ r_\alpha)
            +\frac{1}{5}\delta_{\gamma\alpha}r^2(r^2J_\beta+5\rJ r_\beta)\Big]
            \frac{j_4(kr)}{(kr)^4}
            {\Bigg\}}
        \end{split}\\
        \begin{split} 
            O^\mathrm{m}_{\alpha\beta\gamma} &=
            \frac{35}{8}
            \int\dr
            {\Big[}
            r_\alpha r_\beta \rxJ_\gamma
            +r_\beta r_\gamma \rxJ_\alpha
            +r_\gamma r_\alpha \rxJ_\beta\\
            &\hspace{5mm}
            -\frac{1}{5}\delta_{\alpha\beta}r^2\rxJ_\gamma
            -\frac{1}{5}\delta_{\beta\gamma}r^2\rxJ_\alpha
            -\frac{1}{5}\delta_{\gamma\alpha}r^2\rxJ_\beta
            {\Big]}
            \frac{j_3(kr)}{(kr)^3}
        \end{split}
    \end{align}
    \label{eqn:exact}
    \end{subequations}
    \endgroup
    where \(j_n(z)\) is spherical Bessel function. The familiar approximate Cartesian multipoles and toroidal multipoles can be readily obtained by taking the long wavelength limit \cite{Alaee2018} The expressions for the approximate Cartesian multipoles are presented in Appendix~\ref{sec:approx}, and the expressions for multipole radiations in Appendix~\ref{sec:radiation}.
\end{widetext}

\begin{widetext}
\section{Approximate Cartesian multipoles}\label{sec:approx}
    The exact Cartesian multipoles \emph{exactly} reconstruct multipole radiations of the localized current sources with arbitrary sizes, but their expressions (Eq.~\ref{eqn:exact}) are rather unfamiliar. By taking the long-wavelength limit \cite{Alaee2018}, the expressions of the familiar approximate Cartesian multipoles can be obtained as:
    \begingroup
    \allowdisplaybreaks
    \begin{subequations}
    \begin{align}
        \begin{split} 
            D^\mathrm{e}_\alpha &= 
            -\frac{1}{i\omega}
            \int{\dr J_\alpha}
        \end{split}\\
        \begin{split} 
            D^\mathrm{m}_\alpha &= 
            \frac{1}{2}
            \int{\dr\rxJ_\alpha}
        \end{split}\\
        \begin{split} 
            Q^\mathrm{e}_{\alpha\beta} &= 
            -\frac{1}{i\omega}
            \int\dr
            \Big[r_\alpha J_\beta + r_\beta J_\alpha
            -\frac{2}{3}\delta_{\alpha\beta}\rJ\Big]
        \end{split}\\
        \begin{split} 
            Q^\mathrm{m}_{\alpha\beta} &=
            \frac{1}{2}
            \int{\dr
            [r_\alpha \rxJ_\beta + r_\beta \rxJ_\alpha]
            }
        \end{split}\\
        \begin{split} 
            O^\mathrm{e}_{\alpha\beta\gamma} &= 
            -\frac{1}{2i\omega}
            \int\dr
            \Big[r_\alpha r_\beta J_\gamma +r_\beta r_\gamma J_\alpha +r_\gamma r_\alpha J_\beta \\
            &\hspace{5mm}
            -\frac{1}{5}\delta_{\alpha\beta}(r^2J_\gamma+2\rJ r_\gamma)
            -\frac{1}{5}\delta_{\beta\gamma}(r^2J_\alpha+2\rJ r_\alpha)\\
            &\hspace{5mm}
            -\frac{1}{5}\delta_{\gamma\alpha}(r^2J_\beta+2\rJ r_\beta)\Big]
        \end{split}\\
        \begin{split} 
            O^\mathrm{m}_{\alpha\beta\gamma} &=
            \frac{1}{24}
            \int\dr
            {\Big[}
            r_\alpha r_\beta \rxJ_\gamma
            +r_\beta r_\gamma \rxJ_\alpha
            +r_\gamma r_\alpha \rxJ_\beta\\
            &\hspace{5mm}
            -\frac{1}{5}\delta_{\alpha\beta}r^2\rxJ_\gamma
            -\frac{1}{5}\delta_{\beta\gamma}r^2\rxJ_\alpha
            -\frac{1}{5}\delta_{\gamma\alpha}r^2\rxJ_\beta
            {\Big]}
        \end{split}
    \end{align}
    \label{eqn:approx}
    \end{subequations}
    \endgroup
    Interestingly, toroidal multipoles can also be recovered as the higher-order correction terms \cite{Alaee2018}. 
\end{widetext}

\begin{widetext}
\section{Multipole radiations}\label{sec:radiation}
    It should be noted that the proportionality constants for multipoles (Eq.~\ref{eqn:exact} and \ref{eqn:approx}) may differ from other publications \cite{Evlyukhin2016}, but the reconstructed radiation fields satisfying the Maxwell's equations should be identical. The differences in proportionality constants are compensated by the expressions for the multipole radiations. The radiation fields from the Cartesian multipoles are expressed as:
    \begingroup
    \allowdisplaybreaks
    \begin{subequations}
    \begin{align}
        \begin{split} 
        E_\alpha
        &=\frac{ik^3}{4\pi}
        \Bigg\{
        \frac{1}{\epsilon}\Big[[h_0(kr)-\frac{1}{kr}h_1(kr)]D^\mathrm{e}_\alpha+h_2(kr)n_\alpha n_i D^\mathrm{e}_i\Big]\\
        &\hspace{10mm}+\frac{k}{\epsilon}\Big[[h_1(kr)-\frac{2}{kr}h_2(kr)]{Q^\mathrm{e}_{\alpha i} n_i}+h_3(kr){n_\alpha n_i n_j Q^\mathrm{e}_{ij}}\Big]\\
        &\hspace{10mm}+\frac{k^2}{\epsilon}\Big[[h_2(kr)-\frac{3}{kr}h_3(kr)]{O^\mathrm{e}_{\alpha ij} n_i n_j}+h_4(kr){n_\alpha n_i n_j n_k O^\mathrm{e}_{ijk}}\Big]\\
        &\hspace{10mm}+{i\eta}h_1(kr)\epsilon_{\alpha ij}{D^\mathrm{m}_i n_j}
        +{i\eta}k h_2(kr)\epsilon_{\alpha ij}{Q^\mathrm{m}_{ik} n_j n_k}
        +{i\eta}k^2 h_3(kr)\epsilon_{\alpha ij}{O^\mathrm{m}_{ikl} n_j n_j n_l}
        \Bigg\}
        \end{split}\\
        \begin{split} 
        {i\eta} H_\alpha
        &=\frac{ik^3}{4\pi}
        \Bigg\{
        \frac{1}{\epsilon}h_1(kr)\epsilon_{\alpha ij}{D^\mathrm{e}_i n_j}
        +\frac{k}{\epsilon}h_2(kr)\epsilon_{\alpha ij}{Q^\mathrm{e}_{ik} n_j n_k}
        +\frac{k^2}{\epsilon}h_3(kr)\epsilon_{\alpha ij}{O^\mathrm{e}_{ikl} n_j n_j n_l}\\
        &\hspace{10mm}+{i\eta}\Big[[h_0(kr)-\frac{1}{kr}h_1(kr)]D^\mathrm{m}_\alpha+h_2(kr)n_\alpha n_i D^\mathrm{m}_i\Big]\\
        &\hspace{10mm}+{i\eta}k\Big[[h_1(kr)-\frac{2}{kr}h_2(kr)]{Q^\mathrm{m}_{\alpha i} n_i}+h_3(kr){n_\alpha n_i n_j Q^\mathrm{m}_{ij}}\Big]\\
        &\hspace{10mm}+{i\eta}k^2\Big[[h_2(kr)-\frac{3}{kr}h_3(kr)]{O^\mathrm{m}_{\alpha ij} n_i n_j}+h_4(kr){n_\alpha n_i n_j n_k O^\mathrm{m}_{ijk}}\Big]
        \Bigg\}
        \end{split}
    \end{align}
    \label{eqn:radiationfield}
    \end{subequations}
    \endgroup
    where the Einstein summation notation is used; \(\alpha, \beta, \gamma = x, y, z\); \(\epsilon_{ijk}\) is the Lavi-Civita symbol; and \(h_n(z)\) is the spherical Hankel function. Note that this expression is exact even in the near zone. Also, the proportionality constants are symmetrical for electric and magnetic multipoles, so it is straightforward to transform them to helicity basis \cite{Fruhnert2017}.

    The radiated fields in the far fields are:
    \begin{subequations}
    \begin{align}
        \begin{split} 
            \mathbf{E} &= 
            \frac{1}{\epsilon}\frac{k^2}{4\pi} \frac{e^{ikr}}{r}
            {\Bigg[}
            \mathbf{n} \times (\mathbf{D}^\mathrm{e} \times \mathbf{n})
            -ik\mathbf{n} \times(\hat{Q}^\mathrm{e} \times \mathbf{n})
            -k^2\mathbf{n} \times(\hat{O}^\mathrm{e} \times \mathbf{n})
            {\Big]}\\
            &\hspace{5mm}
            -{i\eta}\frac{ik^2}{4\pi} \frac{e^{ikr}}{r}
            {\Bigg[}
            \mathbf{D}^\mathrm{m} \times \mathbf{n}
            -ik(\hat{Q}^\mathrm{m} \times \mathbf{n})
            -k^2(\hat{O}^\mathrm{m} \times \mathbf{n})
            {\Big]}
        \end{split}\\
        \begin{split} 
            {i\eta}\mathbf{H} &= 
            -\frac{1}{\epsilon}\frac{ik^2}{4\pi} \frac{e^{ikr}}{r}
            {\Bigg[}
            \mathbf{D}^\mathrm{e} \times \mathbf{n}
            -ik(\hat{Q}^\mathrm{e} \times \mathbf{n})
            -k^2(\hat{O}^\mathrm{e} \times \mathbf{n})
            {\Big]}\\
            &\hspace{5mm}
            +{i\eta}\frac{k^2}{4\pi} \frac{e^{ikr}}{r}
            {\Bigg[}
            \mathbf{n} \times (\mathbf{D}^\mathrm{m} \times \mathbf{n})
            -ik\mathbf{n} \times (\hat{Q}^\mathrm{m} \times \mathbf{n})
            -k^2\mathbf{n} \times (\hat{O}^\mathrm{m} \times \mathbf{n})
            {\Big]}
        \end{split}
    \end{align}
    \label{eqn:radiationfarfield}
    \end{subequations}
    where $\hat{Q}^p_\alpha = Q^p_{\alpha\beta}n_\beta$ and $\hat{O}^p_\alpha = O^p_{\alpha\beta\gamma}n_\beta n_\gamma$. 
    
    Finally, the total radiation power is
    \begin{equation}
    \begin{split}
        P &= 
        \frac{1}{\epsilon^2}\frac{k^4}{12 \pi \eta} \sum_{\alpha}{|D^\mathrm{e}_\alpha|^2}
        + \frac{1}{\epsilon^2}\frac{k^6}{40 \pi \eta} \sum_{\alpha\beta}{|Q^\mathrm{e}_{\alpha\beta}|^2}
        + \frac{1}{\epsilon^2}\frac{k^8}{105 \pi \eta} \sum_{\alpha\beta\gamma}{|O^\mathrm{e}_{\alpha\beta\gamma}|^2}\\
        &\hspace{5mm}
        + {\eta^2}\frac{k^4}{12 \pi \eta} \sum_{\alpha}{|D^\mathrm{m}_\alpha|^2}
        + {\eta^2}\frac{k^6}{40 \pi \eta} \sum_{\alpha\beta}{|Q^\mathrm{m}_{\alpha\beta}|^2}
        + {\eta^2}\frac{k^8}{105 \pi \eta} \sum_{\alpha\beta\gamma}{|O^\mathrm{m}_{\alpha\beta\gamma}|^2}
    \end{split}
    \label{eqn:power2}
    \end{equation}
\end{widetext}

\clearpage
\section{Irreducible basis and units}\label{sec:irreducibleunit}
    The normalized irreducible Cartesian basis introduced in this work has several advantages. First, the components of \(\boldsymbol{\alpha}\)-tensor based on the normalized irreducible Cartesian basis have units of volume and their magnitudes are roughly proportional to the scattering intensities. Here, we compare dipolar \(\boldsymbol{\alpha}\)-tensors in SI and non-SI units. The \(\boldsymbol{\alpha}\)-tensor in SI units,
    \begin{equation*}
        \begin{bmatrix}
        \mathbf{D}^\mathrm{e} \\ \mathbf{D}^\mathrm{m}
        \end{bmatrix}
        =
        \begin{bmatrix}
        \alpha^{\mathbf{E}}_{\mathbf{D}^\mathrm{e}} & \alpha^{\mathbf{H}}_{\mathbf{D}^\mathrm{e}} \\
        \alpha^{\mathbf{E}}_{\mathbf{D}^\mathrm{m}} & \alpha^{\mathbf{H}}_{\mathbf{D}^\mathrm{m}}
        \end{bmatrix}
        \cdot
        \begin{bmatrix}
        \mathbf{E} \\ \mathbf{H}
        \end{bmatrix}
    \end{equation*}
    has different units per components, which can differ by several orders, so they cannot be compared directly. On the other hand, \(\boldsymbol{\alpha}\)-tensor expressed using the normalized irreducible basis,
    \begin{equation*}
        \begin{bmatrix}
        \epsilon^{-1}\mathbf{D}^\mathrm{e}\\
        {i\eta}\mathbf{D}^\mathrm{m}
        \end{bmatrix}
        =
        \begin{bmatrix}
        \alpha^{\mathrm{e}1}_{\mathrm{e}1} & \alpha^{\mathrm{m}1}_{\mathrm{e}1} \\
        \alpha^{\mathrm{e}1}_{\mathrm{m}1} & \alpha^{\mathrm{m}1}_{\mathrm{m}1}
        \end{bmatrix}
        \cdot
        \begin{bmatrix}
        \mathbf{E} \\
        {i\eta}\mathbf{H}
        \end{bmatrix},
    \end{equation*}
    has units of volume, and the magnitude of the components are proportional to the scattering power. In addition, the \(\boldsymbol{\alpha}\)-tensor and Green's tensors based on the irreducible basis have many symmetries, which are discussed in the main text and the next section, respectively.

\section{Green's tensors}\label{sec:Green}
    To implement \(\boldsymbol{\alpha}\)-tensors into the multiple-scattering theory, Green's tensors are required. In this section, we present the Green's tensors that relates \(\mathbf{v}^p_n\) in origin to \(\mathbf{u}^{p'}_{n'}\) at \(\mathbf{r}\):
    \begin{equation}
        \mathbf{u}^{p'}_{n'}(k\mathbf{r}) = [G^{pn}_{p'n'}](k\mathbf{r})\cdot\mathbf{v}^p_n
    \end{equation}
    We noted that the constructed Green's tensors have several symmetries:
    \begin{subequations}
    \begin{align}
    \begin{split}
        [G^{\mathrm{e}n}_{\mathrm{e}n}] &= [G^{\mathrm{e}n}_{\mathrm{e}n}]^\top
    \end{split}\\
    \begin{split}
        [G^{\mathrm{m}n}_{\mathrm{m}n}] &= -[G^{\mathrm{m}n}_{\mathrm{m}n}]^\top
    \end{split}\\
    \begin{split}
        [G^{\mathrm{e}n}_{\mathrm{e}n'}] &= (-1)^{n-n'} [G^{\mathrm{e}n'}_{\mathrm{e}n}]^\top
    \end{split}\\
    \begin{split}
        [G^{\mathrm{m}n}_{\mathrm{e}n'}] &= -(-1)^{n-n'} [G^{\mathrm{m}n'}_{\mathrm{e}n}]^\top
    \end{split}\\
    \begin{split}
        [G^{\mathrm{m}n}_{\mathrm{m}n'}] &= [G^{\mathrm{e}n}_{\mathrm{e}n'}]
    \end{split}\\
    \begin{split}
        [G^{\mathrm{m}n}_{\mathrm{e}n'}] &= [G^{\mathrm{e}n}_{\mathrm{m}n'}]
    \end{split}\\
    \begin{split}
        [G^{\mathrm{e}n}_{\mathrm{e}n'}](k\mathbf{r}) &= (-1)^{n-n'}[G^{\mathrm{e}n}_{\mathrm{e}n'}](-k\mathbf{r})
    \end{split}\\
    \begin{split}
        [G^{\mathrm{m}n}_{\mathrm{e}n'}](k\mathbf{r}) &= -(-1)^{n-n'}[G^{\mathrm{m}n}_{\mathrm{e}n'}](-k\mathbf{r})
    \end{split}
    \end{align}
    \label{eqn:Green.sym}
    \end{subequations}
    We only need to calculate some parts (Eq.~\ref{eqn:Green}), from which the other parts can be obtained using symmetries (Eq.~\ref{eqn:Green.sym}).
    \begingroup
    \allowdisplaybreaks
    \footnotesize
    \begin{subequations}
    \begin{align}
    \begin{split}
        [G^\mathrm{e1}_\mathrm{e1}] &= \frac{ik^3}{4\pi}\Bigg\{h_2(kr)\frac{n_1n_1}{r^2}+[h_0(kr)-\frac{1}{kr}h_1(kr)][A^1_1]\Bigg\}
    \end{split}\\
    \begin{split}
        [G^\mathrm{e2}_\mathrm{e1}] &= \frac{ik^3}{4\pi}\Bigg\{h_3(kr)\frac{n_1n_2}{r^3}+[h_1(kr)-\frac{2}{kr}h_2(kr)][A^2_1]\Bigg\}
    \end{split}\\
    \begin{split}
        [G^\mathrm{e3}_\mathrm{e1}] &= \frac{ik^3}{4\pi}
        \Bigg\{h_4(kr)\frac{n_1n_3}{r^4}+[h_2(kr)-\frac{3}{kr}h_3(kr)][A^3_1]\Bigg\}
    \end{split}\\
    \begin{split}
        [G^\mathrm{e2}_\mathrm{e2}] &= -\frac{ik^3}{4\pi}
        \Bigg\{
        h_4(kr)\frac{n_2n_2}{r^4}
        +[h_2(kr)-\frac{4}{kr}h_3(kr)][A^2_2]\\
        &-\frac{1}{kr}[h_1(kr)-\frac{2}{kr}h_2(kr)][A^2_2]'
        \Bigg\}
    \end{split}\\
    \begin{split}
        [G^\mathrm{e3}_\mathrm{e2}] &= -\frac{ik^3}{4\pi}
        \Bigg\{
        h_5(kr)\frac{n_2n_3}{r^5}
        +[h_3(kr)-\frac{6}{kr}h_4(kr)][A^3_2]\\
        &-\frac{1}{kr}[2h_2(kr)-\frac{6}{kr}h_3(kr)][A^3_2]'
        \Bigg\}
    \end{split}\\
    \begin{split}
        [G^\mathrm{e3}_\mathrm{e3}] &= \frac{ik^3}{4\pi}
        \Bigg\{
        h_6(kr)\frac{n_3n_3}{r^6}
        +[h_4(kr)-\frac{9}{kr}h_5(kr)][A^3_3]\\
        &-\frac{1}{kr}[4h_3(kr)-\frac{18}{kr}h_4(kr)][A^3_3]'\\
        &+\frac{1}{(kr)^2}[2h_2(kr)-\frac{6}{kr}h_3(kr)][A^3_3]''
        \Bigg\}
    \end{split}\\
    \begin{split}
        [G^\mathrm{m1}_\mathrm{e1}] &= \frac{ik^3}{4\pi}h_1(kr)[B^1_1]
    \end{split}\\
    \begin{split}
        [G^\mathrm{m2}_\mathrm{e1}] &= \frac{ik^3}{4\pi}h_2(kr)[B^2_1]
    \end{split}\\
    \begin{split}
        [G^\mathrm{m3}_\mathrm{e1}] &= \frac{ik^3}{4\pi}h_3(kr)[B^3_1]
    \end{split}\\
    \begin{split}
        [G^\mathrm{m2}_\mathrm{e2}] &= -\frac{ik^3}{4\pi}
        \Bigg\{h_3(kr)[B^2_2]-\frac{1}{kr}h_2(kr)[B^2_2]'\Bigg\}
    \end{split}\\
    \begin{split}
        [G^\mathrm{m3}_\mathrm{e2}] &= -\frac{ik^3}{4\pi}
        \Bigg\{h_4(kr)[B^3_2]-\frac{2}{kr}h_3(kr)[B^3_2]'\Bigg\}
    \end{split}\\
    \begin{split}
        [G^\mathrm{m3}_\mathrm{e3}] &= \frac{ik^3}{4\pi}
        \Bigg\{
        h_5(kr)[B^3_3]-\frac{4}{kr}h_4(kr)[B^3_3]'\\
        &-\frac{2}{(kr)^2}h_3(kr)[B^3_3]''
        \Bigg\}
    \end{split}
    \end{align}
    \label{eqn:Green}
    \end{subequations}
    \normalsize
    \endgroup
    \vspace{10mm}
    
    \begingroup
    \begin{widetext}
    The matrices are explicitly given as follows. Note that \([B^2_2]\) and \([B^3_3]\) have only half of the components, where the other half can be obtained by noting that they are anti-symmetric.
    
    \small
    \(n_1 = [x,y,z]^\top\),
    \(n_2 = [2xy,2xz,2yz,x^2-z^2,y^2-z^2]^\top\),
    
    \(n_3 = [3xy^2-x^3,3xz^2-x^3,3yz^2-y^3,3yx^2-y^3,3zx^2-z^3,3zy^2-z^3,6xyz]^\top\),
    
    \([A^1_1]
        = \bar{\bar{\mathrm{U}}}_1\bar{\bar{\mathrm{U}}}_1^\dag 
        = \begin{bmatrix}
        1&0&0\\
        0&1&0\\
        0&0&1
        \end{bmatrix}\), %
    \([A^2_1] = 
        \frac{1}{r}
        \begin{bmatrix}
        y&z&0&x&0\\
        x&0&z&0&y\\
        0&x&y&-z&-z
        \end{bmatrix}\), %
    \([A^3_1] = 
        \frac{1}{r^2}
        \begin{bmatrix}
        y^2-x^2&z^2-x^2&0&2xy&2xz&0&2yz\\
        2xy&0&z^2-y^2&x^2-y^2&0&2yz&2xz\\
        0&2xz&2yz&0&x^2-z^2&y^2-z^2&2xy
        \end{bmatrix}\) %
    \([A^2_2] = 
        \frac{1}{r^2}
        \begin{bmatrix}
        x^2+y^2 & yz & xz & xy & xy \\
        yz & x^2+z^2 & xy & 0 & -xz \\
        xz & xy & y^2+z^2 & -yz & 0 \\
        xy & 0 & -yz & x^2+z^2 & z^2 \\
        xy & -xz & 0 & z^2 & y^2+z^2
        \end{bmatrix}\),
     \([A^2_2]' 
        = \bar{\bar{\mathrm{U}}}_2\bar{\bar{\mathrm{U}}}_2^\dag 
        = \begin{bmatrix}
        1&&&&\\
        &1&&&\\
        &&1&&\\
        &&&2&1\\
        &&&1&2
        \end{bmatrix}\), %
        
    \([B^1_1] = 
        \frac{1}{r}
        \begin{bmatrix}
        0 & z & -y \\
        -z & 0 & x \\
        y & -x & 0
        \end{bmatrix}\),
    \([B^2_1] = 
        \frac{1}{r^2}
        \begin{bmatrix}
        2z&-xy&z^2-y^2&yz&2yz\\
        -yz&x^2-z^2&xy&-2xz&-xz\\
        y^2-x^2&yz&-xz&xy&-xy
        \end{bmatrix}\), %
        
    \([B^3_1] = 
        \frac{1}{r^3}
        \begin{bmatrix}
        2xyz&-2xyz&-z(3y^2-z^2)&z(x^2-y^2)&-y(x^2-z^2)&-y(y^2-3z^2)&-2x(y^2-z^2)\\
        z(x^2-y^2)&z(3x^2-z^2)&2xyz&-2xyz&x(x^2-3z^2)&x(y^2-z^2)&2y(x^2-z^2)\\
        -y(3x^2-y^2)&y(z^2-x^2)&x(y^2-z^2)&-x(x^2-3y^2)&2xyz&-2xyz&-2z(x^2-y^2)
        \end{bmatrix}\), %
        
    \([B^2_2] = 
        \frac{1}{r^2}
        \begin{bmatrix}
        0& x(2x^2-r^2) & y(r^2-2y^2) & z(y^2-2x^2) & z(2y^2-x^2) \\
        &0& z(2z^2-r^2) & y(x^2+z^2) & y(2z^2-x^2) \\
        &&0& x(y^2-2z^2) & -x(y^2+z^2) \\
        &&&0& 3xyz \\
        &&&&0
        \end{bmatrix}\),
    \([B^2_2]' = 
        \frac{1}{r}
        \begin{bmatrix}
        0&x&-y&-z&z\\
        &0&z&2y&y\\
        &&0&-x&-2x\\
        &&&0&0\\
        &&&&0
        \end{bmatrix}\), %
    
    \([A^3_2] = 
        \frac{1}{r^3}
        \begin{bmatrix}
        y(x^2+y^2)&y(z^2-x^2)&-x(y^2-z^2)&x(x^2+y^2)&2xyz&2xyz&2z(x^2+y^2)\\
        -z(x^2-y^2)&z(x^2+z^2)&2xyz&2xyz&x(x^2+z^2)&x(y^2-z^2)&2y(x^2+z^2)\\
        2xyz&2xyz&z(y^2+z^2)&z(x^2-y^2)&-y(z^2-x^2)&y(y^2+z^2)&2x(y^2+z^2)\\
        -x(x^2-y^2)&-x(x^2+z^2)&-2yz^2&2x^2y&z(x^2+z^2)&-z(y^2-z^2)&0\\
        2xy^2&-2xz^2&-y(y^2+z^2)&y(x^2-y^2)&z(z^2-x^2)&z(y^2+z^2)&0
        \end{bmatrix}\),
        
    \([A^3_2]' =
        \frac{1}{r}
        \begin{bmatrix}
        2y&0&0&2x&0&0&2z\\
        0&2z&0&0&2x&0&2y\\
        0&0&2z&0&0&2y&2x\\
        -x&-2x&-y&y&2z&z&0\\
        x&-x&-2y&-y&z&2z&0
        \end{bmatrix}\),%
    
    \turnpage
    \scriptsize
    \begin{table}[!ht]
        \def\arraystretch{1.0}
        \begin{tabular}{l}
    \([A^3_3] = 
        \frac{1}{r^4}
        \begin{bmatrix}
        (x^2+y^2)^2&(x^2-y^2)(x^2-z^2)&-2xy(y^2-z^2)&0&-2xz(x^2-y^2)&4xy^2z&2yz(x^2+y^2)\\
        (x^2-y^2)(x^2-z^2)&(x^2+z^2)^2&4xyz^2&2xy(z^2-x^2)&0&2xz(y^2-z^2)&2yz(x^2+z^2)\\
        -2xy(y^2-z^2)&4xyz^2&(y^2+z^2)^2&(x^2-y^2)(z^2-y^2)&2yz(x^2-z^2)&0&2xz(y^2+z^2)\\
        0&2xy(z^2-x^2)&(x^2-y^2)(z^2-y^2)&(x^2+y^2)^2&4x^2yz&2yz(x^2-y^2)&2xz(x^2+y^2)\\
        -2xz(x^2-y^2)&0&2yz(x^2-z^2)&4x^2yz&(x^2+z^2)^2&(x^2-z^2)(y^2-z^2)&2xy(x^2+z^2)\\
        4xy^2z&2xz(y^2-z^2)&0&2yz(x^2-y^2)&(x^2-z^2)(y^2-z^2)&(y^2+z^2)^2&2xy(y^2+z^2)\\
        2yz(x^2+y^2)&2yz(x^2+z^2)&2xz(y^2+z^2)&2xz(x^2+y^2)&2xy(x^2+z^2)&2xy(y^2+z^2)&4(x^2y^2+y^2z^2+z^2x^2)
        \end{bmatrix}\), \\%
    \([A^3_3]' = 
        \frac{1}{r^2}
        \begin{bmatrix}
        2(x^2+y^2)&x^2&-xy&0&-xz&xz&2yz\\
        x^2&2(x^2+z^2)&xy&-xy&0&-xz&2yz\\
        -xy&xy&2(y^2+z^2)&y^2&-yz&0&2xz\\
        0&-xy&xy&2(x^2+y^2)&yz&-yz&2xz\\
        -xz&0&-yz&yz&2(x^2+z^2)&z^2&2xy\\
        xz&-xz&0&-yz&z^2&2(y^2+z^2)&2xy\\
        2yz&2yz&2xz&2xz&2xy&2xy&2(x^2+y^2+z^2)
        \end{bmatrix}\),
    \([A^3_3]'' 
        = \bar{\bar{\mathrm{U}}}_3\bar{\bar{\mathrm{U}}}_3^\dag
        = \begin{bmatrix}
        4&1&&&&&\\
        1&4&&&&&\\
        &&4&1&&&\\
        &&1&4&&&\\
        &&&&4&1&\\
        &&&&1&4&\\
        &&&&&&6
        \end{bmatrix}\), \\%
    \([B^3_2] = 
        \frac{1}{r^4}
        \begin{bmatrix}
        xz(x^2+y^2)&xz(3x^2-2y^2-z^2)&yz(2x^2-3y^2+z^2)&-yz(x^2+y^2)&\inalign{x^2(x^2-3z^2)}{+y^2(z^2-x^2)}&\inalign{y^2(3z^2-y^2)}{+x^2(y^2-z^2)}&2xy(x^2-y^2)\\
        xy(-3x^2+y^2+2z^2)&-xy(x^2+z^2)&\inalign{z^2(z^2-3y^2)}{+x^2(y^2-z^2)}&\inalign{x^2(3y^2-x^2)}{+z^2(x^2-y^2)}&yz(x^2+z^2)&yz(-2x^2-y^2+3z^2)&2xz(z^2-x^2)\\
        \inalign{y^2(y^2-3x^2)}{+z^2(x^2-y^2)}&\inalign{z^2(3x^2-z^2)}{+y^2(z^2-x^2)}&xy(y^2+z^2)&xy(-x^2+3y^2-2z^2)&xz(x^2+2y^2-3z^2)&-xz(y^2+z^2)&2yz(y^2-z^2)\\
        yz(5x^2-y^2)&-yz(x^2+z^2)&-2xz(2y^2-z^2)&2xz(x^2-2y^2)&-xy(x^2+z^2)&-xy(y^2-5z^2)&\inalign{2x^2(z^2-y^2)}{+2z^2(x^2+y^2)}\\
        2yz(2x^2-y^2)&-2yz(z^2-2x^2)&xz(y^2+z^2)&xz(x^2-5y^2)&-xy(5z^2-x^2)&xy(y^2+z^2)&\inalign{2y^2(x^2-z^2)}{+2z^2(x^2-y^2)}
        \end{bmatrix}\), \\%
    \([B^3_2]' = 
        \frac{1}{r^2}
        \begin{bmatrix}
        2xz&2xz&-2yz&-2yz&x^2-z^2&z^2-y^2&0\\
        -2xy&-2xy&z^2-y^2&y^2-x^2&2yz&2yz&0\\
        y^2-x^2&x^2-z^2&2xy&2xy&-2xz&-2xz&0\\
        yz&-2yz&xz&xz&-2xy&xy&r^2-3y^2\\
        -yz&-yz&2xz&-xz&-xy&2xy&3x^2-r^2
        \end{bmatrix}\), \\%
    \([B^3_3] = 
        \frac{1}{r^5}
        \begin{bmatrix}
        0&-2xyz(r^2-5x^2)&\inalign{y^2z(7x^2-3y^2)}{-z^3(x^2-y^2)}&-z(x^2+y^2)^2&\inalign{-x^2y(7z^2-3x^2)}{+y^3(z^2-x^2)}&-5x^2yz^2-4y^5+3y^3r^2&\inalign{-2xz^2(x^2+y^2)}{-2xy^2(y^2-3x^2)}\\
        &0&5x^2y^2z+4z^5-3z^3r^2&\inalign{x^2z(7y^2-3x^2)}{+z^3(x^2-y^2)}&y(x^2+z^2)^2&\inalign{-yz^2(7x^2-3z^2)}{+y^3(x^2-z^2)}&\inalign{2xy^2(x^2+z^2)}{+2xz^2(z^2-3x^2)}\\
        &&0&2xyz(5y^2-r^2)&\inalign{xz^2(7y^2-3z^2)}{+x^3(z^2-y^2)}&-x(y^2+z^2)^2&\inalign{-2x^2y(y^2+z^2)}{-2yz^2(z^2-3y^2)}\\
        &&&0&5xy^2z^2+4x^5-3x^3r^2&\inalign{xy^2(7z^2-3y^2)}{+x^3(y^2-z^2)}&\inalign{2yz^2(x^2+y^2)}{+2x^2y(x^2-3y^2)}\\
        &&&&0&-2xyz(r^2-5z^2)&\inalign{-2y^2z(x^2+z^2)}{-2x^2z(x^2-3z^2)}\\
        &&&&&0&\inalign{2x^2z(y^2+z^2)}{+2y^2z(y^2-3z^2)}\\
        &&&&&&0
        \end{bmatrix}\), \\%
    \([B^3_3]' = 
        \frac{1}{r^3}
        \begin{bmatrix}
        0&3xyz&z(x^2-2y^2)&-2z(x^2+y^2)&y(2x^2-z^2)&y(r^2-2y^2)&x(r^2-3z^2)\\
        &0&-z(r^2-2z^2)&-z(2x^2-y^2)&2y(x^2+z^2)&-y(x^2-2z^2)&-x(r^2-3y^2)\\
        &&0&3xyz&x(y^2-2z^2)&-2x(y^2+z^2)&y(r^2-3x^2)\\
        &&&0&-x(r^2-2x^2)&-x(2y^2-z^2)&-y(r^2-3z^2)\\
        &&&&0&3xyz&z(r^2-3y^2)\\
        &&&&&0&-z(r^2-3x^2)\\
        &&&&&&0
        \end{bmatrix}\), %
    \([B^3_3]'' = 
        \frac{1}{r}
        \begin{bmatrix}
        0&0&-z&-4z&y&-y&2x\\
        &0&z&-z&4y&y&-2x\\
        &&0&0&-x&-4x&2y\\
        &&&0&x&-x&-2y\\
        &&&&0&0&2z\\
        &&&&&0&-2z\\
        &&&&&&0
        \end{bmatrix}\) %
        \end{tabular}
    \end{table}
    \end{widetext}
    \endgroup

\clearpage
\section{Spherical multipoles}\label{sec:sphericalmultipole}
    The spherical multipoles can be calculated from the localized current sources \cite{Jackson1999, Grahn2012} or from the scattered fields using the orthogonality of the vector spherical wave functions (VSWFs). Note that expressions for VSWFs differ by publications, and our expression is
    \begin{subequations}
    \begin{align}
        \begin{split}
            \mathbf{M}^{(i)}_{nm}(k\mathbf{r}) &= i\gamma_{nm}z_n^{(i)}(kr)(i\pi_{nm}\hat{e}_\theta-\tau_{nm}\hat{e}_\phi)e^{im\phi}
        \end{split}\\
        \begin{split}
            \mathbf{N}^{(i)}_{nm}(k\mathbf{r}) &= i\gamma_{nm}\Big[n(n+1)\frac{z_n^{(i)}(kr)}{kr}P_n^m(\cos{\theta})\hat{e}_r\\
            &+\frac{1}{kr}\frac{d[kr z_n^{(i)}(kr)]}{d(kr)}(\tau_{nm}\hat{e}_\theta+i\pi_{nm}\hat{e}_\phi)\Big]e^{im\phi}
        \end{split}\\
        \begin{split}
            \tau_{nm}(\theta) &= \frac{d}{d\theta} P_n^m(\cos{\theta})
        \end{split}\\
        \begin{split}
            \pi_{nm}(\theta) &= \frac{m}{\sin{\theta}}P_n^m(\cos{\theta})
        \end{split}\\
        \begin{split}
            \gamma_{nm} &= \sqrt{\frac{(2n+1)(n-m)!}{4\pi n(n+1)(n+m)!}}
        \end{split}
    \end{align}
    \end{subequations}
    where superscripts (1) and (+) refer to the regular and singular spherical waves, respectively; $z_n^{(+)}(kr) = h_n^{(1)}(kr)$ and $z_n^{(1)}(kr) = j_n(kr)$. 
    The incident and radiation fields are reconstructed as
    \begin{subequations}
    \begin{equation} 
        \mathbf{E}_\mathrm{sca} = E_0\sum_{n=1}^{\infty}\sum_{m=-n}^{n}{[b^\mathrm{e}_{nm}\mathbf{N}^{(+)}_{nm}(k\mathbf{r})+b^\mathrm{m}_{nm}\mathbf{M}^{(+)}_{nm}(k\mathbf{r})]}
    \end{equation}
    \begin{equation} 
        \mathbf{E}_\mathrm{inc} = E_0\sum_{n=1}^{\infty}\sum_{m=-n}^{n}{[a^\mathrm{e}_{nm}\mathbf{N}^{(1)}_{nm}(k\mathbf{r})+a^\mathrm{m}_{nm}\mathbf{M}^{(1)}_{nm}(k\mathbf{r})]}
    \end{equation}
    \end{subequations}
    
\section{T-matrix and polarizability-tensors of meta-atoms}\label{sec:tmatrix}
\subsection{T-matrix retrieval}
    Up to arbitrary multipole orders, \(\mathbf{T}\)-matrix of several particle systems can be analytically calculated including notably spheres from Mie coefficients, chiral spheres \cite{Wu2012}, homogeneous anisotropic spheres \cite{Stout2007}, and even nonlocal spheres and coreshells \cite{David2011}, and \(\mathbf{T}\)-matrix of a system of multiple particles can also be defined \cite{Mishchenko2010}. In general, \(\mathbf{T}\)-matrix of nonspherical particles should be calculated numerically using Extended Boundary Condition Method \cite{Mishchenko2002}, Discrete-Sources Null-Field Method \cite{Doicu2006}, or FEM \cite{Fruhnert2017}. In this work, we used FEM to extract \(\mathbf{T}\)-matrix of meta-atoms for its convenient implementation. However, it should be noted that FEM is very costly for calculating \(\mathbf{T}\)-matrix compared to surface-integral methods \cite{Mishchenko2002, Doicu2006}, because they require only particle surfaces to be discretized, whereas FEM requires larger simulation domains including the PML and the spacer between the particle surface and the PML.
    
    The retrieved \(\mathbf{T}\)-matrix is then used to obtain exact higher-order \(\boldsymbol{\alpha}\)-tensor. The retrieved \(\mathbf{T}\)-matrix (or \(\boldsymbol{\alpha}\)-tensor) can be analytically treated to efficiently calculate orientation-averaged optical responses, or inserted in the multiple-scattering theory to calculate coherently coupled optical responses between discrete scattering objects.
\subsection{Meta-atoms in the main text}
\begin{figure}[!h]
        \centering
        \includegraphics{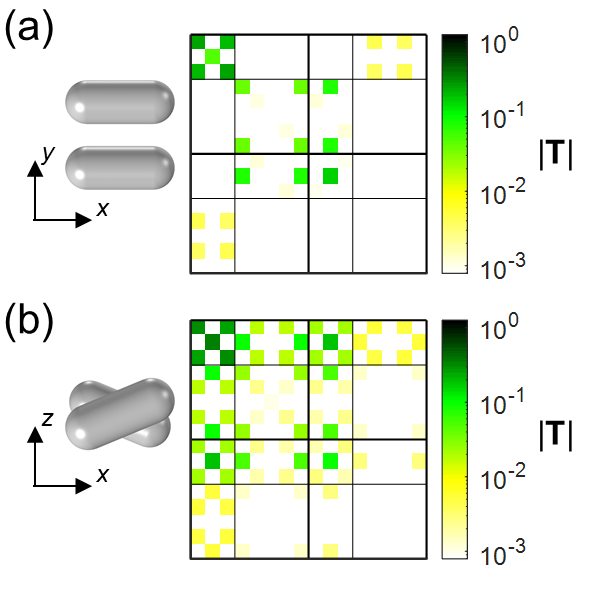}
        \caption{\(\mathbf{T}\)-matrices of the meta-atoms discussed in the main text: (a) plasmonic double-bar in Fig.~1 and (b) twisted double-bar in Fig.~2}
        \label{fig:main}
    \end{figure}

    Here, we present \(\mathbf{T}\)-matrices of meta-atoms, whose \(\boldsymbol{\alpha}\)-tensors are presented in the main text. Compared to the \(\boldsymbol{\alpha}\)-tensors with only a few components are visible (Fig.~1b and 2b), \(\mathbf{T}\)-matrices show more complicated structures, which are difficult to interpret (Fig.~\ref{fig:main}a,b) because of the spherical basis. In the following sections, we demonstrate analysis of meta-atoms from their retrieved \(\boldsymbol{\alpha}\)-tensors.
    
\subsection{Plasmonic nanorod} 
\begin{figure}[!h]
        \centering
        \includegraphics[width=80mm]{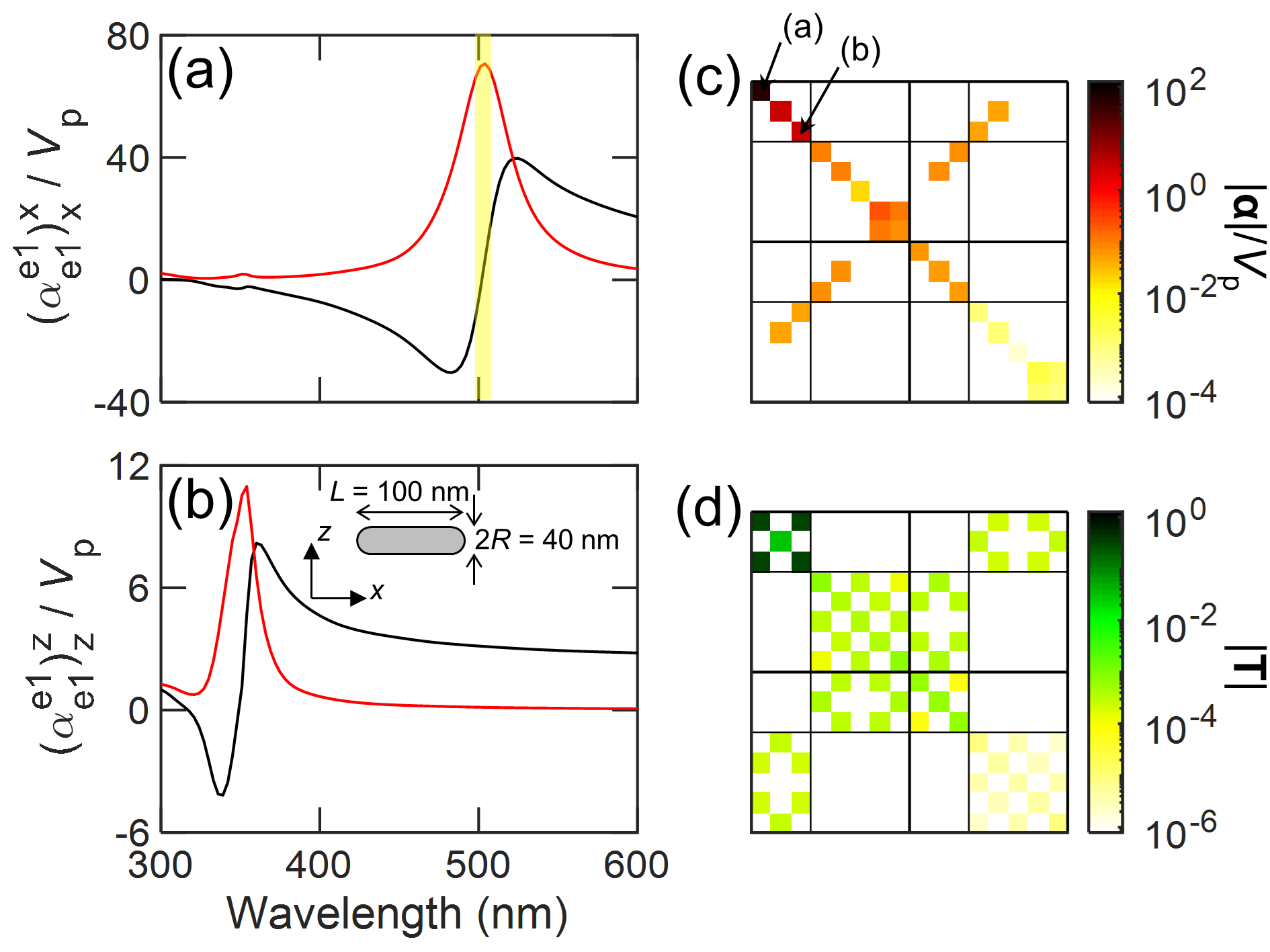}
        \caption{(a) \((\alpha_{e1}^{e1})_{x}^{x}\) and (b) \((\alpha_{e1}^{e1})_{z}^{z}\) of a plasmonic nanorod oriented in \(x\)-direction (shown in inset), and its (c) \(\boldsymbol{\alpha}\)-tensor and (d) \(\mathbf{T}\)-matrix at \(\lambda\) = 660~nm.}
        \label{fig:nanorod}
    \end{figure}

    First, we analyze a plasmonic nanorod with strongly anisotropic response, which is excited when the incident electric field is parallel to the nanorod axis. This long-axis mode is also known to be strongly redshifted compared to the short-axis mode. The retrieved \(\boldsymbol{\alpha}\)-tensor of a Ag nanorod clearly demonstrates this feature (Fig.~\ref{fig:nanorod}a,b). ED response in nanorod-axis (\(x\)) direction is red-shifted and stronger as can be seen from \((\alpha_{e1}^{e1})_{x}^{x}\) than in short-axis direction shown in \((\alpha_{e1}^{e1})_{z}^{z}\), which is blue-shifted and far weaker than \((\alpha_{e1}^{e1})_{x}^{x}\).
    In addition, a plasmonic nanorod has only one dominant \(\boldsymbol{\alpha}\)-tensor component (note that the colorbar is in logarithmic scale). This strongly anisotropic response of a plasmonic nanorod allows it to be safely approximated as a point polarizable anisotropic element with \((\alpha_{e1}^{e1})_{x}^{x}\) present as in Fig.~3 in the main text.
    
\subsection{High-index dielectric spheres} 
\begin{figure}[!ht]
        \centering
        \includegraphics[width=84mm]{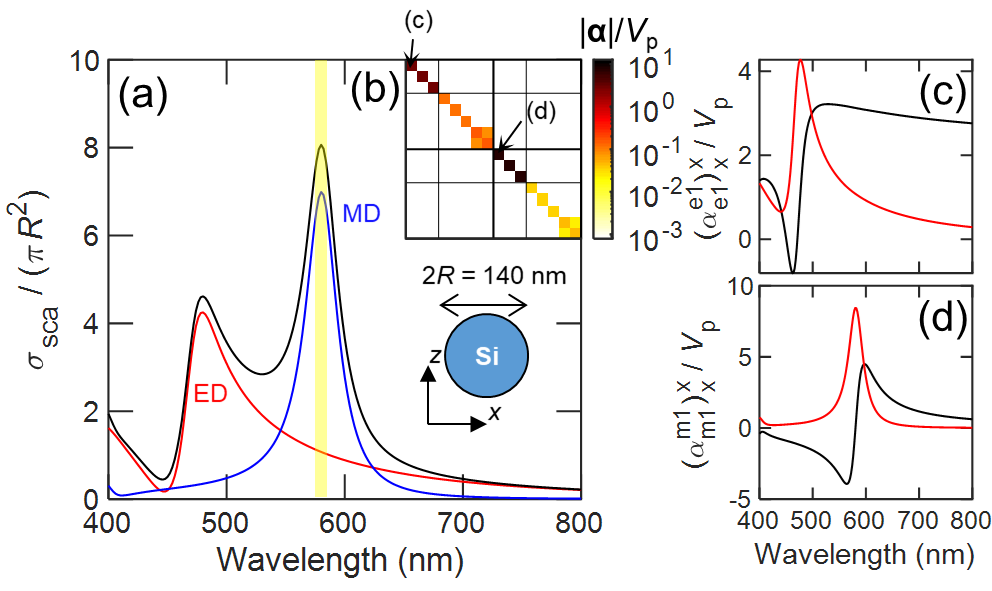}
        \caption{(a) Multipole-decomposed scattering cross-sections of a Si sphere with $R=70~\text{nm}$. (b) \(\boldsymbol{\alpha}\)-tensor at $\lambda=580~\text{nm}$, and spectra of (c) $(\alpha_{e1}^{e1})_{x}^{x}$ and (d) $(\alpha_{m1}^{m1})_{x}^{x}$.}
        \label{fig:dielectric_sphere}
    \end{figure}

    Recently, high-refractive-index dielectric nanoparticles have been noted for their low loss and higher-order multipole modes coming from Mie-like resonances \cite{Kuznetsov2012}. A small Si sphere with $R=70~\text{nm}$ by planewave incidence shows strong MD radiation at 580~nm (Fig.~\ref{fig:dielectric_sphere}a). 
    The ED and MD resonances originates from \(\alpha_{e1}^{e1}\) and \(\alpha_{m1}^{m1}\), respectively. Due to the isotropic response coming from the spherical symmetry, \(\apnm{e1}{x}{e1}{x} = \apnm{e1}{y}{e1}{y} = \apnm{e1}{y}{e1}{y}\) and \(\apnm{e2}{xy}{e2}{xy} = \apnm{e2}{xz}{e2}{xz} = \apnm{e2}{yz}{e2}{yz} = \frac{3}{4}\apnm{e2}{xx}{e2}{xx} = \frac{3}{4}\apnm{e2}{yy}{e2}{yy} = -\frac{3}{2}\apnm{e2}{yy}{e2}{xx} = -\frac{3}{2}\apnm{e2}{xx}{e2}{yy}\) (Fig.~\ref{fig:dielectric_sphere}b), and the same goes for magnetic-magnetic transitions.
    
\subsection{Split-ring resonators} 
\begin{figure}[!h]
        \centering
        \includegraphics[width=84mm]{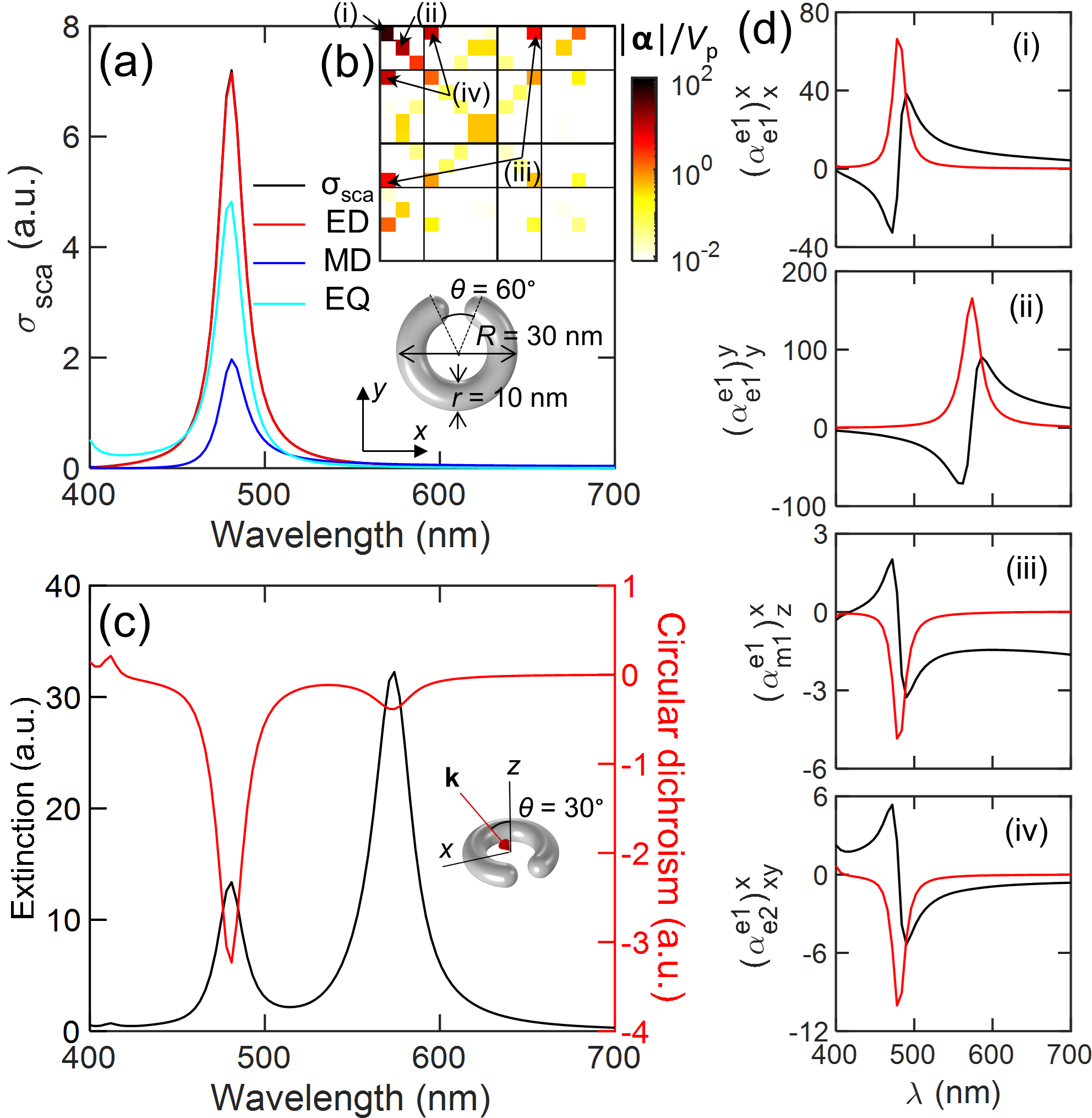}
        \caption{Ag split-ring resonator. (a) Multipole-decomposed scattering cross-section at x-polarized planewave incidence propagating in z-direction. (c) Retrieved \(\boldsymbol{\alpha}\)-tensor at 480~nm. (b) Extinction = \(\sigma_\mathrm{ext}^{+}+\sigma_\mathrm{ext}^{-}\) and Circular dichroism = \(\sigma_\mathrm{ext}^{+}-\sigma_\mathrm{ext}^{-}\), where \(\sigma_\mathrm{ext}^{\pm}\) are extinction cross-sections at obliquely incident left- and right-circularly-polarized planewaves. (d) Spectra of \((\alpha_{e1}^{e1})_{x}^{x}\), \((\alpha_{e1}^{e1})_{y}^{y}\), \((\alpha_{m1}^{e1})_{z}^{x}\), and \((\alpha_{e2}^{e1})_{xy}^{x}\). $R=30~\text{nm}$, $r=10~\text{nm}$, $\theta=60^\circ$.}
        \label{fig:SRR}
    \end{figure}

    A split-ring resonator (SRR) is one of the most widely studied elements for achieving optically magnetic responses. Interestingly, the origin of this magnetic dipole mode can be explained from the retrieved \(\boldsymbol{\alpha}\)-tensor (Fig.~\ref{fig:SRR}b). Incident \(x\)-polarized electric field on SRR generates current loop in \(xy\)-plane, which corresponds to magnetic dipole moment oriented in z-direction. This transition is visible in the $(\alpha_{m1}^{e1})_{z}^{x}$ component, which shows transition from $E_x$ into $D^m_z$.
    However, it should be noted that the magnetic response of SRR is rather weak in visible regime due to large Ohmic losses \cite{Dolling2005}, as can be seen from the weak scattering cross-section intensities (Fig.~\ref{fig:SRR}a). 
    
    Another interesting property from SRR is extrinsic chirality, or helicity dependent chiral response from a geometrically achiral structure at obliquely incident field \cite{Sersic2012}. It is counter-intuitive that an achiral structure can undergo chiral interaction. Extrinsic chirality occurs because incident field has a defined wavevector, which is not included in the mirror plane of the system.
    At oblique incidence, two resonances at 480~nm and 680~nm are observed (Fig.~\ref{fig:SRR}c), which corresponds to dominantly $(\alpha_{e1}^{e1})_{x}^{x}$ and $(\alpha_{e1}^{e1})_{y}^{y}$, respectively. 
    However, the two modes cannot explain extrinsic chirality at the oblique incidence (Fig.~\ref{fig:SRR}d). From the retrieved \(\boldsymbol{\alpha}\)-tensor, this extrinsic chirality comes from magneto-electric coupling term $(\alpha_{m1}^{e1})_{z}^{x}$
    \cite{Sersic2012, Hu2016}. It should be noted that this magneto-electric coupling term disappears for asymmetric SRR \cite{Liu2016} due to even-parity.

\subsection{Helicoids} 
\begin{figure}[!h]
        \centering
        \includegraphics[width=84mm]{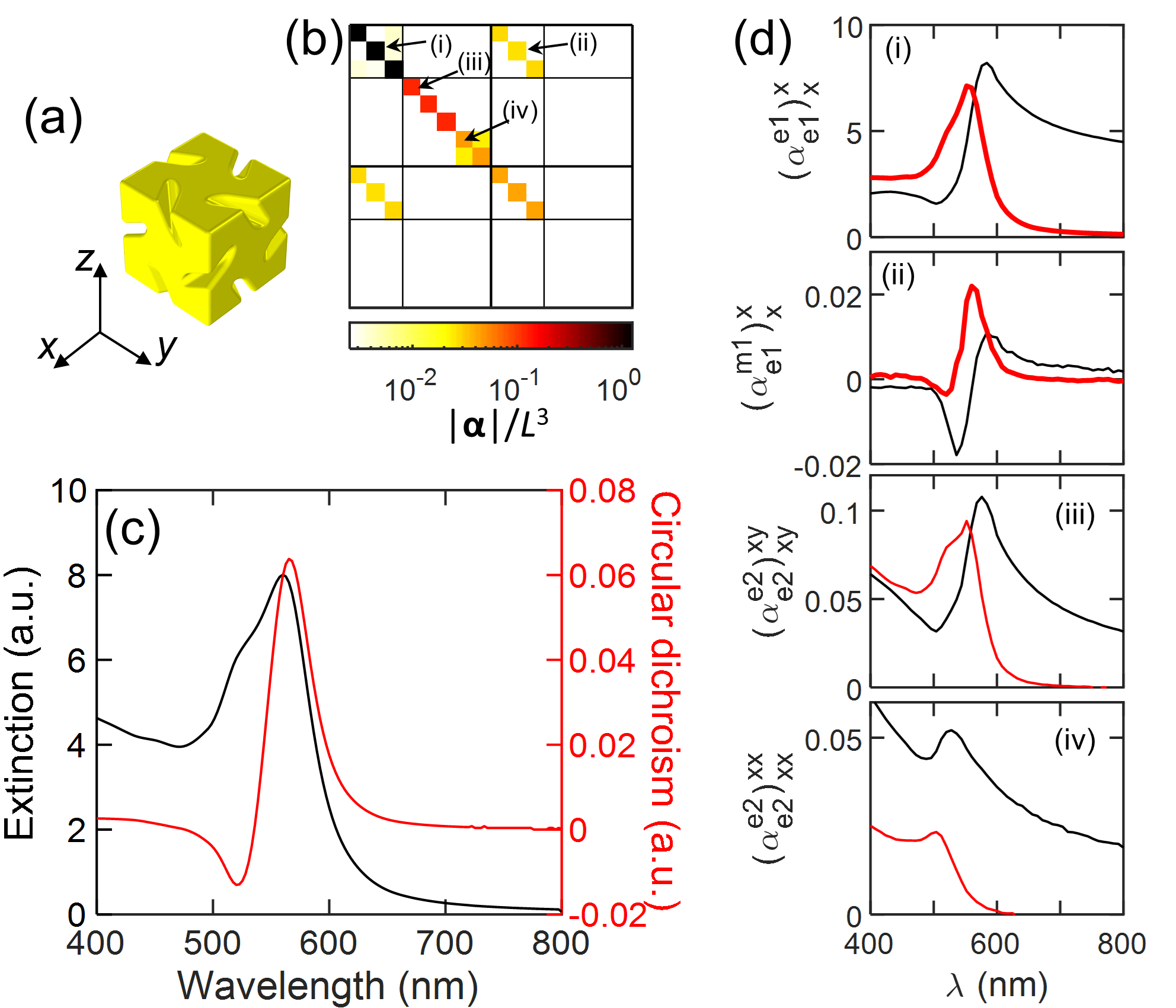}
        \caption{(a) The schematics of Au helicoid with side length 50~nm. (b) The retrieved \(\boldsymbol{\alpha}\)-tensor at 580~nm. (c) Extinction and circular dichorism. (d) Spectra of \((\alpha_{e1}^{e1})_{x}^{x}\), \((\alpha_{e1}^{m1})_{x}^{x}\), \((\alpha_{e2}^{e2})_{xx}^{xx}\), and \((\alpha_{e2}^{e2})_{xy}^{xy}\).}
        \label{fig:helicoid}
    \end{figure}

    Finally, we present \(\boldsymbol{\alpha}\)-tensor of an interesting system with 4-fold rotational symmetry without inversion symmetry. Such system was recently demonstrated in Au helicoids synthesized in solution-phase \cite{Lee2018}. Due to the 4-fold rotational symmetry in 3D space, only diagonal terms appear in dipole order with degenerate  \(xx\), \(yy\), and \(zz\) components; that is, \(\apnm{e1}{x}{e1}{x} = \apnm{e1}{y}{e1}{y} = \apnm{e1}{y}{e1}{y}\), and the same goes for \(\apn{e1}{m1}, \apn{m1}{e1}, and \apn{m1}{m1}\).
    Notably, chiral response is preserved even with this high symmetry. Due to the small size of this particle, dipolar order is sufficient to describe both achiral and chiral responses (Fig.~\ref{fig:helicoid}c) from the imaginary parts of \((\alpha_{e1}^{e1})_{x}^{x}\) and \((\alpha_{e1}^{m1})_{x}^{x}\), respectively (Fig.~\ref{fig:helicoid}d). Finally, EQ--EQ transition terms also appear and may become nonnegligible at larger sizes \cite{Lee2018}.
    Due to the symmetry, 
    \(\apnm{e2}{xy}{e2}{xy} = \apnm{e2}{xz}{e2}{xz} = \apnm{e2}{yz}{e2}{yz}\)
    and \(\apnm{e2}{xx}{e2}{xx} = \apnm{e2}{yy}{e2}{yy} = -2\apnm{e2}{yy}{e2}{xx} = -2\apnm{e2}{xx}{e2}{yy}\).
    
\section{Quasistatic polarizablities}\label{sec:quasistatic}
    The multipole approach has been successful in providing simple analytic form for describing small nanoparticles. This allowed modelling small dielectric, plasmonic, or chiral nanospheres for many different phenomena including plasmon-enhanced scattering, optical trapping, and chiral optical forces, to name a few. 
    In this section, we compare quasistatic polarizability expressions with the exact polarizability.
    The quasistatic expressions can be applied to very small nanoparticles and do not describe higher-order multipole contributions that arise in near-field interactions or larger particles or clusters. 
    
    \begin{figure}[!h]
        \centering
        \includegraphics{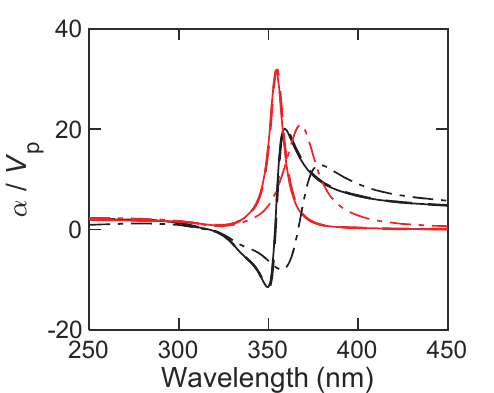}
        \caption{Polarizability of a Ag sphere normalized by its volume. Solid line: quasistatic; dashed line: exact, small (\(R\) = 5~nm); dot-dashed line: exact, large (\(R\) = 30~nm); black: real; red: imaginary.}
        \label{fig:qs_nanosphere}
    \end{figure}
    Notably, subwavelength nanospheres have often been expressed by the quasistatic polarizability
    \begin{equation}
        \alpha = 4\pi R^3 \frac{\epsilon_r-1}{\epsilon_r+2},
    \end{equation}
    where $\epsilon_r$ is the relative permittivity. 
    Note that the quasistatic expression normalized by the sphere volume is independent of the radius. For a small Ag sphere with $R$ = 10~nm, the quasistatic limit (solid line) shows good agreement with the exact polarizability (dashed line), but for a larger Ag sphere with \(R\) = 30~nm, the exact polarizability shows red-shifted and broadened resonance due to the larger radiative damping, which the quasistatic model cannot incorporate (Fig.~\ref{fig:qs_nanosphere}).

    \begin{figure}[!h]
        \centering
        \includegraphics[width=80mm]{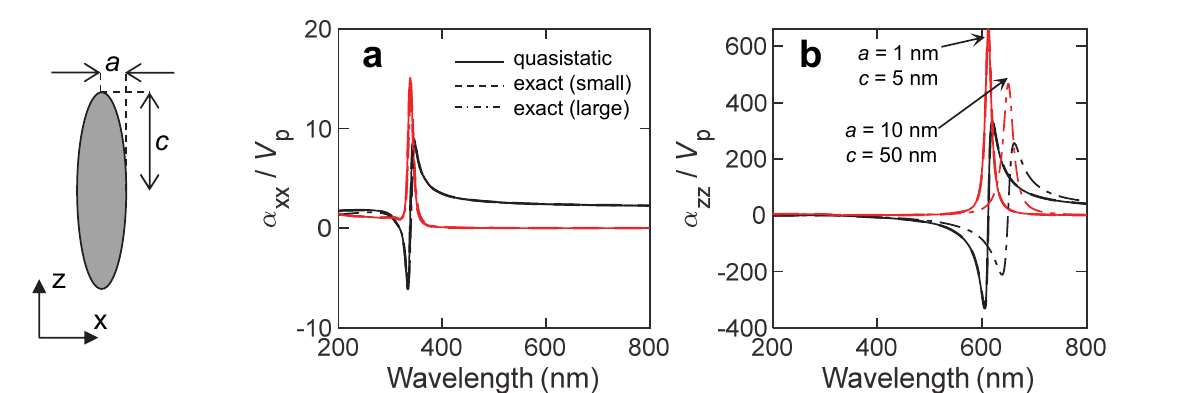}
        \caption{Polarizabilities (a) \(\alpha_{xx}\) and (b) \(\alpha_{zz}\) of an Ag nanorod. Solid line: quasistatic; dashed line: exact, small (\(a\) = 1~nm and \(c\) = 5~nm); dot-dashed line: exact, large (\(a\) = 10~nm and \(c\) = 50~nm); black: real; red: imaginary.}
        \label{fig:qs_nanorod}
    \end{figure}
    The quasistatic expression for polarizability of subwavelength nanorods was only recently reported as simple analytic form \cite{Moroz2009, Majic2017}
    \begin{subequations}
    \begin{align}
        \begin{split}
            \alpha_{x}^{x} &= 4\pi a^2c \frac{\epsilon_r-1}{3+3L_x(\epsilon_r-1)}
        \end{split}\\
        \begin{split}
            \alpha_{z}^{z} &= 4\pi a^2c \frac{\epsilon_r-1}{3+3L_z(\epsilon_r-1)}
        \end{split}\\
        \begin{split}
            L_z &= \frac{1-e^2}{e^2}{\Big[}\frac{1}{2e}\ln{\frac{1+e}{1-e}}-1{\Big]}
        \end{split}\\
        \begin{split}
            L_x &= (1-L_z)/2
        \end{split}
    \end{align}
    \end{subequations}
    where the focal length \(f=\sqrt{c^2-a^2}\) and the eccentricity \(e=f/c\). Note that the quasistatic expression for nanorod normalized by its volume also is independent of the size, but only depends on the eccentricity. For the short-axis mode, the quasistatic and exact polarizabilities agree well (Fig.~\ref{fig:qs_nanorod}a). For the long-axis mode, quasistatic polarizability (solid line) agrees well with the exact polarizability of the small nanorod (dashed line), but the exact polarizability of the larger nanorod shows red-shifted and broadened response due to the larger radiative damping (Fig.~\ref{fig:qs_nanorod}b).

    \begin{figure}[!h]
        \centering
        \includegraphics[width=80mm]{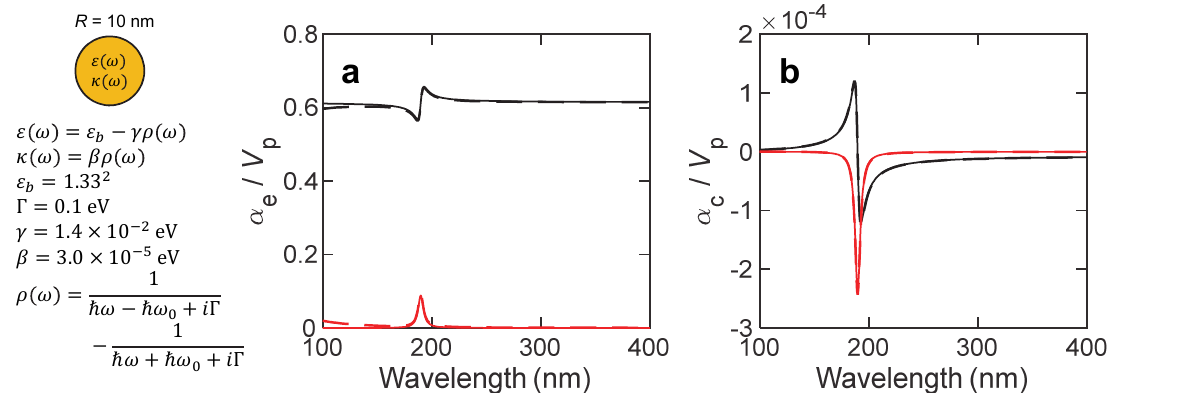}
        \caption{Polarizabilities (a) $\alpha_e$ and (b) $\alpha_c$ of a chiral sphere with $R=10~\text{nm}$. Solid line: quasistatic; dashed line: exact; black: real; red: imaginary.}
        \label{fig:qs_chiral}
    \end{figure}    
    Beyond the ED approximation, the quasistatic polarizabilities of a chiral sphere is given as \cite{Canaguier-Durand2015}
    \begin{subequations}
        \begin{equation}
            \alpha_{e}= 4\pi R^3 \frac{(\epsilon_r-1)(\mu_r+2)-\kappa^2}{(\epsilon_r+2)(\mu_r+2)-\kappa^2}
        \end{equation}
        \begin{equation}
            \alpha_{m}= 4\pi R^3 \frac{(\mu_r-1)(\epsilon_r+2)-\kappa^2}{(\mu_r+2)(\epsilon_r+2)-\kappa^2}
        \end{equation}
        \begin{equation}
            \alpha_{c}= 4\pi R^3 \frac{3\kappa}{(\epsilon_r+2)(\mu_r+2)-\kappa^2}
        \end{equation}
    \end{subequations}
    where $\mu_r$ and $\kappa$ are relative relative permeability and chirality parameter of the chiral sphere, respectively. The quasistatic and exact polarizabilities of a small nonmagnetic chiral sphere agree very well (Fig.~\ref{fig:qs_chiral}).

\begin{thebibliography}{82}%
\makeatletter
\providecommand \@ifxundefined [1]{%
 \@ifx{#1\undefined}
}%
\providecommand \@ifnum [1]{%
 \ifnum #1\expandafter \@firstoftwo
 \else \expandafter \@secondoftwo
 \fi
}%
\providecommand \@ifx [1]{%
 \ifx #1\expandafter \@firstoftwo
 \else \expandafter \@secondoftwo
 \fi
}%
\providecommand \natexlab [1]{#1}%
\providecommand \enquote  [1]{``#1''}%
\providecommand \bibnamefont  [1]{#1}%
\providecommand \bibfnamefont [1]{#1}%
\providecommand \citenamefont [1]{#1}%
\providecommand \href@noop [0]{\@secondoftwo}%
\providecommand \href [0]{\begingroup \@sanitize@url \@href}%
\providecommand \@href[1]{\@@startlink{#1}\@@href}%
\providecommand \@@href[1]{\endgroup#1\@@endlink}%
\providecommand \@sanitize@url [0]{\catcode `\\12\catcode `\$12\catcode
  `\&12\catcode `\#12\catcode `\^12\catcode `\_12\catcode `\%12\relax}%
\providecommand \@@startlink[1]{}%
\providecommand \@@endlink[0]{}%
\providecommand \url  [0]{\begingroup\@sanitize@url \@url }%
\providecommand \@url [1]{\endgroup\@href {#1}{\urlprefix }}%
\providecommand \urlprefix  [0]{URL }%
\providecommand \Eprint [0]{\href }%
\providecommand \doibase [0]{https://doi.org/}%
\providecommand \selectlanguage [0]{\@gobble}%
\providecommand \bibinfo  [0]{\@secondoftwo}%
\providecommand \bibfield  [0]{\@secondoftwo}%
\providecommand \translation [1]{[#1]}%
\providecommand \BibitemOpen [0]{}%
\providecommand \bibitemStop [0]{}%
\providecommand \bibitemNoStop [0]{.\EOS\space}%
\providecommand \EOS [0]{\spacefactor3000\relax}%
\providecommand \BibitemShut  [1]{\csname bibitem#1\endcsname}%
\let\auto@bib@innerbib\@empty
\bibitem [{\citenamefont {Mühlig}\ \emph {et~al.}(2011)\citenamefont
  {Mühlig}, \citenamefont {Menzel}, \citenamefont {Rockstuhl},\ and\
  \citenamefont {Lederer}}]{Muhlig2011}%
  \BibitemOpen
  \bibfield  {author} {\bibinfo {author} {\bibfnamefont {S.}~\bibnamefont
  {Mühlig}}, \bibinfo {author} {\bibfnamefont {C.}~\bibnamefont {Menzel}},
  \bibinfo {author} {\bibfnamefont {C.}~\bibnamefont {Rockstuhl}},\ and\
  \bibinfo {author} {\bibfnamefont {F.}~\bibnamefont {Lederer}},\ }\bibfield
  {title} {\bibinfo {title} {Multipole analysis of meta-atoms},\ }\href
  {https://doi.org/https://doi.org/10.1016/j.metmat.2011.03.003} {\bibfield
  {journal} {\bibinfo  {journal} {Metamaterials}\ }\textbf {\bibinfo {volume}
  {5}},\ \bibinfo {pages} {64 } (\bibinfo {year} {2011})}\BibitemShut {NoStop}%
\bibitem [{\citenamefont {Grahn}\ \emph {et~al.}(2012)\citenamefont {Grahn},
  \citenamefont {Shevchenko},\ and\ \citenamefont {Kaivola}}]{Grahn2012}%
  \BibitemOpen
  \bibfield  {author} {\bibinfo {author} {\bibfnamefont {P.}~\bibnamefont
  {Grahn}}, \bibinfo {author} {\bibfnamefont {A.}~\bibnamefont {Shevchenko}},\
  and\ \bibinfo {author} {\bibfnamefont {M.}~\bibnamefont {Kaivola}},\
  }\bibfield  {title} {\bibinfo {title} {Electromagnetic multipole theory for
  optical nanomaterials},\ }\href
  {https://doi.org/10.1088/1367-2630/14/9/093033} {\bibfield  {journal}
  {\bibinfo  {journal} {New J. Phys}\ }\textbf {\bibinfo {volume} {14}},\
  \bibinfo {pages} {093033} (\bibinfo {year} {2012})}\BibitemShut {NoStop}%
\bibitem [{\citenamefont {Liu}\ and\ \citenamefont {Kivshar}(2017)}]{Liu2017}%
  \BibitemOpen
  \bibfield  {author} {\bibinfo {author} {\bibfnamefont {W.}~\bibnamefont
  {Liu}}\ and\ \bibinfo {author} {\bibfnamefont {Y.~S.}\ \bibnamefont
  {Kivshar}},\ }\bibfield  {title} {\bibinfo {title} {Multipolar interference
  effects in nanophotonics},\ }\href {https://doi.org/10.1098/rsta.2016.0317}
  {\bibfield  {journal} {\bibinfo  {journal} {Philos. Trans. R. Soc. A}\
  }\textbf {\bibinfo {volume} {375}},\ \bibinfo {pages} {20160317} (\bibinfo
  {year} {2017})}\BibitemShut {NoStop}%
\bibitem [{\citenamefont {Liu}\ and\ \citenamefont {Kivshar}(2018)}]{Liu2018}%
  \BibitemOpen
  \bibfield  {author} {\bibinfo {author} {\bibfnamefont {W.}~\bibnamefont
  {Liu}}\ and\ \bibinfo {author} {\bibfnamefont {Y.~S.}\ \bibnamefont
  {Kivshar}},\ }\bibfield  {title} {\bibinfo {title} {Generalized kerker
  effects in nanophotonics and meta-optics [invited]},\ }\href
  {https://doi.org/10.1364/OE.26.013085} {\bibfield  {journal} {\bibinfo
  {journal} {Opt. Express}\ }\textbf {\bibinfo {volume} {26}},\ \bibinfo
  {pages} {13085} (\bibinfo {year} {2018})}\BibitemShut {NoStop}%
\bibitem [{\citenamefont {Babicheva}\ and\ \citenamefont
  {Evlyukhin}(2017)}]{Babicheva2017}%
  \BibitemOpen
  \bibfield  {author} {\bibinfo {author} {\bibfnamefont {V.~E.}\ \bibnamefont
  {Babicheva}}\ and\ \bibinfo {author} {\bibfnamefont {A.~B.}\ \bibnamefont
  {Evlyukhin}},\ }\bibfield  {title} {\bibinfo {title} {Resonant lattice kerker
  effect in metasurfaces with electric and magnetic optical responses},\ }\href
  {https://doi.org/10.1002/lpor.201700132} {\bibfield  {journal} {\bibinfo
  {journal} {Laser Photonics Rev.}\ }\textbf {\bibinfo {volume} {11}},\
  \bibinfo {pages} {1700132} (\bibinfo {year} {2017})}\BibitemShut {NoStop}%
\bibitem [{\citenamefont {Gurvitz}\ \emph {et~al.}(2019)\citenamefont
  {Gurvitz}, \citenamefont {Ladutenko}, \citenamefont {Dergachev},
  \citenamefont {Evlyukhin}, \citenamefont {Miroshnichenko},\ and\
  \citenamefont {Shalin}}]{Gurvitz2019}%
  \BibitemOpen
  \bibfield  {author} {\bibinfo {author} {\bibfnamefont {E.~A.}\ \bibnamefont
  {Gurvitz}}, \bibinfo {author} {\bibfnamefont {K.~S.}\ \bibnamefont
  {Ladutenko}}, \bibinfo {author} {\bibfnamefont {P.~A.}\ \bibnamefont
  {Dergachev}}, \bibinfo {author} {\bibfnamefont {A.~B.}\ \bibnamefont
  {Evlyukhin}}, \bibinfo {author} {\bibfnamefont {A.~E.}\ \bibnamefont
  {Miroshnichenko}},\ and\ \bibinfo {author} {\bibfnamefont {A.~S.}\
  \bibnamefont {Shalin}},\ }\bibfield  {title} {\bibinfo {title} {The
  high-order toroidal moments and anapole states in all-dielectric photonics},\
  }\href {https://doi.org/10.1002/lpor.201800266} {\bibfield  {journal}
  {\bibinfo  {journal} {Laser Photonics Rev.}\ }\textbf {\bibinfo {volume}
  {13}},\ \bibinfo {pages} {1800266} (\bibinfo {year} {2019})}\BibitemShut
  {NoStop}%
\bibitem [{\citenamefont {Baryshnikova}\ \emph {et~al.}(2019)\citenamefont
  {Baryshnikova}, \citenamefont {Smirnova}, \citenamefont {Luk'yanchuk},\ and\
  \citenamefont {Kivshar}}]{Baryshnikova2019}%
  \BibitemOpen
  \bibfield  {author} {\bibinfo {author} {\bibfnamefont {K.~V.}\ \bibnamefont
  {Baryshnikova}}, \bibinfo {author} {\bibfnamefont {D.~A.}\ \bibnamefont
  {Smirnova}}, \bibinfo {author} {\bibfnamefont {B.~S.}\ \bibnamefont
  {Luk'yanchuk}},\ and\ \bibinfo {author} {\bibfnamefont {Y.~S.}\ \bibnamefont
  {Kivshar}},\ }\bibfield  {title} {\bibinfo {title} {Optical anapoles:
  Concepts and applications},\ }\href {https://doi.org/10.1002/adom.201801350}
  {\bibfield  {journal} {\bibinfo  {journal} {Adv. Opt. Mater}\ }\textbf
  {\bibinfo {volume} {7}},\ \bibinfo {pages} {1801350} (\bibinfo {year}
  {2019})}\BibitemShut {NoStop}%
\bibitem [{\citenamefont {Terekhov}\ \emph {et~al.}(2019)\citenamefont
  {Terekhov}, \citenamefont {Babicheva}, \citenamefont {Baryshnikova},
  \citenamefont {Shalin}, \citenamefont {Karabchevsky},\ and\ \citenamefont
  {Evlyukhin}}]{Terekhov2019}%
  \BibitemOpen
  \bibfield  {author} {\bibinfo {author} {\bibfnamefont {P.~D.}\ \bibnamefont
  {Terekhov}}, \bibinfo {author} {\bibfnamefont {V.~E.}\ \bibnamefont
  {Babicheva}}, \bibinfo {author} {\bibfnamefont {K.~V.}\ \bibnamefont
  {Baryshnikova}}, \bibinfo {author} {\bibfnamefont {A.~S.}\ \bibnamefont
  {Shalin}}, \bibinfo {author} {\bibfnamefont {A.}~\bibnamefont
  {Karabchevsky}},\ and\ \bibinfo {author} {\bibfnamefont {A.~B.}\ \bibnamefont
  {Evlyukhin}},\ }\bibfield  {title} {\bibinfo {title} {Multipole analysis of
  dielectric metasurfaces composed of nonspherical nanoparticles and lattice
  invisibility effect},\ }\href {https://doi.org/10.1103/PhysRevB.99.045424}
  {\bibfield  {journal} {\bibinfo  {journal} {Phys. Rev. B}\ }\textbf {\bibinfo
  {volume} {99}},\ \bibinfo {pages} {045424} (\bibinfo {year}
  {2019})}\BibitemShut {NoStop}%
\bibitem [{\citenamefont {Gallinet}\ and\ \citenamefont
  {Martin}(2011)}]{Gallinet2011a}%
  \BibitemOpen
  \bibfield  {author} {\bibinfo {author} {\bibfnamefont {B.}~\bibnamefont
  {Gallinet}}\ and\ \bibinfo {author} {\bibfnamefont {O.~J.~F.}\ \bibnamefont
  {Martin}},\ }\bibfield  {title} {\bibinfo {title} {Ab initio theory of fano
  resonances in plasmonic nanostructures and metamaterials},\ }\href@noop {}
  {\bibfield  {journal} {\bibinfo  {journal} {Phys. Rev. B}\ }\textbf {\bibinfo
  {volume} {83}},\ \bibinfo {pages} {235427} (\bibinfo {year}
  {2011})}\BibitemShut {NoStop}%
\bibitem [{\citenamefont {Suryadharma}\ \emph {et~al.}(2019)\citenamefont
  {Suryadharma}, \citenamefont {Rockstuhl}, \citenamefont {Martin},\ and\
  \citenamefont {Fernandez-Corbaton}}]{Suryadharma2019}%
  \BibitemOpen
  \bibfield  {author} {\bibinfo {author} {\bibfnamefont {R.~N.~S.}\
  \bibnamefont {Suryadharma}}, \bibinfo {author} {\bibfnamefont
  {C.}~\bibnamefont {Rockstuhl}}, \bibinfo {author} {\bibfnamefont {O.~J.~F.}\
  \bibnamefont {Martin}},\ and\ \bibinfo {author} {\bibfnamefont
  {I.}~\bibnamefont {Fernandez-Corbaton}},\ }\bibfield  {title} {\bibinfo
  {title} {Quantifying fano properties in self-assembled metamaterials},\
  }\href@noop {} {\bibfield  {journal} {\bibinfo  {journal} {Phys. Rev. B}\
  }\textbf {\bibinfo {volume} {99}},\ \bibinfo {pages} {195416} (\bibinfo
  {year} {2019})}\BibitemShut {NoStop}%
\bibitem [{\citenamefont {Smirnova}\ and\ \citenamefont
  {Kivshar}(2016)}]{Smirnova2016}%
  \BibitemOpen
  \bibfield  {author} {\bibinfo {author} {\bibfnamefont {D.}~\bibnamefont
  {Smirnova}}\ and\ \bibinfo {author} {\bibfnamefont {Y.~S.}\ \bibnamefont
  {Kivshar}},\ }\bibfield  {title} {\bibinfo {title} {Multipolar nonlinear
  nanophotonics},\ }\href {https://doi.org/10.1364/OPTICA.3.001241} {\bibfield
  {journal} {\bibinfo  {journal} {Optica}\ }\textbf {\bibinfo {volume} {3}},\
  \bibinfo {pages} {1241} (\bibinfo {year} {2016})}\BibitemShut {NoStop}%
\bibitem [{\citenamefont {Mishchenko}(2008)}]{Mishchenko2008}%
  \BibitemOpen
  \bibfield  {author} {\bibinfo {author} {\bibfnamefont {M.~I.}\ \bibnamefont
  {Mishchenko}},\ }\bibfield  {title} {\bibinfo {title} {Multiple scattering,
  radiative transfer, and weak localization in discrete random media: Unified
  microphysical approach},\ }\href {https://doi.org/10.1029/2007RG000230}
  {\bibfield  {journal} {\bibinfo  {journal} {Rev. Geophys}\ }\textbf {\bibinfo
  {volume} {46}} (\bibinfo {year} {2008})}\BibitemShut {NoStop}%
\bibitem [{\citenamefont {Pocock}\ \emph {et~al.}(2018)\citenamefont {Pocock},
  \citenamefont {Xiao}, \citenamefont {Huidobro},\ and\ \citenamefont
  {Giannini}}]{Pocock2018}%
  \BibitemOpen
  \bibfield  {author} {\bibinfo {author} {\bibfnamefont {S.~R.}\ \bibnamefont
  {Pocock}}, \bibinfo {author} {\bibfnamefont {X.}~\bibnamefont {Xiao}},
  \bibinfo {author} {\bibfnamefont {P.~A.}\ \bibnamefont {Huidobro}},\ and\
  \bibinfo {author} {\bibfnamefont {V.}~\bibnamefont {Giannini}},\ }\bibfield
  {title} {\bibinfo {title} {Topological plasmonic chain with retardation and
  radiative effects},\ }\href {https://doi.org/10.1021/acsphotonics.8b00117}
  {\bibfield  {journal} {\bibinfo  {journal} {ACS Photonics}\ }\textbf
  {\bibinfo {volume} {5}},\ \bibinfo {pages} {2271} (\bibinfo {year}
  {2018})}\BibitemShut {NoStop}%
\bibitem [{\citenamefont {Sadrieva}\ \emph {et~al.}(2019)\citenamefont
  {Sadrieva}, \citenamefont {Frizyuk}, \citenamefont {Petrov}, \citenamefont
  {Kivshar},\ and\ \citenamefont {Bogdanov}}]{Sadrieva2019}%
  \BibitemOpen
  \bibfield  {author} {\bibinfo {author} {\bibfnamefont {Z.}~\bibnamefont
  {Sadrieva}}, \bibinfo {author} {\bibfnamefont {K.}~\bibnamefont {Frizyuk}},
  \bibinfo {author} {\bibfnamefont {M.}~\bibnamefont {Petrov}}, \bibinfo
  {author} {\bibfnamefont {Y.}~\bibnamefont {Kivshar}},\ and\ \bibinfo {author}
  {\bibfnamefont {A.}~\bibnamefont {Bogdanov}},\ }\bibfield  {title} {\bibinfo
  {title} {Multipolar origin of bound states in the continuum},\ }\href@noop {}
  {\bibfield  {journal} {\bibinfo  {journal} {Phys. Rev. B}\ }\textbf {\bibinfo
  {volume} {100}},\ \bibinfo {pages} {115303} (\bibinfo {year}
  {2019})}\BibitemShut {NoStop}%
\bibitem [{\citenamefont {Jackson}(1999)}]{Jackson1999}%
  \BibitemOpen
  \bibfield  {author} {\bibinfo {author} {\bibfnamefont {J.~D.}\ \bibnamefont
  {Jackson}},\ }\href@noop {} {\emph {\bibinfo {title} {Classical
  electrodynamics}}},\ \bibinfo {edition} {3rd}\ ed.\ (\bibinfo  {publisher}
  {Wiley},\ \bibinfo {address} {New York},\ \bibinfo {year} {1999})\BibitemShut
  {NoStop}%
\bibitem [{\citenamefont {Alaee}\ \emph {et~al.}(2018)\citenamefont {Alaee},
  \citenamefont {Rockstuhl},\ and\ \citenamefont
  {Fernandez-Corbaton}}]{Alaee2018}%
  \BibitemOpen
  \bibfield  {author} {\bibinfo {author} {\bibfnamefont {R.}~\bibnamefont
  {Alaee}}, \bibinfo {author} {\bibfnamefont {C.}~\bibnamefont {Rockstuhl}},\
  and\ \bibinfo {author} {\bibfnamefont {I.}~\bibnamefont
  {Fernandez-Corbaton}},\ }\bibfield  {title} {\bibinfo {title} {An
  electromagnetic multipole expansion beyond the long-wavelength
  approximation},\ }\href
  {https://doi.org/https://doi.org/10.1016/j.optcom.2017.08.064} {\bibfield
  {journal} {\bibinfo  {journal} {Opt. Commun.}\ }\textbf {\bibinfo {volume}
  {407}},\ \bibinfo {pages} {17} (\bibinfo {year} {2018})}\BibitemShut
  {NoStop}%
\bibitem [{\citenamefont {Fruhnert}\ \emph {et~al.}(2017)\citenamefont
  {Fruhnert}, \citenamefont {Fernandez-Corbaton}, \citenamefont {Yannopapas},\
  and\ \citenamefont {Rockstuhl}}]{Fruhnert2017}%
  \BibitemOpen
  \bibfield  {author} {\bibinfo {author} {\bibfnamefont {M.}~\bibnamefont
  {Fruhnert}}, \bibinfo {author} {\bibfnamefont {I.}~\bibnamefont
  {Fernandez-Corbaton}}, \bibinfo {author} {\bibfnamefont {V.}~\bibnamefont
  {Yannopapas}},\ and\ \bibinfo {author} {\bibfnamefont {C.}~\bibnamefont
  {Rockstuhl}},\ }\bibfield  {title} {\bibinfo {title} {{Computing the T-matrix
  of a scattering object with multiple plane wave illuminations}},\ }\href
  {https://doi.org/10.3762/bjnano.8.66} {\bibfield  {journal} {\bibinfo
  {journal} {Beilstein J. Nanotechnol.}\ }\textbf {\bibinfo {volume} {8}},\
  \bibinfo {pages} {614} (\bibinfo {year} {2017})}\BibitemShut {NoStop}%
\bibitem [{\citenamefont {Savinov}\ \emph {et~al.}(2019)\citenamefont
  {Savinov}, \citenamefont {Papasimakis}, \citenamefont {Tsai},\ and\
  \citenamefont {Zheludev}}]{Savinov2019}%
  \BibitemOpen
  \bibfield  {author} {\bibinfo {author} {\bibfnamefont {V.}~\bibnamefont
  {Savinov}}, \bibinfo {author} {\bibfnamefont {N.}~\bibnamefont
  {Papasimakis}}, \bibinfo {author} {\bibfnamefont {D.~P.}\ \bibnamefont
  {Tsai}},\ and\ \bibinfo {author} {\bibfnamefont {N.~I.}\ \bibnamefont
  {Zheludev}},\ }\bibfield  {title} {\bibinfo {title} {{Optical anapoles}},\
  }\href {https://doi.org/10.1038/s42005-019-0167-z} {\bibfield  {journal}
  {\bibinfo  {journal} {Commun. Phys.}\ }\textbf {\bibinfo {volume} {2}},\
  \bibinfo {pages} {69} (\bibinfo {year} {2019})}\BibitemShut {NoStop}%
\bibitem [{\citenamefont {de~Vries}\ \emph {et~al.}(1998)\citenamefont
  {de~Vries}, \citenamefont {van Coevorden},\ and\ \citenamefont
  {Lagendijk}}]{DeVries1998}%
  \BibitemOpen
  \bibfield  {author} {\bibinfo {author} {\bibfnamefont {P.}~\bibnamefont
  {de~Vries}}, \bibinfo {author} {\bibfnamefont {D.~V.}\ \bibnamefont {van
  Coevorden}},\ and\ \bibinfo {author} {\bibfnamefont {A.}~\bibnamefont
  {Lagendijk}},\ }\bibfield  {title} {\bibinfo {title} {Point scatterers for
  classical waves},\ }\href {https://doi.org/10.1103/RevModPhys.70.447}
  {\bibfield  {journal} {\bibinfo  {journal} {Rev. Mod. Phys.}\ }\textbf
  {\bibinfo {volume} {70}},\ \bibinfo {pages} {447} (\bibinfo {year}
  {1998})}\BibitemShut {NoStop}%
\bibitem [{\citenamefont {Mishchenko}\ \emph {et~al.}(2010)\citenamefont
  {Mishchenko}, \citenamefont {Travis},\ and\ \citenamefont
  {Mackowski}}]{Mishchenko2010}%
  \BibitemOpen
  \bibfield  {author} {\bibinfo {author} {\bibfnamefont {M.~I.}\ \bibnamefont
  {Mishchenko}}, \bibinfo {author} {\bibfnamefont {L.~D.}\ \bibnamefont
  {Travis}},\ and\ \bibinfo {author} {\bibfnamefont {D.~W.}\ \bibnamefont
  {Mackowski}},\ }\bibfield  {title} {\bibinfo {title} {T-matrix method and its
  applications to electromagnetic scattering by particles: A current
  perspective},\ }\href@noop {} {\bibfield  {journal} {\bibinfo  {journal} {J.
  Quant. Spectrosc. Radiat.}\ }\textbf {\bibinfo {volume} {111}},\ \bibinfo
  {pages} {1700 } (\bibinfo {year} {2010})}\BibitemShut {NoStop}%
\bibitem [{\citenamefont {Suryadharma}\ \emph {et~al.}(2017)\citenamefont
  {Suryadharma}, \citenamefont {Fruhnert}, \citenamefont {Fernandez-Corbaton},\
  and\ \citenamefont {Rockstuhl}}]{Suryadharma2017}%
  \BibitemOpen
  \bibfield  {author} {\bibinfo {author} {\bibfnamefont {R.~N.~S.}\
  \bibnamefont {Suryadharma}}, \bibinfo {author} {\bibfnamefont
  {M.}~\bibnamefont {Fruhnert}}, \bibinfo {author} {\bibfnamefont
  {I.}~\bibnamefont {Fernandez-Corbaton}},\ and\ \bibinfo {author}
  {\bibfnamefont {C.}~\bibnamefont {Rockstuhl}},\ }\bibfield  {title} {\bibinfo
  {title} {Studying plasmonic resonance modes of hierarchical self-assembled
  meta-atoms based on their transfer matrix},\ }\href@noop {} {\bibfield
  {journal} {\bibinfo  {journal} {Phys. Rev. B}\ }\textbf {\bibinfo {volume}
  {96}},\ \bibinfo {pages} {045406} (\bibinfo {year} {2017})}\BibitemShut
  {NoStop}%
\bibitem [{\citenamefont {Garc\'{\i}a~de Abajo}(1999)}]{DeAbajo1999}%
  \BibitemOpen
  \bibfield  {author} {\bibinfo {author} {\bibfnamefont {F.~J.}\ \bibnamefont
  {Garc\'{\i}a~de Abajo}},\ }\bibfield  {title} {\bibinfo {title} {Multiple
  scattering of radiation in clusters of dielectrics},\ }\href
  {https://doi.org/10.1103/PhysRevB.60.6086} {\bibfield  {journal} {\bibinfo
  {journal} {Phys. Rev. B}\ }\textbf {\bibinfo {volume} {60}},\ \bibinfo
  {pages} {6086} (\bibinfo {year} {1999})}\BibitemShut {NoStop}%
\bibitem [{\citenamefont {Stout}\ \emph {et~al.}(2008)\citenamefont {Stout},
  \citenamefont {Auger},\ and\ \citenamefont {Devilez}}]{Stout2008}%
  \BibitemOpen
  \bibfield  {author} {\bibinfo {author} {\bibfnamefont {B.}~\bibnamefont
  {Stout}}, \bibinfo {author} {\bibfnamefont {J.~C.}\ \bibnamefont {Auger}},\
  and\ \bibinfo {author} {\bibfnamefont {A.}~\bibnamefont {Devilez}},\
  }\bibfield  {title} {\bibinfo {title} {Recursive t matrix algorithm for
  resonant multiple scattering: applications to localized plasmon
  excitations},\ }\href {https://doi.org/10.1364/JOSAA.25.002549} {\bibfield
  {journal} {\bibinfo  {journal} {J. Opt. Soc. Am. A}\ }\textbf {\bibinfo
  {volume} {25}},\ \bibinfo {pages} {2549} (\bibinfo {year}
  {2008})}\BibitemShut {NoStop}%
\bibitem [{\citenamefont {Stout}\ \emph {et~al.}(2011)\citenamefont {Stout},
  \citenamefont {Devilez}, \citenamefont {Rolly},\ and\ \citenamefont
  {Bonod}}]{Stout2011}%
  \BibitemOpen
  \bibfield  {author} {\bibinfo {author} {\bibfnamefont {B.}~\bibnamefont
  {Stout}}, \bibinfo {author} {\bibfnamefont {A.}~\bibnamefont {Devilez}},
  \bibinfo {author} {\bibfnamefont {B.}~\bibnamefont {Rolly}},\ and\ \bibinfo
  {author} {\bibfnamefont {N.}~\bibnamefont {Bonod}},\ }\bibfield  {title}
  {\bibinfo {title} {Multipole methods for nanoantennas design: applications to
  yagi-uda configurations},\ }\href {https://doi.org/10.1364/JOSAB.28.001213}
  {\bibfield  {journal} {\bibinfo  {journal} {J. Opt. Soc. Am. B}\ }\textbf
  {\bibinfo {volume} {28}},\ \bibinfo {pages} {1213} (\bibinfo {year}
  {2011})}\BibitemShut {NoStop}%
\bibitem [{\citenamefont {Garc\'{\i}a~de Abajo}(2007)}]{DeAbajo2007}%
  \BibitemOpen
  \bibfield  {author} {\bibinfo {author} {\bibfnamefont {F.~J.}\ \bibnamefont
  {Garc\'{\i}a~de Abajo}},\ }\bibfield  {title} {\bibinfo {title} {Colloquium:
  Light scattering by particle and hole arrays},\ }\href
  {https://doi.org/10.1103/RevModPhys.79.1267} {\bibfield  {journal} {\bibinfo
  {journal} {Rev. Mod. Phys.}\ }\textbf {\bibinfo {volume} {79}},\ \bibinfo
  {pages} {1267} (\bibinfo {year} {2007})}\BibitemShut {NoStop}%
\bibitem [{\citenamefont {Baur}\ \emph {et~al.}(2018)\citenamefont {Baur},
  \citenamefont {Sanders},\ and\ \citenamefont {Manjavacas}}]{Baur2018}%
  \BibitemOpen
  \bibfield  {author} {\bibinfo {author} {\bibfnamefont {S.}~\bibnamefont
  {Baur}}, \bibinfo {author} {\bibfnamefont {S.}~\bibnamefont {Sanders}},\ and\
  \bibinfo {author} {\bibfnamefont {A.}~\bibnamefont {Manjavacas}},\ }\bibfield
   {title} {\bibinfo {title} {Hybridization of lattice resonances},\ }\href
  {https://doi.org/10.1021/acsnano.7b08206} {\bibfield  {journal} {\bibinfo
  {journal} {ACS Nano}\ }\textbf {\bibinfo {volume} {12}},\ \bibinfo {pages}
  {1618} (\bibinfo {year} {2018})}\BibitemShut {NoStop}%
\bibitem [{\citenamefont {Evlyukhin}\ \emph {et~al.}(2010)\citenamefont
  {Evlyukhin}, \citenamefont {Reinhardt}, \citenamefont {Seidel}, \citenamefont
  {Luk'yanchuk},\ and\ \citenamefont {Chichkov}}]{Evlyukhin2010}%
  \BibitemOpen
  \bibfield  {author} {\bibinfo {author} {\bibfnamefont {A.~B.}\ \bibnamefont
  {Evlyukhin}}, \bibinfo {author} {\bibfnamefont {C.}~\bibnamefont
  {Reinhardt}}, \bibinfo {author} {\bibfnamefont {A.}~\bibnamefont {Seidel}},
  \bibinfo {author} {\bibfnamefont {B.~S.}\ \bibnamefont {Luk'yanchuk}},\ and\
  \bibinfo {author} {\bibfnamefont {B.~N.}\ \bibnamefont {Chichkov}},\
  }\bibfield  {title} {\bibinfo {title} {Optical response features of
  si-nanoparticle arrays},\ }\href {https://doi.org/10.1103/PhysRevB.82.045404}
  {\bibfield  {journal} {\bibinfo  {journal} {Phys. Rev. B}\ }\textbf {\bibinfo
  {volume} {82}},\ \bibinfo {pages} {045404} (\bibinfo {year}
  {2010})}\BibitemShut {NoStop}%
\bibitem [{\citenamefont {Babicheva}\ and\ \citenamefont
  {Evlyukhin}(2018)}]{Babicheva2018}%
  \BibitemOpen
  \bibfield  {author} {\bibinfo {author} {\bibfnamefont {V.~E.}\ \bibnamefont
  {Babicheva}}\ and\ \bibinfo {author} {\bibfnamefont {A.~B.}\ \bibnamefont
  {Evlyukhin}},\ }\bibfield  {title} {\bibinfo {title} {Metasurfaces with
  electric quadrupole and magnetic dipole resonant coupling},\ }\href
  {https://doi.org/10.1021/acsphotonics.7b01520} {\bibfield  {journal}
  {\bibinfo  {journal} {ACS Photonics}\ }\textbf {\bibinfo {volume} {5}},\
  \bibinfo {pages} {2022} (\bibinfo {year} {2018})}\BibitemShut {NoStop}%
\bibitem [{\citenamefont {Babicheva}\ and\ \citenamefont
  {Evlyukhin}(2019)}]{Babicheva2019}%
  \BibitemOpen
  \bibfield  {author} {\bibinfo {author} {\bibfnamefont {V.~E.}\ \bibnamefont
  {Babicheva}}\ and\ \bibinfo {author} {\bibfnamefont {A.~B.}\ \bibnamefont
  {Evlyukhin}},\ }\bibfield  {title} {\bibinfo {title} {Analytical model of
  resonant electromagnetic dipole-quadrupole coupling in nanoparticle arrays},\
  }\href {https://doi.org/10.1103/PhysRevB.99.195444} {\bibfield  {journal}
  {\bibinfo  {journal} {Phys. Rev. B}\ }\textbf {\bibinfo {volume} {99}},\
  \bibinfo {pages} {195444} (\bibinfo {year} {2019})}\BibitemShut {NoStop}%
\bibitem [{\citenamefont {Salary}\ \emph {et~al.}(2017)\citenamefont {Salary},
  \citenamefont {Forouzmand},\ and\ \citenamefont
  {Mosallaei}}]{MahdiSalary2017}%
  \BibitemOpen
  \bibfield  {author} {\bibinfo {author} {\bibfnamefont {M.~M.}\ \bibnamefont
  {Salary}}, \bibinfo {author} {\bibfnamefont {A.}~\bibnamefont {Forouzmand}},\
  and\ \bibinfo {author} {\bibfnamefont {H.}~\bibnamefont {Mosallaei}},\
  }\bibfield  {title} {\bibinfo {title} {Model order reduction of large-scale
  metasurfaces using a hierarchical dipole approximation},\ }\href
  {https://doi.org/10.1021/acsphotonics.6b00568} {\bibfield  {journal}
  {\bibinfo  {journal} {ACS Photonics}\ }\textbf {\bibinfo {volume} {4}},\
  \bibinfo {pages} {63} (\bibinfo {year} {2017})}\BibitemShut {NoStop}%
\bibitem [{\citenamefont {Watson}\ \emph {et~al.}(2017)\citenamefont {Watson},
  \citenamefont {Jenkins},\ and\ \citenamefont {Ruostekoski}}]{Watson2017}%
  \BibitemOpen
  \bibfield  {author} {\bibinfo {author} {\bibfnamefont {D.~W.}\ \bibnamefont
  {Watson}}, \bibinfo {author} {\bibfnamefont {S.~D.}\ \bibnamefont
  {Jenkins}},\ and\ \bibinfo {author} {\bibfnamefont {J.}~\bibnamefont
  {Ruostekoski}},\ }\bibfield  {title} {\bibinfo {title} {Point dipole and
  quadrupole scattering approximation to collectively responding resonator
  systems},\ }\href {https://doi.org/10.1103/PhysRevB.96.035403} {\bibfield
  {journal} {\bibinfo  {journal} {Phys. Rev. B}\ }\textbf {\bibinfo {volume}
  {96}},\ \bibinfo {pages} {035403} (\bibinfo {year} {2017})}\BibitemShut
  {NoStop}%
\bibitem [{\citenamefont {Rahimzadegan}\ \emph {et~al.}(2019)\citenamefont
  {Rahimzadegan}, \citenamefont {Arslan}, \citenamefont {Suryadharma},
  \citenamefont {Fasold}, \citenamefont {Falkner}, \citenamefont {Pertsch},
  \citenamefont {Staude},\ and\ \citenamefont {Rockstuhl}}]{Rahimzadegan2019}%
  \BibitemOpen
  \bibfield  {author} {\bibinfo {author} {\bibfnamefont {A.}~\bibnamefont
  {Rahimzadegan}}, \bibinfo {author} {\bibfnamefont {D.}~\bibnamefont
  {Arslan}}, \bibinfo {author} {\bibfnamefont {R.~N.~S.}\ \bibnamefont
  {Suryadharma}}, \bibinfo {author} {\bibfnamefont {S.}~\bibnamefont {Fasold}},
  \bibinfo {author} {\bibfnamefont {M.}~\bibnamefont {Falkner}}, \bibinfo
  {author} {\bibfnamefont {T.}~\bibnamefont {Pertsch}}, \bibinfo {author}
  {\bibfnamefont {I.}~\bibnamefont {Staude}},\ and\ \bibinfo {author}
  {\bibfnamefont {C.}~\bibnamefont {Rockstuhl}},\ }\bibfield  {title} {\bibinfo
  {title} {Disorder-induced phase transitions in the transmission of dielectric
  metasurfaces},\ }\href {https://doi.org/10.1103/PhysRevLett.122.015702}
  {\bibfield  {journal} {\bibinfo  {journal} {Phys. Rev. Lett.}\ }\textbf
  {\bibinfo {volume} {122}},\ \bibinfo {pages} {015702} (\bibinfo {year}
  {2019})}\BibitemShut {NoStop}%
\bibitem [{\citenamefont {Jenkins}\ \emph {et~al.}(2018)\citenamefont
  {Jenkins}, \citenamefont {Papasimakis}, \citenamefont {Savo}, \citenamefont
  {Zheludev},\ and\ \citenamefont {Ruostekoski}}]{Jenkins2018}%
  \BibitemOpen
  \bibfield  {author} {\bibinfo {author} {\bibfnamefont {S.~D.}\ \bibnamefont
  {Jenkins}}, \bibinfo {author} {\bibfnamefont {N.}~\bibnamefont
  {Papasimakis}}, \bibinfo {author} {\bibfnamefont {S.}~\bibnamefont {Savo}},
  \bibinfo {author} {\bibfnamefont {N.~I.}\ \bibnamefont {Zheludev}},\ and\
  \bibinfo {author} {\bibfnamefont {J.}~\bibnamefont {Ruostekoski}},\
  }\bibfield  {title} {\bibinfo {title} {Strong interactions and subradiance in
  disordered metamaterials},\ }\href
  {https://doi.org/10.1103/PhysRevB.98.245136} {\bibfield  {journal} {\bibinfo
  {journal} {Phys. Rev. B}\ }\textbf {\bibinfo {volume} {98}},\ \bibinfo
  {pages} {245136} (\bibinfo {year} {2018})}\BibitemShut {NoStop}%
\bibitem [{\citenamefont {Pattelli}\ \emph {et~al.}(2018)\citenamefont
  {Pattelli}, \citenamefont {Egel}, \citenamefont {Lemmer},\ and\ \citenamefont
  {Wiersma}}]{Pattelli2018}%
  \BibitemOpen
  \bibfield  {author} {\bibinfo {author} {\bibfnamefont {L.}~\bibnamefont
  {Pattelli}}, \bibinfo {author} {\bibfnamefont {A.}~\bibnamefont {Egel}},
  \bibinfo {author} {\bibfnamefont {U.}~\bibnamefont {Lemmer}},\ and\ \bibinfo
  {author} {\bibfnamefont {D.~S.}\ \bibnamefont {Wiersma}},\ }\bibfield
  {title} {\bibinfo {title} {Role of packing density and spatial correlations
  in strongly scattering 3d systems},\ }\href
  {https://doi.org/10.1364/OPTICA.5.001037} {\bibfield  {journal} {\bibinfo
  {journal} {Optica}\ }\textbf {\bibinfo {volume} {5}},\ \bibinfo {pages}
  {1037} (\bibinfo {year} {2018})}\BibitemShut {NoStop}%
\bibitem [{\citenamefont {Govorov}\ \emph {et~al.}(2010)\citenamefont
  {Govorov}, \citenamefont {Fan}, \citenamefont {Hernandez}, \citenamefont
  {Slocik},\ and\ \citenamefont {Naik}}]{Govorov2010}%
  \BibitemOpen
  \bibfield  {author} {\bibinfo {author} {\bibfnamefont {A.~O.}\ \bibnamefont
  {Govorov}}, \bibinfo {author} {\bibfnamefont {Z.}~\bibnamefont {Fan}},
  \bibinfo {author} {\bibfnamefont {P.}~\bibnamefont {Hernandez}}, \bibinfo
  {author} {\bibfnamefont {J.~M.}\ \bibnamefont {Slocik}},\ and\ \bibinfo
  {author} {\bibfnamefont {R.~R.}\ \bibnamefont {Naik}},\ }\bibfield  {title}
  {\bibinfo {title} {Theory of circular dichroism of nanomaterials comprising
  chiral molecules and nanocrystals: Plasmon enhancement, dipole interactions,
  and dielectric effects},\ }\href {https://doi.org/10.1021/nl100010v}
  {\bibfield  {journal} {\bibinfo  {journal} {Nano Lett.}\ }\textbf {\bibinfo
  {volume} {10}},\ \bibinfo {pages} {1374} (\bibinfo {year}
  {2010})}\BibitemShut {NoStop}%
\bibitem [{\citenamefont {Wu}\ \emph {et~al.}(2015)\citenamefont {Wu},
  \citenamefont {Wang},\ and\ \citenamefont {Zhang}}]{Wu2015}%
  \BibitemOpen
  \bibfield  {author} {\bibinfo {author} {\bibfnamefont {T.}~\bibnamefont
  {Wu}}, \bibinfo {author} {\bibfnamefont {R.}~\bibnamefont {Wang}},\ and\
  \bibinfo {author} {\bibfnamefont {X.}~\bibnamefont {Zhang}},\ }\bibfield
  {title} {\bibinfo {title} {{Plasmon-induced strong interaction between chiral
  molecules and orbital angular momentum of light}},\ }\href
  {https://doi.org/10.1038/srep18003} {\bibfield  {journal} {\bibinfo
  {journal} {Sci. Rep.}\ }\textbf {\bibinfo {volume} {5}},\ \bibinfo {pages}
  {18003} (\bibinfo {year} {2015})}\BibitemShut {NoStop}%
\bibitem [{\citenamefont {Alaee}\ \emph {et~al.}(2019)\citenamefont {Alaee},
  \citenamefont {Rockstuhl},\ and\ \citenamefont
  {Fernandez-Corbaton}}]{Alaee2019}%
  \BibitemOpen
  \bibfield  {author} {\bibinfo {author} {\bibfnamefont {R.}~\bibnamefont
  {Alaee}}, \bibinfo {author} {\bibfnamefont {C.}~\bibnamefont {Rockstuhl}},\
  and\ \bibinfo {author} {\bibfnamefont {I.}~\bibnamefont
  {Fernandez-Corbaton}},\ }\bibfield  {title} {\bibinfo {title} {Exact
  multipolar decompositions with applications in nanophotonics},\ }\href
  {https://doi.org/10.1002/adom.201800783} {\bibfield  {journal} {\bibinfo
  {journal} {Adv. Opt. Mater}\ }\textbf {\bibinfo {volume} {7}},\ \bibinfo
  {pages} {1800783} (\bibinfo {year} {2019})}\BibitemShut {NoStop}%
\bibitem [{\citenamefont {Evlyukhin}\ and\ \citenamefont
  {Chichkov}(2019)}]{Evlyukhin2019}%
  \BibitemOpen
  \bibfield  {author} {\bibinfo {author} {\bibfnamefont {A.~B.}\ \bibnamefont
  {Evlyukhin}}\ and\ \bibinfo {author} {\bibfnamefont {B.~N.}\ \bibnamefont
  {Chichkov}},\ }\bibfield  {title} {\bibinfo {title} {Multipole decompositions
  for directional light scattering},\ }\href
  {https://doi.org/10.1103/PhysRevB.100.125415} {\bibfield  {journal} {\bibinfo
   {journal} {Phys. Rev. B}\ }\textbf {\bibinfo {volume} {100}},\ \bibinfo
  {pages} {125415} (\bibinfo {year} {2019})}\BibitemShut {NoStop}%
\bibitem [{\citenamefont {Talebi}\ \emph {et~al.}(2018)\citenamefont {Talebi},
  \citenamefont {Guo},\ and\ \citenamefont {{Van Aken}}}]{Talebi2018}%
  \BibitemOpen
  \bibfield  {author} {\bibinfo {author} {\bibfnamefont {N.}~\bibnamefont
  {Talebi}}, \bibinfo {author} {\bibfnamefont {S.}~\bibnamefont {Guo}},\ and\
  \bibinfo {author} {\bibfnamefont {P.~A.}\ \bibnamefont {{Van Aken}}},\
  }\bibfield  {title} {\bibinfo {title} {{Theory and applications of toroidal
  moments in electrodynamics: Their emergence, characteristics, and
  technological relevance}},\ }\href {https://doi.org/10.1515/nanoph-2017-0017}
  {\bibfield  {journal} {\bibinfo  {journal} {Nanophotonics}\ }\textbf
  {\bibinfo {volume} {7}},\ \bibinfo {pages} {93} (\bibinfo {year}
  {2018})}\BibitemShut {NoStop}%
\bibitem [{\citenamefont {Evlyukhin}\ \emph {et~al.}(2016)\citenamefont
  {Evlyukhin}, \citenamefont {Fischer}, \citenamefont {Reinhardt},\ and\
  \citenamefont {Chichkov}}]{Evlyukhin2016}%
  \BibitemOpen
  \bibfield  {author} {\bibinfo {author} {\bibfnamefont {A.~B.}\ \bibnamefont
  {Evlyukhin}}, \bibinfo {author} {\bibfnamefont {T.}~\bibnamefont {Fischer}},
  \bibinfo {author} {\bibfnamefont {C.}~\bibnamefont {Reinhardt}},\ and\
  \bibinfo {author} {\bibfnamefont {B.~N.}\ \bibnamefont {Chichkov}},\
  }\bibfield  {title} {\bibinfo {title} {Optical theorem and multipole
  scattering of light by arbitrarily shaped nanoparticles},\ }\href
  {https://doi.org/10.1103/PhysRevB.94.205434} {\bibfield  {journal} {\bibinfo
  {journal} {Phys. Rev. B}\ }\textbf {\bibinfo {volume} {94}},\ \bibinfo
  {pages} {205434} (\bibinfo {year} {2016})}\BibitemShut {NoStop}%
\bibitem [{\citenamefont {Fernandez-Corbaton}\ \emph
  {et~al.}(2017)\citenamefont {Fernandez-Corbaton}, \citenamefont {Nanz},\ and\
  \citenamefont {Rockstuhl}}]{Fernandez-Corbaton2017}%
  \BibitemOpen
  \bibfield  {author} {\bibinfo {author} {\bibfnamefont {I.}~\bibnamefont
  {Fernandez-Corbaton}}, \bibinfo {author} {\bibfnamefont {S.}~\bibnamefont
  {Nanz}},\ and\ \bibinfo {author} {\bibfnamefont {C.}~\bibnamefont
  {Rockstuhl}},\ }\bibfield  {title} {\bibinfo {title} {On the dynamic toroidal
  multipoles from localized electric current distributions},\ }\href@noop {}
  {\bibfield  {journal} {\bibinfo  {journal} {Sci. Rep.}\ }\textbf {\bibinfo
  {volume} {7}},\ \bibinfo {pages} {7527} (\bibinfo {year} {2017})}\BibitemShut
  {NoStop}%
\bibitem [{\citenamefont {Fernandez-Corbaton}\ \emph
  {et~al.}(2015)\citenamefont {Fernandez-Corbaton}, \citenamefont {Nanz},
  \citenamefont {Alaee},\ and\ \citenamefont
  {Rockstuhl}}]{Fernandez-Corbaton2015}%
  \BibitemOpen
  \bibfield  {author} {\bibinfo {author} {\bibfnamefont {I.}~\bibnamefont
  {Fernandez-Corbaton}}, \bibinfo {author} {\bibfnamefont {S.}~\bibnamefont
  {Nanz}}, \bibinfo {author} {\bibfnamefont {R.}~\bibnamefont {Alaee}},\ and\
  \bibinfo {author} {\bibfnamefont {C.}~\bibnamefont {Rockstuhl}},\ }\bibfield
  {title} {\bibinfo {title} {Exact dipolar moments of a localized electric
  current distribution},\ }\href {https://doi.org/10.1364/OE.23.033044}
  {\bibfield  {journal} {\bibinfo  {journal} {Opt. Express}\ }\textbf {\bibinfo
  {volume} {23}},\ \bibinfo {pages} {33044} (\bibinfo {year}
  {2015})}\BibitemShut {NoStop}%
\bibitem [{\citenamefont {Arango}\ and\ \citenamefont
  {Koenderink}(2013)}]{Arango2013}%
  \BibitemOpen
  \bibfield  {author} {\bibinfo {author} {\bibfnamefont {F.~B.}\ \bibnamefont
  {Arango}}\ and\ \bibinfo {author} {\bibfnamefont {A.~F.}\ \bibnamefont
  {Koenderink}},\ }\bibfield  {title} {\bibinfo {title} {Polarizability tensor
  retrieval for magnetic and plasmonic antenna design},\ }\href
  {https://doi.org/10.1088/1367-2630/15/7/073023} {\bibfield  {journal}
  {\bibinfo  {journal} {New J. Phys}\ }\textbf {\bibinfo {volume} {15}},\
  \bibinfo {pages} {073023} (\bibinfo {year} {2013})}\BibitemShut {NoStop}%
\bibitem [{\citenamefont {Asadchy}\ \emph {et~al.}(2014)\citenamefont
  {Asadchy}, \citenamefont {Faniayeu}, \citenamefont {Ra’di},\ and\
  \citenamefont {Tretyakov}}]{Asadchy2014}%
  \BibitemOpen
  \bibfield  {author} {\bibinfo {author} {\bibfnamefont {V.~S.}\ \bibnamefont
  {Asadchy}}, \bibinfo {author} {\bibfnamefont {I.~A.}\ \bibnamefont
  {Faniayeu}}, \bibinfo {author} {\bibfnamefont {Y.}~\bibnamefont {Ra’di}},\
  and\ \bibinfo {author} {\bibfnamefont {S.~A.}\ \bibnamefont {Tretyakov}},\
  }\bibfield  {title} {\bibinfo {title} {Determining polarizability tensors for
  an arbitrary small electromagnetic scatterer},\ }\href
  {https://doi.org/https://doi.org/10.1016/j.photonics.2014.04.004} {\bibfield
  {journal} {\bibinfo  {journal} {Photonics Nanostructures: Fundam. Appl}\
  }\textbf {\bibinfo {volume} {12}},\ \bibinfo {pages} {298 } (\bibinfo {year}
  {2014})}\BibitemShut {NoStop}%
\bibitem [{\citenamefont {{Liu}}\ \emph {et~al.}(2016)\citenamefont {{Liu}},
  \citenamefont {{Zhao}},\ and\ \citenamefont {{Alù}}}]{Liu2016}%
  \BibitemOpen
  \bibfield  {author} {\bibinfo {author} {\bibfnamefont {X.}~\bibnamefont
  {{Liu}}}, \bibinfo {author} {\bibfnamefont {Y.}~\bibnamefont {{Zhao}}},\ and\
  \bibinfo {author} {\bibfnamefont {A.}~\bibnamefont {{Alù}}},\ }\bibfield
  {title} {\bibinfo {title} {Polarizability tensor retrieval for subwavelength
  particles of arbitrary shape},\ }\href
  {https://doi.org/10.1109/TAP.2016.2546958} {\bibfield  {journal} {\bibinfo
  {journal} {IEEE Trans. Antennas Propag.}\ }\textbf {\bibinfo {volume} {64}},\
  \bibinfo {pages} {2301} (\bibinfo {year} {2016})}\BibitemShut {NoStop}%
\bibitem [{\citenamefont {Bernal~Arango}\ \emph {et~al.}(2014)\citenamefont
  {Bernal~Arango}, \citenamefont {Coenen},\ and\ \citenamefont
  {Koenderink}}]{Arango2014}%
  \BibitemOpen
  \bibfield  {author} {\bibinfo {author} {\bibfnamefont {F.}~\bibnamefont
  {Bernal~Arango}}, \bibinfo {author} {\bibfnamefont {T.}~\bibnamefont
  {Coenen}},\ and\ \bibinfo {author} {\bibfnamefont {A.~F.}\ \bibnamefont
  {Koenderink}},\ }\bibfield  {title} {\bibinfo {title} {Underpinning
  hybridization intuition for complex nanoantennas by magnetoelectric
  quadrupolar polarizability retrieval},\ }\href
  {https://doi.org/10.1021/ph5000133} {\bibfield  {journal} {\bibinfo
  {journal} {ACS Photonics}\ }\textbf {\bibinfo {volume} {1}},\ \bibinfo
  {pages} {444} (\bibinfo {year} {2014})}\BibitemShut {NoStop}%
\bibitem [{\citenamefont {Mishchenko}\ \emph {et~al.}(1996)\citenamefont
  {Mishchenko}, \citenamefont {Travis},\ and\ \citenamefont
  {Mackowski}}]{Mishchenko1996}%
  \BibitemOpen
  \bibfield  {author} {\bibinfo {author} {\bibfnamefont {M.~I.}\ \bibnamefont
  {Mishchenko}}, \bibinfo {author} {\bibfnamefont {L.~D.}\ \bibnamefont
  {Travis}},\ and\ \bibinfo {author} {\bibfnamefont {D.~W.}\ \bibnamefont
  {Mackowski}},\ }\bibfield  {title} {\bibinfo {title} {T-matrix computations
  of light scattering by nonspherical particles: A review},\ }\href@noop {}
  {\bibfield  {journal} {\bibinfo  {journal} {J. Quant. Spectrosc. Radiat.}\
  }\textbf {\bibinfo {volume} {55}},\ \bibinfo {pages} {535 } (\bibinfo {year}
  {1996})}\BibitemShut {NoStop}%
\bibitem [{\citenamefont {Sersic}\ \emph {et~al.}(2011)\citenamefont {Sersic},
  \citenamefont {Tuambilangana}, \citenamefont {Kampfrath},\ and\ \citenamefont
  {Koenderink}}]{Sersic2011}%
  \BibitemOpen
  \bibfield  {author} {\bibinfo {author} {\bibfnamefont {I.}~\bibnamefont
  {Sersic}}, \bibinfo {author} {\bibfnamefont {C.}~\bibnamefont
  {Tuambilangana}}, \bibinfo {author} {\bibfnamefont {T.}~\bibnamefont
  {Kampfrath}},\ and\ \bibinfo {author} {\bibfnamefont {A.~F.}\ \bibnamefont
  {Koenderink}},\ }\bibfield  {title} {\bibinfo {title} {Magnetoelectric point
  scattering theory for metamaterial scatterers},\ }\href
  {https://doi.org/10.1103/PhysRevB.83.245102} {\bibfield  {journal} {\bibinfo
  {journal} {Phys. Rev. B}\ }\textbf {\bibinfo {volume} {83}},\ \bibinfo
  {pages} {245102} (\bibinfo {year} {2011})}\BibitemShut {NoStop}%
\bibitem [{\citenamefont {Waterman}(1971)}]{Waterman1971}%
  \BibitemOpen
  \bibfield  {author} {\bibinfo {author} {\bibfnamefont {P.~C.}\ \bibnamefont
  {Waterman}},\ }\bibfield  {title} {\bibinfo {title} {Symmetry, unitarity, and
  geometry in electromagnetic scattering},\ }\href
  {https://doi.org/10.1103/PhysRevD.3.825} {\bibfield  {journal} {\bibinfo
  {journal} {Phys. Rev. D}\ }\textbf {\bibinfo {volume} {3}},\ \bibinfo {pages}
  {825} (\bibinfo {year} {1971})}\BibitemShut {NoStop}%
\bibitem [{\citenamefont {Barron}(1986)}]{Barron1986}%
  \BibitemOpen
  \bibfield  {author} {\bibinfo {author} {\bibfnamefont {L.}~\bibnamefont
  {Barron}},\ }\bibfield  {title} {\bibinfo {title} {True and false chirality
  and parity violation},\ }\href@noop {} {\bibfield  {journal} {\bibinfo
  {journal} {Chem. Phys. Lett.}\ }\textbf {\bibinfo {volume} {123}},\ \bibinfo
  {pages} {423 } (\bibinfo {year} {1986})}\BibitemShut {NoStop}%
\bibitem [{\citenamefont {Asadchy}\ and\ \citenamefont
  {Tretyakov}(2019)}]{Asadchy2019}%
  \BibitemOpen
  \bibfield  {author} {\bibinfo {author} {\bibfnamefont {V.~S.}\ \bibnamefont
  {Asadchy}}\ and\ \bibinfo {author} {\bibfnamefont {S.~A.}\ \bibnamefont
  {Tretyakov}},\ }\bibfield  {title} {\bibinfo {title} {Modular analysis of
  arbitrary dipolar scatterers},\ }\href@noop {} {\bibfield  {journal}
  {\bibinfo  {journal} {Phys. Rev. Applied}\ }\textbf {\bibinfo {volume}
  {12}},\ \bibinfo {pages} {024059} (\bibinfo {year} {2019})}\BibitemShut
  {NoStop}%
\bibitem [{\citenamefont {Kuznetsov}\ \emph {et~al.}(2012)\citenamefont
  {Kuznetsov}, \citenamefont {Miroshnichenko}, \citenamefont {Fu},
  \citenamefont {Zhang},\ and\ \citenamefont {Luk`yanchukl}}]{Kuznetsov2012}%
  \BibitemOpen
  \bibfield  {author} {\bibinfo {author} {\bibfnamefont {A.~I.}\ \bibnamefont
  {Kuznetsov}}, \bibinfo {author} {\bibfnamefont {A.~E.}\ \bibnamefont
  {Miroshnichenko}}, \bibinfo {author} {\bibfnamefont {Y.~H.}\ \bibnamefont
  {Fu}}, \bibinfo {author} {\bibfnamefont {J.}~\bibnamefont {Zhang}},\ and\
  \bibinfo {author} {\bibfnamefont {B.}~\bibnamefont {Luk`yanchukl}},\
  }\bibfield  {title} {\bibinfo {title} {{Magnetic light}},\ }\href
  {https://doi.org/10.1038/srep00492} {\bibfield  {journal} {\bibinfo
  {journal} {Sci. Rep.}\ }\textbf {\bibinfo {volume} {2}},\ \bibinfo {pages}
  {492} (\bibinfo {year} {2012})}\BibitemShut {NoStop}%
\bibitem [{\citenamefont {Dolling}\ \emph {et~al.}(2005)\citenamefont
  {Dolling}, \citenamefont {Enkrich}, \citenamefont {Wegener}, \citenamefont
  {Zhou}, \citenamefont {Soukoulis},\ and\ \citenamefont
  {Linden}}]{Dolling2005}%
  \BibitemOpen
  \bibfield  {author} {\bibinfo {author} {\bibfnamefont {G.}~\bibnamefont
  {Dolling}}, \bibinfo {author} {\bibfnamefont {C.}~\bibnamefont {Enkrich}},
  \bibinfo {author} {\bibfnamefont {M.}~\bibnamefont {Wegener}}, \bibinfo
  {author} {\bibfnamefont {J.~F.}\ \bibnamefont {Zhou}}, \bibinfo {author}
  {\bibfnamefont {C.~M.}\ \bibnamefont {Soukoulis}},\ and\ \bibinfo {author}
  {\bibfnamefont {S.}~\bibnamefont {Linden}},\ }\bibfield  {title} {\bibinfo
  {title} {{Cut-wire pairs and plate pairs as magnetic atoms for optical
  metamaterials}},\ }\href {https://doi.org/10.1364/ol.30.003198} {\bibfield
  {journal} {\bibinfo  {journal} {Opt. Lett.}\ }\textbf {\bibinfo {volume}
  {30}},\ \bibinfo {pages} {3198} (\bibinfo {year} {2005})}\BibitemShut
  {NoStop}%
\bibitem [{\citenamefont {Auguié}\ \emph {et~al.}(2011)\citenamefont
  {Auguié}, \citenamefont {Alonso-Gómez}, \citenamefont
  {Guerrero-Martínez},\ and\ \citenamefont {Liz-Marzán}}]{Auguie2011}%
  \BibitemOpen
  \bibfield  {author} {\bibinfo {author} {\bibfnamefont {B.}~\bibnamefont
  {Auguié}}, \bibinfo {author} {\bibfnamefont {J.~L.}\ \bibnamefont
  {Alonso-Gómez}}, \bibinfo {author} {\bibfnamefont {A.}~\bibnamefont
  {Guerrero-Martínez}},\ and\ \bibinfo {author} {\bibfnamefont {L.~M.}\
  \bibnamefont {Liz-Marzán}},\ }\bibfield  {title} {\bibinfo {title} {Fingers
  crossed: Optical activity of a chiral dimer of plasmonic nanorods},\ }\href
  {https://doi.org/10.1021/jz200279x} {\bibfield  {journal} {\bibinfo
  {journal} {J. Phys. Chem. Lett.}\ }\textbf {\bibinfo {volume} {2}},\ \bibinfo
  {pages} {846} (\bibinfo {year} {2011})}\BibitemShut {NoStop}%
\bibitem [{\citenamefont {Mun}\ and\ \citenamefont {Rho}(2019)}]{Mun2019}%
  \BibitemOpen
  \bibfield  {author} {\bibinfo {author} {\bibfnamefont {J.}~\bibnamefont
  {Mun}}\ and\ \bibinfo {author} {\bibfnamefont {J.}~\bibnamefont {Rho}},\
  }\bibfield  {title} {\bibinfo {title} {{Importance of higher-order multipole
  transitions on chiral nearfield interactions}},\ }\href
  {https://doi.org/10.1515/nanoph-2019-0046} {\bibfield  {journal} {\bibinfo
  {journal} {Nanophotonics}\ }\textbf {\bibinfo {volume} {8}},\ \bibinfo
  {pages} {941} (\bibinfo {year} {2019})}\BibitemShut {NoStop}%
\bibitem [{\citenamefont {Augui\'e}\ and\ \citenamefont
  {Barnes}(2008)}]{Auguie2008}%
  \BibitemOpen
  \bibfield  {author} {\bibinfo {author} {\bibfnamefont {B.}~\bibnamefont
  {Augui\'e}}\ and\ \bibinfo {author} {\bibfnamefont {W.~L.}\ \bibnamefont
  {Barnes}},\ }\bibfield  {title} {\bibinfo {title} {Collective resonances in
  gold nanoparticle arrays},\ }\href
  {https://doi.org/10.1103/PhysRevLett.101.143902} {\bibfield  {journal}
  {\bibinfo  {journal} {Phys. Rev. Lett.}\ }\textbf {\bibinfo {volume} {101}},\
  \bibinfo {pages} {143902} (\bibinfo {year} {2008})}\BibitemShut {NoStop}%
\bibitem [{\citenamefont {Swiecicki}\ and\ \citenamefont
  {Sipe}(2017)}]{Swiecicki2017}%
  \BibitemOpen
  \bibfield  {author} {\bibinfo {author} {\bibfnamefont {S.~D.}\ \bibnamefont
  {Swiecicki}}\ and\ \bibinfo {author} {\bibfnamefont {J.~E.}\ \bibnamefont
  {Sipe}},\ }\bibfield  {title} {\bibinfo {title} {Surface-lattice resonances
  in two-dimensional arrays of spheres: Multipolar interactions and a mode
  analysis},\ }\href {https://doi.org/10.1103/PhysRevB.95.195406} {\bibfield
  {journal} {\bibinfo  {journal} {Phys. Rev. B}\ }\textbf {\bibinfo {volume}
  {95}},\ \bibinfo {pages} {195406} (\bibinfo {year} {2017})}\BibitemShut
  {NoStop}%
\bibitem [{\citenamefont {Martikainen}\ \emph {et~al.}(2017)\citenamefont
  {Martikainen}, \citenamefont {Moilanen},\ and\ \citenamefont
  {Törmä}}]{Martikainen2017}%
  \BibitemOpen
  \bibfield  {author} {\bibinfo {author} {\bibfnamefont {J.-P.}\ \bibnamefont
  {Martikainen}}, \bibinfo {author} {\bibfnamefont {A.~J.}\ \bibnamefont
  {Moilanen}},\ and\ \bibinfo {author} {\bibfnamefont {P.}~\bibnamefont
  {Törmä}},\ }\bibfield  {title} {\bibinfo {title} {Coupled dipole
  approximation across the \(\gamma\)-point in a finite-sized nanoparticle
  array},\ }\href {https://doi.org/10.1098/rsta.2016.0316} {\bibfield
  {journal} {\bibinfo  {journal} {Philos. Trans. R. Soc. A}\ }\textbf {\bibinfo
  {volume} {375}},\ \bibinfo {pages} {20160316} (\bibinfo {year}
  {2017})}\BibitemShut {NoStop}%
\bibitem [{\citenamefont {Draine}\ and\ \citenamefont
  {Flatau}(1994)}]{Draine1994}%
  \BibitemOpen
  \bibfield  {author} {\bibinfo {author} {\bibfnamefont {B.~T.}\ \bibnamefont
  {Draine}}\ and\ \bibinfo {author} {\bibfnamefont {P.~J.}\ \bibnamefont
  {Flatau}},\ }\bibfield  {title} {\bibinfo {title} {{Discrete-Dipole
  Approximation For Scattering Calculations}},\ }\href
  {https://doi.org/10.1364/josaa.11.001491} {\bibfield  {journal} {\bibinfo
  {journal} {J. Opt. Soc. Am. A}\ }\textbf {\bibinfo {volume} {11}},\ \bibinfo
  {pages} {1491} (\bibinfo {year} {1994})}\BibitemShut {NoStop}%
\bibitem [{\citenamefont {Kwadrin}\ and\ \citenamefont
  {Koenderink}(2014)}]{Kwadrin2014}%
  \BibitemOpen
  \bibfield  {author} {\bibinfo {author} {\bibfnamefont {A.}~\bibnamefont
  {Kwadrin}}\ and\ \bibinfo {author} {\bibfnamefont {A.~F.}\ \bibnamefont
  {Koenderink}},\ }\bibfield  {title} {\bibinfo {title} {Diffractive stacks of
  metamaterial lattices with a complex unit cell: Self-consistent long-range
  bianisotropic interactions in experiment and theory},\ }\href
  {https://doi.org/10.1103/PhysRevB.89.045120} {\bibfield  {journal} {\bibinfo
  {journal} {Phys. Rev. B}\ }\textbf {\bibinfo {volume} {89}},\ \bibinfo
  {pages} {045120} (\bibinfo {year} {2014})}\BibitemShut {NoStop}%
\bibitem [{\citenamefont {Liu}\ \emph {et~al.}(2008)\citenamefont {Liu},
  \citenamefont {Guo}, \citenamefont {Fu}, \citenamefont {Kaiser},
  \citenamefont {Schweizer},\ and\ \citenamefont {Giessen}}]{Liu2008}%
  \BibitemOpen
  \bibfield  {author} {\bibinfo {author} {\bibfnamefont {N.}~\bibnamefont
  {Liu}}, \bibinfo {author} {\bibfnamefont {H.}~\bibnamefont {Guo}}, \bibinfo
  {author} {\bibfnamefont {L.}~\bibnamefont {Fu}}, \bibinfo {author}
  {\bibfnamefont {S.}~\bibnamefont {Kaiser}}, \bibinfo {author} {\bibfnamefont
  {H.}~\bibnamefont {Schweizer}},\ and\ \bibinfo {author} {\bibfnamefont
  {H.}~\bibnamefont {Giessen}},\ }\bibfield  {title} {\bibinfo {title}
  {{Three-dimensional photonic metamaterials at optical frequencies}},\ }\href
  {https://doi.org/10.1038/nmat2072} {\bibfield  {journal} {\bibinfo  {journal}
  {Nat. Mater.}\ }\textbf {\bibinfo {volume} {7}},\ \bibinfo {pages} {31}
  (\bibinfo {year} {2008})}\BibitemShut {NoStop}%
\bibitem [{\citenamefont {Kim}\ \emph {et~al.}(2017)\citenamefont {Kim},
  \citenamefont {Yoo}, \citenamefont {Huh}, \citenamefont {Park},\ and\
  \citenamefont {Lee}}]{Kim2017}%
  \BibitemOpen
  \bibfield  {author} {\bibinfo {author} {\bibfnamefont {K.}~\bibnamefont
  {Kim}}, \bibinfo {author} {\bibfnamefont {S.}~\bibnamefont {Yoo}}, \bibinfo
  {author} {\bibfnamefont {J.-H.}\ \bibnamefont {Huh}}, \bibinfo {author}
  {\bibfnamefont {Q.-H.}\ \bibnamefont {Park}},\ and\ \bibinfo {author}
  {\bibfnamefont {S.}~\bibnamefont {Lee}},\ }\bibfield  {title} {\bibinfo
  {title} {Limitations and opportunities for optical metafluids to achieve an
  unnatural refractive index},\ }\href
  {https://doi.org/10.1021/acsphotonics.7b00546} {\bibfield  {journal}
  {\bibinfo  {journal} {ACS Photonics}\ }\textbf {\bibinfo {volume} {4}},\
  \bibinfo {pages} {2298} (\bibinfo {year} {2017})}\BibitemShut {NoStop}%
\bibitem [{\citenamefont {Nordlander}\ \emph {et~al.}(2004)\citenamefont
  {Nordlander}, \citenamefont {Oubre}, \citenamefont {Prodan}, \citenamefont
  {Li},\ and\ \citenamefont {Stockman}}]{Nordlander2004}%
  \BibitemOpen
  \bibfield  {author} {\bibinfo {author} {\bibfnamefont {P.}~\bibnamefont
  {Nordlander}}, \bibinfo {author} {\bibfnamefont {C.}~\bibnamefont {Oubre}},
  \bibinfo {author} {\bibfnamefont {E.}~\bibnamefont {Prodan}}, \bibinfo
  {author} {\bibfnamefont {K.}~\bibnamefont {Li}},\ and\ \bibinfo {author}
  {\bibfnamefont {M.~I.}\ \bibnamefont {Stockman}},\ }\bibfield  {title}
  {\bibinfo {title} {Plasmon hybridization in nanoparticle dimers},\ }\href
  {https://doi.org/10.1021/nl049681c} {\bibfield  {journal} {\bibinfo
  {journal} {Nano Lett.}\ }\textbf {\bibinfo {volume} {4}},\ \bibinfo {pages}
  {899} (\bibinfo {year} {2004})}\BibitemShut {NoStop}%
\bibitem [{\citenamefont {Park}(2014)}]{Park2014}%
  \BibitemOpen
  \bibfield  {author} {\bibinfo {author} {\bibfnamefont {W.}~\bibnamefont
  {Park}},\ }\bibfield  {title} {\bibinfo {title} {{Optical interactions in
  plasmonic nanostructures}},\ }\href
  {https://doi.org/10.1186/s40580-014-0002-x} {\bibfield  {journal} {\bibinfo
  {journal} {Nano Converg.}\ }\textbf {\bibinfo {volume} {1}},\ \bibinfo
  {pages} {2} (\bibinfo {year} {2014})}\BibitemShut {NoStop}%
\bibitem [{\citenamefont {Pinheiro}\ \emph {et~al.}(2017)\citenamefont
  {Pinheiro}, \citenamefont {Fedotov}, \citenamefont {Papasimakis},\ and\
  \citenamefont {Zheludev}}]{Pinheiro2017}%
  \BibitemOpen
  \bibfield  {author} {\bibinfo {author} {\bibfnamefont {F.~A.}\ \bibnamefont
  {Pinheiro}}, \bibinfo {author} {\bibfnamefont {V.~A.}\ \bibnamefont
  {Fedotov}}, \bibinfo {author} {\bibfnamefont {N.}~\bibnamefont
  {Papasimakis}},\ and\ \bibinfo {author} {\bibfnamefont {N.~I.}\ \bibnamefont
  {Zheludev}},\ }\bibfield  {title} {\bibinfo {title} {Spontaneous natural
  optical activity in disordered media},\ }\href
  {https://doi.org/10.1103/PhysRevB.95.220201} {\bibfield  {journal} {\bibinfo
  {journal} {Phys. Rev. B}\ }\textbf {\bibinfo {volume} {95}},\ \bibinfo
  {pages} {220201} (\bibinfo {year} {2017})}\BibitemShut {NoStop}%
\bibitem [{\citenamefont {Forestiere}\ \emph {et~al.}(2012)\citenamefont
  {Forestiere}, \citenamefont {Pasquale}, \citenamefont {Capretti},
  \citenamefont {Miano}, \citenamefont {Tamburrino}, \citenamefont {Lee},
  \citenamefont {Reinhard},\ and\ \citenamefont {Dal~Negro}}]{Forestiere2012}%
  \BibitemOpen
  \bibfield  {author} {\bibinfo {author} {\bibfnamefont {C.}~\bibnamefont
  {Forestiere}}, \bibinfo {author} {\bibfnamefont {A.~J.}\ \bibnamefont
  {Pasquale}}, \bibinfo {author} {\bibfnamefont {A.}~\bibnamefont {Capretti}},
  \bibinfo {author} {\bibfnamefont {G.}~\bibnamefont {Miano}}, \bibinfo
  {author} {\bibfnamefont {A.}~\bibnamefont {Tamburrino}}, \bibinfo {author}
  {\bibfnamefont {S.~Y.}\ \bibnamefont {Lee}}, \bibinfo {author} {\bibfnamefont
  {B.~M.}\ \bibnamefont {Reinhard}},\ and\ \bibinfo {author} {\bibfnamefont
  {L.}~\bibnamefont {Dal~Negro}},\ }\bibfield  {title} {\bibinfo {title}
  {Genetically engineered plasmonic nanoarrays},\ }\href
  {https://doi.org/10.1021/nl300140g} {\bibfield  {journal} {\bibinfo
  {journal} {Nano Lett.}\ }\textbf {\bibinfo {volume} {12}},\ \bibinfo {pages}
  {2037} (\bibinfo {year} {2012})}\BibitemShut {NoStop}%
\bibitem [{\citenamefont {So}\ \emph {et~al.}(2019)\citenamefont {So},
  \citenamefont {Mun},\ and\ \citenamefont {Rho}}]{So2019}%
  \BibitemOpen
  \bibfield  {author} {\bibinfo {author} {\bibfnamefont {S.}~\bibnamefont
  {So}}, \bibinfo {author} {\bibfnamefont {J.}~\bibnamefont {Mun}},\ and\
  \bibinfo {author} {\bibfnamefont {J.}~\bibnamefont {Rho}},\ }\bibfield
  {title} {\bibinfo {title} {Simultaneous inverse design of materials and
  structures via deep learning: Demonstration of dipole resonance engineering
  using core–shell nanoparticles},\ }\href
  {https://doi.org/10.1021/acsami.9b05857} {\bibfield  {journal} {\bibinfo
  {journal} {ACS Appl. Mater. Interfaces}\ }\textbf {\bibinfo {volume} {11}},\
  \bibinfo {pages} {24264} (\bibinfo {year} {2019})}\BibitemShut {NoStop}%
\bibitem [{\citenamefont {Novotny}\ and\ \citenamefont
  {Hecht}(2006)}]{Novotny2009}%
  \BibitemOpen
  \bibfield  {author} {\bibinfo {author} {\bibfnamefont {L.}~\bibnamefont
  {Novotny}}\ and\ \bibinfo {author} {\bibfnamefont {B.}~\bibnamefont
  {Hecht}},\ }\href {https://doi.org/10.1017/CBO9780511813535} {\emph {\bibinfo
  {title} {Principles of Nano-Optics}}}\ (\bibinfo  {publisher} {Cambridge
  University Press},\ \bibinfo {year} {2006})\BibitemShut {NoStop}%
\bibitem [{\citenamefont {Neves}\ \emph {et~al.}(2006)\citenamefont {Neves},
  \citenamefont {Fontes}, \citenamefont {Padilha}, \citenamefont {Rodriguez},
  \citenamefont {de~Brito~Cruz}, \citenamefont {Barbosa},\ and\ \citenamefont
  {Cesar}}]{AlvaroRanhaNeves2006}%
  \BibitemOpen
  \bibfield  {author} {\bibinfo {author} {\bibfnamefont {A.~A.~R.}\
  \bibnamefont {Neves}}, \bibinfo {author} {\bibfnamefont {A.}~\bibnamefont
  {Fontes}}, \bibinfo {author} {\bibfnamefont {L.~A.}\ \bibnamefont {Padilha}},
  \bibinfo {author} {\bibfnamefont {E.}~\bibnamefont {Rodriguez}}, \bibinfo
  {author} {\bibfnamefont {C.~H.}\ \bibnamefont {de~Brito~Cruz}}, \bibinfo
  {author} {\bibfnamefont {L.~C.}\ \bibnamefont {Barbosa}},\ and\ \bibinfo
  {author} {\bibfnamefont {C.~L.}\ \bibnamefont {Cesar}},\ }\bibfield  {title}
  {\bibinfo {title} {Exact partial wave expansion of optical beams with respect
  to an arbitrary origin},\ }\href {https://doi.org/10.1364/OL.31.002477}
  {\bibfield  {journal} {\bibinfo  {journal} {Opt. Lett.}\ }\textbf {\bibinfo
  {volume} {31}},\ \bibinfo {pages} {2477} (\bibinfo {year}
  {2006})}\BibitemShut {NoStop}%
\bibitem [{\citenamefont {Wang}\ \emph {et~al.}(2017)\citenamefont {Wang},
  \citenamefont {Wriedt}, \citenamefont {Mädler}, \citenamefont {Han},\ and\
  \citenamefont {Hartmann}}]{Wang2017}%
  \BibitemOpen
  \bibfield  {author} {\bibinfo {author} {\bibfnamefont {J.~J.}\ \bibnamefont
  {Wang}}, \bibinfo {author} {\bibfnamefont {T.}~\bibnamefont {Wriedt}},
  \bibinfo {author} {\bibfnamefont {L.}~\bibnamefont {Mädler}}, \bibinfo
  {author} {\bibfnamefont {Y.~P.}\ \bibnamefont {Han}},\ and\ \bibinfo {author}
  {\bibfnamefont {P.}~\bibnamefont {Hartmann}},\ }\bibfield  {title} {\bibinfo
  {title} {Multipole expansion of circularly symmetric bessel beams of
  arbitrary order for scattering calculations},\ }\href@noop {} {\bibfield
  {journal} {\bibinfo  {journal} {Opt. Commun.}\ }\textbf {\bibinfo {volume}
  {387}},\ \bibinfo {pages} {102} (\bibinfo {year} {2017})}\BibitemShut
  {NoStop}%
\bibitem [{\citenamefont {Fernandez-Corbaton}\ \emph
  {et~al.}(2016)\citenamefont {Fernandez-Corbaton}, \citenamefont {Fruhnert},\
  and\ \citenamefont {Rockstuhl}}]{Fernandez-Corbaton2016}%
  \BibitemOpen
  \bibfield  {author} {\bibinfo {author} {\bibfnamefont {I.}~\bibnamefont
  {Fernandez-Corbaton}}, \bibinfo {author} {\bibfnamefont {M.}~\bibnamefont
  {Fruhnert}},\ and\ \bibinfo {author} {\bibfnamefont {C.}~\bibnamefont
  {Rockstuhl}},\ }\bibfield  {title} {\bibinfo {title} {Objects of maximum
  electromagnetic chirality},\ }\href
  {https://doi.org/10.1103/PhysRevX.6.031013} {\bibfield  {journal} {\bibinfo
  {journal} {Phys. Rev. X}\ }\textbf {\bibinfo {volume} {6}},\ \bibinfo {pages}
  {031013} (\bibinfo {year} {2016})}\BibitemShut {NoStop}%
\bibitem [{\citenamefont {Wu}\ \emph {et~al.}(2012)\citenamefont {Wu},
  \citenamefont {Shang},\ and\ \citenamefont {Li}}]{Wu2012}%
  \BibitemOpen
  \bibfield  {author} {\bibinfo {author} {\bibfnamefont {Z.~S.}\ \bibnamefont
  {Wu}}, \bibinfo {author} {\bibfnamefont {Q.~C.}\ \bibnamefont {Shang}},\ and\
  \bibinfo {author} {\bibfnamefont {Z.~J.}\ \bibnamefont {Li}},\ }\bibfield
  {title} {\bibinfo {title} {{Calculation of electromagnetic scattering by a
  large chiral sphere}},\ }\href {https://doi.org/10.1364/AO.51.006661}
  {\bibfield  {journal} {\bibinfo  {journal} {Appl. Opt.}\ }\textbf {\bibinfo
  {volume} {51}},\ \bibinfo {pages} {6661} (\bibinfo {year}
  {2012})}\BibitemShut {NoStop}%
\bibitem [{\citenamefont {Stout}\ \emph {et~al.}(2007)\citenamefont {Stout},
  \citenamefont {Nevi{\`{e}}re},\ and\ \citenamefont {Popov}}]{Stout2007}%
  \BibitemOpen
  \bibfield  {author} {\bibinfo {author} {\bibfnamefont {B.}~\bibnamefont
  {Stout}}, \bibinfo {author} {\bibfnamefont {M.}~\bibnamefont
  {Nevi{\`{e}}re}},\ and\ \bibinfo {author} {\bibfnamefont {E.}~\bibnamefont
  {Popov}},\ }\bibfield  {title} {\bibinfo {title} {{T matrix of the
  homogeneous anisotropic sphere: applications to orientation-averaged resonant
  scattering}},\ }\href {https://doi.org/10.1364/josaa.24.001120} {\bibfield
  {journal} {\bibinfo  {journal} {J. Opt. Soc. Am. A}\ }\textbf {\bibinfo
  {volume} {24}},\ \bibinfo {pages} {1120} (\bibinfo {year}
  {2007})}\BibitemShut {NoStop}%
\bibitem [{\citenamefont {David}\ and\ \citenamefont {García~de
  Abajo}(2011)}]{David2011}%
  \BibitemOpen
  \bibfield  {author} {\bibinfo {author} {\bibfnamefont {C.}~\bibnamefont
  {David}}\ and\ \bibinfo {author} {\bibfnamefont {F.~J.}\ \bibnamefont
  {García~de Abajo}},\ }\bibfield  {title} {\bibinfo {title} {Spatial
  nonlocality in the optical response of metal nanoparticles},\ }\href
  {https://doi.org/10.1021/jp204261u} {\bibfield  {journal} {\bibinfo
  {journal} {J. Phys. Chem. C.}\ }\textbf {\bibinfo {volume} {115}},\ \bibinfo
  {pages} {19470} (\bibinfo {year} {2011})}\BibitemShut {NoStop}%
\bibitem [{\citenamefont {Bohren}\ and\ \citenamefont
  {Huffman}(1998)}]{Mishchenko2002}%
  \BibitemOpen
  \bibfield  {author} {\bibinfo {author} {\bibfnamefont {C.}~\bibnamefont
  {Bohren}}\ and\ \bibinfo {author} {\bibfnamefont {D.~R.}\ \bibnamefont
  {Huffman}},\ }\href@noop {} {\emph {\bibinfo {title} {Absorption and
  Scattering of Light by Small Particles}}}\ (\bibinfo  {publisher}
  {Wiley-VCH},\ \bibinfo {address} {Weinheim},\ \bibinfo {year}
  {1998})\BibitemShut {NoStop}%
\bibitem [{\citenamefont {Doicu}\ \emph {et~al.}(2006)\citenamefont {Doicu},
  \citenamefont {Wriedt},\ and\ \citenamefont {Eremin}}]{Doicu2006}%
  \BibitemOpen
  \bibfield  {author} {\bibinfo {author} {\bibfnamefont {A.}~\bibnamefont
  {Doicu}}, \bibinfo {author} {\bibfnamefont {T.}~\bibnamefont {Wriedt}},\ and\
  \bibinfo {author} {\bibfnamefont {Y.~A.}\ \bibnamefont {Eremin}},\
  }\href@noop {} {\emph {\bibinfo {title} {{Light Scattering by Systems of
  Particles}}}}\ (\bibinfo  {publisher} {Springer},\ \bibinfo {address}
  {Berlin},\ \bibinfo {year} {2006})\BibitemShut {NoStop}%
\bibitem [{\citenamefont {Sersic}\ \emph {et~al.}(2012)\citenamefont {Sersic},
  \citenamefont {van~de Haar}, \citenamefont {Arango},\ and\ \citenamefont
  {Koenderink}}]{Sersic2012}%
  \BibitemOpen
  \bibfield  {author} {\bibinfo {author} {\bibfnamefont {I.}~\bibnamefont
  {Sersic}}, \bibinfo {author} {\bibfnamefont {M.~A.}\ \bibnamefont {van~de
  Haar}}, \bibinfo {author} {\bibfnamefont {F.~B.}\ \bibnamefont {Arango}},\
  and\ \bibinfo {author} {\bibfnamefont {A.~F.}\ \bibnamefont {Koenderink}},\
  }\bibfield  {title} {\bibinfo {title} {Ubiquity of optical activity in planar
  metamaterial scatterers},\ }\href
  {https://doi.org/10.1103/PhysRevLett.108.223903} {\bibfield  {journal}
  {\bibinfo  {journal} {Phys. Rev. Lett.}\ }\textbf {\bibinfo {volume} {108}},\
  \bibinfo {pages} {223903} (\bibinfo {year} {2012})}\BibitemShut {NoStop}%
\bibitem [{\citenamefont {Hu}\ \emph {et~al.}(2016)\citenamefont {Hu},
  \citenamefont {Tian}, \citenamefont {Huang}, \citenamefont {Fang},\ and\
  \citenamefont {Fang}}]{Hu2016}%
  \BibitemOpen
  \bibfield  {author} {\bibinfo {author} {\bibfnamefont {L.}~\bibnamefont
  {Hu}}, \bibinfo {author} {\bibfnamefont {X.}~\bibnamefont {Tian}}, \bibinfo
  {author} {\bibfnamefont {Y.}~\bibnamefont {Huang}}, \bibinfo {author}
  {\bibfnamefont {L.}~\bibnamefont {Fang}},\ and\ \bibinfo {author}
  {\bibfnamefont {Y.}~\bibnamefont {Fang}},\ }\bibfield  {title} {\bibinfo
  {title} {{Quantitatively analyzing the mechanism of giant circular dichroism
  in extrinsic plasmonic chiral nanostructures by tracking the interplay of
  electric and magnetic dipoles}},\ }\href {https://doi.org/10.1039/c5nr08527f}
  {\bibfield  {journal} {\bibinfo  {journal} {Nanoscale}\ }\textbf {\bibinfo
  {volume} {8}},\ \bibinfo {pages} {3720} (\bibinfo {year} {2016})}\BibitemShut
  {NoStop}%
\bibitem [{\citenamefont {Lee}\ \emph {et~al.}(2018)\citenamefont {Lee},
  \citenamefont {Ahn}, \citenamefont {Mun}, \citenamefont {Lee}, \citenamefont
  {Kim}, \citenamefont {Cho}, \citenamefont {Chang}, \citenamefont {Kim},
  \citenamefont {Rho},\ and\ \citenamefont {Nam}}]{Lee2018}%
  \BibitemOpen
  \bibfield  {author} {\bibinfo {author} {\bibfnamefont {H.~E.}\ \bibnamefont
  {Lee}}, \bibinfo {author} {\bibfnamefont {H.~Y.}\ \bibnamefont {Ahn}},
  \bibinfo {author} {\bibfnamefont {J.}~\bibnamefont {Mun}}, \bibinfo {author}
  {\bibfnamefont {Y.~Y.}\ \bibnamefont {Lee}}, \bibinfo {author} {\bibfnamefont
  {M.}~\bibnamefont {Kim}}, \bibinfo {author} {\bibfnamefont {N.~H.}\
  \bibnamefont {Cho}}, \bibinfo {author} {\bibfnamefont {K.}~\bibnamefont
  {Chang}}, \bibinfo {author} {\bibfnamefont {W.~S.}\ \bibnamefont {Kim}},
  \bibinfo {author} {\bibfnamefont {J.}~\bibnamefont {Rho}},\ and\ \bibinfo
  {author} {\bibfnamefont {K.~T.}\ \bibnamefont {Nam}},\ }\bibfield  {title}
  {\bibinfo {title} {{Amino-acid- and peptide-directed synthesis of chiral
  plasmonic gold nanoparticles}},\ }\href
  {https://doi.org/10.1038/s41586-018-0034-1} {\bibfield  {journal} {\bibinfo
  {journal} {Nature}\ }\textbf {\bibinfo {volume} {556}},\ \bibinfo {pages}
  {360} (\bibinfo {year} {2018})}\BibitemShut {NoStop}%
\bibitem [{\citenamefont {Moroz}(2009)}]{Moroz2009}%
  \BibitemOpen
  \bibfield  {author} {\bibinfo {author} {\bibfnamefont {A.}~\bibnamefont
  {Moroz}},\ }\bibfield  {title} {\bibinfo {title} {{Depolarization field of
  spheroidal particles}},\ }\href {https://doi.org/10.1364/josab.26.000517}
  {\bibfield  {journal} {\bibinfo  {journal} {J. Opt. Soc. Am. B}\ }\textbf
  {\bibinfo {volume} {26}},\ \bibinfo {pages} {517} (\bibinfo {year}
  {2009})}\BibitemShut {NoStop}%
\bibitem [{\citenamefont {Maji{\'{c}}}\ \emph {et~al.}(2017)\citenamefont
  {Maji{\'{c}}}, \citenamefont {Gray}, \citenamefont {Augui{\'{e}}},\ and\
  \citenamefont {Ru}}]{Majic2017}%
  \BibitemOpen
  \bibfield  {author} {\bibinfo {author} {\bibfnamefont {M.~R.}\ \bibnamefont
  {Maji{\'{c}}}}, \bibinfo {author} {\bibfnamefont {F.}~\bibnamefont {Gray}},
  \bibinfo {author} {\bibfnamefont {B.}~\bibnamefont {Augui{\'{e}}}},\ and\
  \bibinfo {author} {\bibfnamefont {E.~C.}\ \bibnamefont {Ru}},\ }\bibfield
  {title} {\bibinfo {title} {{Electrostatic limit of the T-matrix for
  electromagnetic scattering: Exact results for spheroidal particles}},\ }\href
  {https://doi.org/10.1016/j.jqsrt.2017.05.031} {\bibfield  {journal} {\bibinfo
   {journal} {J. Quant. Spectrosc. Radiat.}\ }\textbf {\bibinfo {volume}
  {200}},\ \bibinfo {pages} {50} (\bibinfo {year} {2017})}\BibitemShut
  {NoStop}%
\bibitem [{\citenamefont {Canaguier-Durand}\ and\ \citenamefont
  {Genet}(2015)}]{Canaguier-Durand2015}%
  \BibitemOpen
  \bibfield  {author} {\bibinfo {author} {\bibfnamefont {A.}~\bibnamefont
  {Canaguier-Durand}}\ and\ \bibinfo {author} {\bibfnamefont {C.}~\bibnamefont
  {Genet}},\ }\bibfield  {title} {\bibinfo {title} {Chiral route to pulling
  optical forces and left-handed optical torques},\ }\href
  {https://doi.org/10.1103/PhysRevA.92.043823} {\bibfield  {journal} {\bibinfo
  {journal} {Phys. Rev. A}\ }\textbf {\bibinfo {volume} {92}},\ \bibinfo
  {pages} {043823} (\bibinfo {year} {2015})}\BibitemShut {NoStop}%
\end{thebibliography}%


%
\end{document}